\documentclass[fleqn,usenatbib,useAMS]{Papers}
\usepackage{newtxtext,newtxmath}
\usepackage[T1]{fontenc}
\usepackage{subfloat}
\usepackage{makecell}

\usepackage[authoryear]{natbib}

\usepackage{graphicx}	
\usepackage{amsmath}	
\usepackage{lipsum}
\usepackage{caption}
\usepackage{comment}
\usepackage{caption}
\usepackage{lineno}

\title[Universal spectrum and scaling laws]{Universal spectrum and scaling laws for halo mass function and structure}

\author[Z. Xu]{Zhijie (Jay) Xu,$^{1}$\thanks{E-mail: \href{mailto:zhijie.xu@pnnl.gov}{zhijie.xu@pnnl.gov}; \href{mailto:zhijiexu@hotmail.com}{zhijiexu@hotmail.com}}
\\
$^{1}$Physical and Computational Sciences Directorate, Pacific Northwest National Laboratory; Richland, WA 99354, USA\\
}

\date{Accepted XXX. Received YYY; in original form ZZZ}

\pubyear{2025}

\begin{document}
\label{firstpage}
\pagerange{\pageref{firstpage}--\pageref{lastpage}}
\maketitle
\begin{abstract}
We propose a universal transition range between the linear and nonlinear regimes, centered on a characteristic halo mass $m_h^*\propto t$, within which gravitational dynamics self-organize the matter field into a spectrum with an effective index n=-1. In a bottom-up hierarchy, low-mass halos collapse early and retain strong primordial imprints, while prolonged assembly of large halos near $m_h^*$ tends to erase that memory and establishes a universal spectrum. A generalized "cascade", rather than in a viscous turbulence, is used to describe the redistribution of mass and energy across scales, which enables the universality and universal scaling laws in this range. Globally, a cascade redistributes mass among haloes, leading to a random walk of haloes with a mass-dependent waiting time $\tau_g\propto m_h^{-\lambda}$. A universal mass function is naturally given by the Fokker-Planck equation as a stretched Gaussian with $f_M\propto m_h^{-\lambda}$ and $\lambda=2/3$. Locally, within nonequilibrium haloes, a radially directed cascade sets the particle distribution, leading to a random walk of particles with a waiting time $\tau_{gr}\propto r^{-\gamma}$. Analytical solutions suggest a universal density $\rho_r\propto r^{-2\gamma}$ with $\gamma=2/3$ for haloes near $m_h^*$. On both levels, the cascade drives the nonequilibrium system toward a statistically steady state that continuously releases energy and maximizes entropy. A defining property of that state is scale-independent rates $\varepsilon_m\propto t^{-2/3}$ and $\varepsilon_u\propto t^0$, such that no net accumulation of mass or energy occurs on any intermediate scale. The slopes of both the mass function and the density are directly related to the exponents $\lambda$ and $\gamma$ for the waiting time. The dominance of the primordial spectrum and gravity on different scales implies two effective values of $\lambda$ and $\gamma$, leading to a double-$\lambda$ mass function and a double-$\gamma$ density that agrees with the simulations very well. A small-scale permanence is identified, where the density of different halo masses and redshifts converges to the same -4/3 scaling. Using Illustris and Virgo simulations, we show that kinetic energy is transferred inversely from small to large scales at a rate of $\varepsilon_u \approx -10^{-7}$m$^2$/s$^3$, while potential energy undergoes a direct cascade from large to small scales at $-1.4\varepsilon_u$. The global decline of total energy is enabled by the cascade: the energy is injected around $m_h^*$, transferred from large to small scales, and ultimately "dissipated" at a rate of -0.4$\varepsilon_u$ through the merging of haloes and particle migrations. The dependence of waiting time and step length on the particle mass suggests new mass constraints near $10^{12}$GeV.

\end{abstract}

\begin{keywords}
\vspace*{-19pt}
Dark matter; N-body simulations; Theoretical models; Universal spectrum; Scaling laws;
\end{keywords}

\begingroup
\let\clearpage\relax
\tableofcontents
\endgroup
\newpage

\section{Introduction}
\label{sec:1}

Collisionless systems often have properties that suggest common physical principles governing their motion and evolution. The self-gravitating collisionless fluid dynamics (SG-CFD) is the study of these principles for the motion of collisionless matter under the influence of gravity. A typical example is the gravitational collapse of collisionless dark matter, a non-equilibrium self-gravitating collisionless flow problem \citep{Lukic:2007-The-halo-mass-function--High-r}. Gravitational instability leads to the self-organization of collisionless dark matter particles and the structure formation and evolution. Within a CDM paradigm (cold dark matter), the initial density fluctuation has a larger amplitude at smaller scales \citep{Blumenthal:1984-Formation-of-Galaxies-and-Larg}. The formation of structures starts from the gravitational collapse of small-scale density fluctuations and proceeds hierarchically so that small structures coalesce into large structures in a "bottom-up" fashion. 

The hierarchical structure formation has been studied in great detail since the 1980s using cosmological N-body simulations of increasing resolution and fidelity. The vast majority of these simulations adopt Gaussian initial conditions with a linear power spectrum of cold, warm, or hot dark matter variants. Evolution from these initial conditions is governed by the coupled collisionless Boltzmann (Vlasov) and Poisson equations. Consequently, any statistical properties are primarily consequences of the Gaussian statistics together with the shape of the initial linear power spectrum. This is reflected in most analytic and/or semi-analytical approaches, such as Press–Schechter formalisms and merger trees, where the abundance and internal structures of halos are direct reflections of the initial spectrum and cosmological conditions.

In the linear regime, gravity drives the linear amplification of density perturbations. The growth rate of perturbations is scale-independent, i.e., each mode grows independently and proportionally to the linear spectrum as $\propto a^2$, where $a$ is the scale factor. Consequently, the abundance of the halo and the internal structure retain quantitative memory of the initial distribution of fluctuations. From this perspective, structure formation reflects a continuous interplay between gravitational dynamics and the statistical imprint of the initial power spectrum: gravity governs the dynamics of collapse, while the initial spectrum determines the sequence and characteristic scales at which the collapse unfolds. 

In the nonlinear regime, because of the "bottom-up" nature, smaller haloes form earlier under conditions that are more directly reflective of the initial spectrum. Massive haloes are formed later through the continuous hierarchical merging and smooth accretion of matter. This prolonged and violent assembly history, which includes many mergers and phases of violent relaxation, acts to "wash out" the imprints of the initial spectrum. Therefore, small scales retain more memory of the shape of the initial spectrum than large scales. Merging enabled mass and energy redistribution (or "cascade") across different scales progressively erases the imprint of the primordial spectrum. In this paper, we thereby propose that with increasing scale or mass of haloes, gravitational dynamics ultimately dominate, establishing a quasi-universal range where the shape of the spectrum is fully determined by gravity and no longer depends on the initial conditions. The enabling process, a "cascade" of mass and energy across different scales, is also proposed and formulated.

\begin{figure}
\includegraphics*[width=\columnwidth]{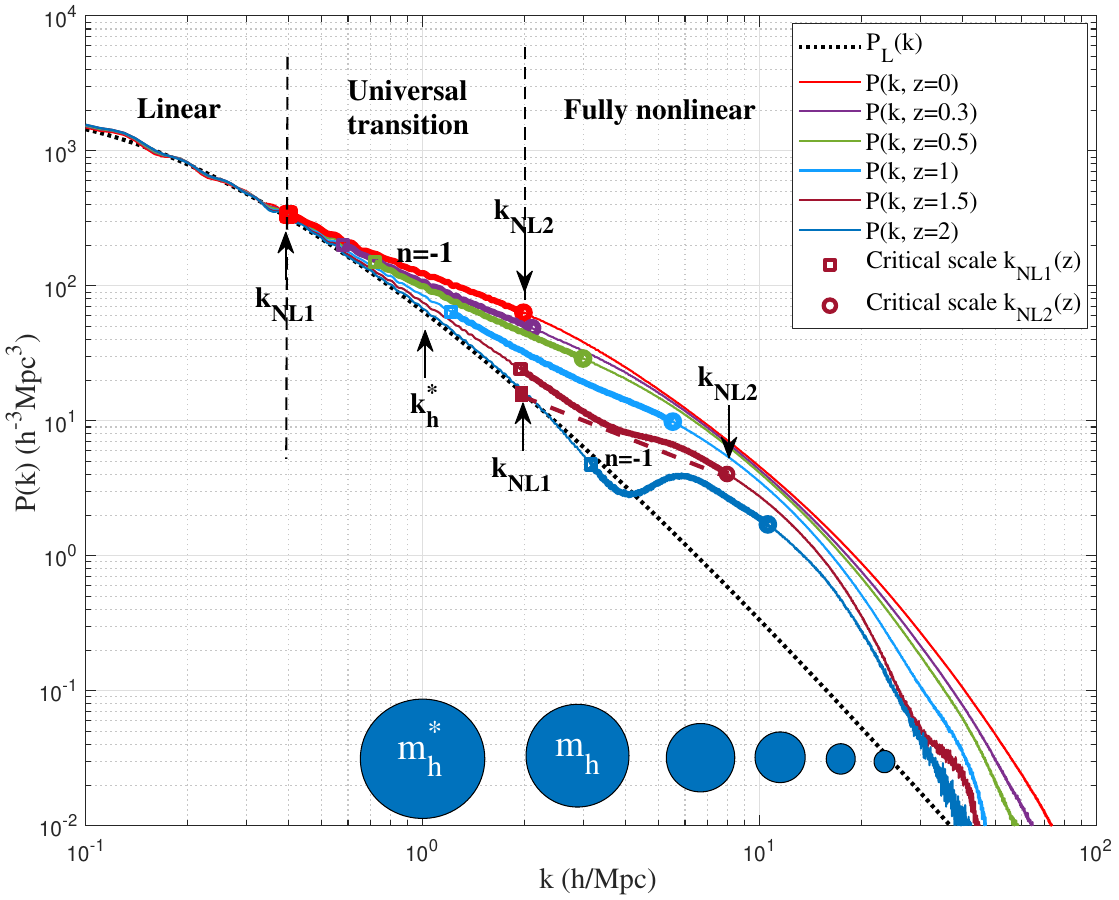}
\caption{The density power spectrum $P(k,z)$ with comoving wavenumber $k$ at different redshifts \textit{z} from Virgo SCDM simulations. The dotted line denotes the linear power spectrum $P_L(k)$. The figure presents three distinct regimes and two critical scales: 1) a linear regime on scales $k<k_{NL1}$, where scale $k_{NL1}$ denotes the deviation from the linear theory; 2) a universal transition range (thick solid lines) on scales $k_{NL1}<k<k_{NL2}$ with $P(k)\propto k^{-1}$, where the gravitational dynamics dominate over the the initial spectrum; 3) a fully nonlinear range on scales $k>k_{NL2}$ that retains the memory of the initial spectrum. While both scales decrease with time, the ratio $k_{NL2}/k_{NL1}$ increases with time, reflecting the expanding universal range. The scale $k_h^*\approx$ 1h/Mpc corresponds to the size of haloes of characteristic mass $m^*_h$ at z=0, such that these haloes are primarily dominated by the gravitational dynamics, not the initial conditions. This paper focuses on the universal range and associated mass and energy cascade across scales, and their effects on structure formation and evolution.} 
\label{fig:999}
\end{figure}

To better demonstrate the idea, Figure \ref{fig:999} presents the evolution of the density power spectrum $P(k,z)$ with redshifts \textit{z} from the Virgo SCDM simulations \citep{Frenk:2000-Public-Release-of-N-body-simul}. The linear power spectrum $P_L(k)$ is plotted as a dotted line. Three distinct regimes and two critical scales can be identified: 1) a linear regime on scales $k<k_{NL1}$, where scale $k_{NL1}$ is determined by the linear dimensionless power spectrum $\Delta_L^2(k_{NL1})=1$, at which fluctuations are just turning nonlinear; 2) a universal transition range (thick solid lines) on scales $k_{NL1}<k<k_{NL2}$ with a spectrum index $n\approx$-1, where gravitational dynamics dominate over the effects of the initial spectrum. The critical scale $k_{NL2}$ is determined by extending the spectrum $n$=-1 (dashed line) from $k_{NL1}$ to intersect with the actual spectrum $P(k,z)$ at $k_{NL2}$; 3) a fully nonlinear range on scales $k>k_{NL2}$ that retains the full memory of the shape of the initial spectrum. The universal range with $n$=-1 is gradually formed at a high redshift and expands with time, with an increasing ratio $k_{NL2}/k_{NL1}\propto a^{1/2}$. The scale $k_h^*\approx$ 1h/Mpc corresponds to the size of the halo of the characteristic mass $m^*_h$ at z=0, which is located in the universal transition range with $k_{NL1}<k_h^*<k_{NL2}$. The haloes of mass $m_h^*$ rapidly collapse, merge, and virialize, with the most intense energy and mass transfer. This is the most gravitationally violent scale that is primarily dominated by gravitational dynamics, leading to a universal spectrum index $n$=-1 around this scale.

The existence of such a universal and expanding transition range is a direct result of the mass and energy cascade across scales. As the mass and energy are transferred across scales through many nonlinear couplings, the detailed structure of the initial spectrum becomes statistically irrelevant. Therefore, scales in the universal range no longer “remember” the shape of the initial spectrum, leading to universal abundance and structure for haloes near $m_h^*$ in this range. Thus, the loss of memory of the initial spectrum is a direct consequence of the mass/energy cascade. Therefore, we also aim to develop a cascade theory leading to this quasi-universal range, which is distinct from the Press–Schechter formalisms. To facilitate the discussion, we start with a brief overview of the current state of knowledge.

Highly localized, over-dense, and virialized haloes are major manifestations of the nonlinear gravitational collapse \citep{Neyman:1952-A-Theory-of-the-Spatial-Distri,Cooray:2002-Halo-models-of-large-scale-str} and the building blocks of large-scale structures, whose abundance and internal structures have been extensively studied over the last several decades. The abundance of haloes is often described by a halo mass function, one of the most fundamental quantities to probe large-scale structures and model the formation and evolution of the structure. The first landmark might be the Press-Schechter (PS) formalism \citep{Press:1974-Formation-of-Galaxies-and-Clus,Bond:1991-Excursion-Set-Mass-Functions-f}. The distribution of halo mass is determined by postulating that the probability of forming haloes is related to the amplitude of density fluctuations on that scale. Haloes will form on some mass scale once the smoothed linear density contrast on that scale exceeds a threshold value. This threshold value must be analytically derived by examining the nonlinear collapse of a spherical top hat over-density \citep{Tomita:1969-Formation-of-Gravitationally-B,Gunn:1972-Infall-of-Matter-into-Clusters}. Alternative derivations using an excursion set approach (EPS) put the PS formalism on a firmer footing by removing the fudge factor in the original PS model \citep{Bond:1991-Excursion-Set-Mass-Functions-f}. This was further extended to the excursion set with correlated steps \citep{Musso:2012-One-step-beyond-the-excursion-set-approach,Paranjape:2012-Peaks-theory-and-the-excursion-set-approach,Maggiore:2010-THE-HALO-MASS-FUNCTION-FROM-EXCURSION-SET-THEORY}. Although mathematically less rigorous, PS formalism and its extensions are still very useful and allow one to compute many different structural properties. Examples are the halo mass function, merging rates, and clustering properties. 

The internal structures of the halo are generally characterized by the halo density profile \citep{Del_Popolo:2009-Density-profiles-of-dark-matte}, 
which can be analyzed both analytically and numerically through N-body simulations \citep{Moore:1998-Resolving-the-structure-of-col,Klypin:2001-Resolving-the-structure-of-col}. Since the first study on spherical collapse \citep{Gunn:1972-Infall-of-Matter-into-Clusters}, the power-law density profile was proposed using the self-similar approximation \citep{Bertschinger:1985-Self-Similar-Secondary-Infall-,Fillmore:1984-Self-Similar-Gravitational-Col}. High-resolution N-body simulations have revealed a nearly universal profile exhibiting a cuspy density—less steep than the isothermal profile at smaller radii and becomes steeper at larger radii \citep{Navarro:1997-A-universal-density-profile-fr,Navarro:2004-The-inner-structure-of-ACDM-ha}. However, there has yet to be a consensus on the exact slope value of the inner density slope $s$ obtained by N-body simulations. Since the first slope $s = -1.0$ described by the NFW profile \citep{Navarro:1997-A-universal-density-profile-fr}, inner slopes of the simulated haloes have varied in a wide range, including values such as $s > -1.0$ \citep{Navarro:2010-The-diversity-and-similarity-of-simulated}, $s = -1.2$ \citep{Diemand:2011-The-Structure-and-Evolution-of-Cold-Dark}, and $s \approx -1.3$ \citep{Governato:2010-Bulgeless-dwarf-galaxies-and-dark-matter-cores,McKeown:2022-Amplified-J-factors-in-the-Galactic-Centre,Lazar:2020-A-dark-matter-profile-to-model-diverse}. Compared with observations, the predicted inner density of the cuspy tends to be higher \citep{Blok:2002GALAXIES:-STRUCTURE-GALAXIES,Blok:2003-Simulating-observations-of-dark-matter,Swaters:2002-The-Central-mass-distribution-in-dwarf,Naray:2011-Recovering-cores-and-cusps-in-dark-matter,Salucci:2019-The-distribution-of-dark-matter}. This gives rise to the cusp-core problem, one of the so-called small scale challenges. 

To our knowledge, the exact origin of the nearly universal density profile and the different inner density slopes of the simulated haloes is not yet fully understood \citep{Cooray:2002-Halo-models-of-large-scale-str}. This paper treats the halo mass function and density profiles from a different perspective. We first establish the existence of a universal range and mass and energy cascade in the collisionless dark matter flow. The mass cascade describes the mass transfer between haloes across different mass scales, which leads to the distribution of haloes with respect to the halo mass, i.e., the halo mass function. The energy cascade in haloes describes the energy flow along the halo radial direction, which suggests the distribution of particles, i.e., the halo density profile. This new perspective offers a theory for nearly universal halo mass functions and density profiles. To illustrate and compare, we begin by revisiting the cascade phenomenon in turbulence. 
 
Turbulence is a typical non-equilibrium system, a long-standing and probably the last unresolved problem in classical physics. 
Constant chaotic fluid motion and energy flowing through different scales prevent the turbulence from reaching a stable equilibrium state. The energy cascade is fundamental to turbulence. At high Reynolds numbers, turbulence consists of a random collection of eddies (building blocks) at different length scales that interact with each other and dynamically change, which can be described by a famous poem \citep{Richardson:1922-Weather-Prediction-by-Numerica}: 
\smallbreak
\centerline{"Big whirls have little whirls, That feed on their velocity;}
\centerline{And little whirls have lesser whirls, And so on to viscosity."} 
\smallbreak
\noindent The poem describes a conceptual picture that large eddies feed smaller eddies, which feed even smaller eddies and then lead to viscous dissipation at the smallest scale, i.e., the concept of a direct energy cascade. There is a broad spectrum of eddy sizes within fully developed turbulence. Large eddies are usually created by the instability of large-scale mean flow at an integral scale $L$. They rapidly break up and pass their kinetic energy to smaller eddies as a result of the inertial force. Smaller eddies are transient and, in turn, pass their energy to even smaller eddies. The cascade continues down the scale and stops operating in the smallest eddies (dissipation scale $\eta$), where the viscous force becomes dominant over the inertial force. At high Reynolds numbers, there exists a range of length scales in which the viscous force is negligible and the inertial force is dominant. The rate $\varepsilon$ of energy passing through the cascade should be scale-independent in this range and match exactly the rate of energy dissipation at the smallest scale. The direct energy cascade is a dominant feature of three-dimensional turbulence. However, an inverse energy cascade was predicted for two-dimensional turbulence, where kinetic energy is transferred from small to large scales \citep{Kraichnan:1967-Inertial-Ranges-in-2-Dimension}. 

From this brief description, turbulence has two key features that enable the cascade phenomenon: i) a broad spectrum of eddies that mediate the energy cascade across different scales; and ii) a viscous force operating on small scales to dissipate the system energy into heat (radiation) and maintain a steady energy cascade at a constant rate. Through energy cascade and dissipation, the turbulence, as a typical nonequilibrium system, reaches a statistically steady state to continuously release system energy and maximize entropy. This intermediate statistical equilibrium state is manifested by a scale-independent rate of the energy cascade $\varepsilon$. 

Apparently, the self-gravitating collisionless dark matter flow (SG-CFD) is different from fluid turbulence. Although SG-CFD is incompressible in full six-dimensional phase space, its three-dimensional fluid projection lacks the dissipative mechanisms and the vortex mechanism to enable the same energy cascade as turbulence. Therefore, SG-CFD cannot directly admit an eddy-mediated energy cascade in a viscous fluid. However, collisionless dark matter flow also shares striking similarities with turbulence, allowing a different,  halo-mediated mass and energy "cascade". Note that the term “cascade” is used in a generalized sense to denote the redistribution of mass and energy across scales, rather than a turbulent viscous process.

First, there also exists a broad spectrum of halo sizes to maximize the entropy of the system \citep{Xu:2023-Maximum-entropy-distributions-of-dark-matter}. Small haloes are created by gravitational instability and interacting and merging with other haloes. Haloes pass their mass onto larger and larger haloes. Consequently, we expect a continuous cascade of mass from small to large scales. By simply replacing "whirls" in that poem with "haloes," a new poem now describes the inverse mass cascade from small to large scales, which is consistent with hierarchical structure formation:
\smallbreak
\centerline{"Little haloes have big haloes, That feed on their mass;} 
\centerline{And big haloes have greater haloes, And so on to growth."} 
\smallbreak

Second, based on Noether's theorem, energy is not globally conserved in an expanding universe because constantly stretching space is not symmetric in time. The total energy of collisionless dark matter constantly decreases with time during gravitational collapse in an expanding background, as shown by the cosmic energy evolution (Section \ref{sec:2-2}). Since there is no viscous force operating in collisionless dark matter, the decrease in total energy is caused by the Hubble expansion (Eq. \eqref{eq:4}). In this regard, the Hubble parameter $H$ plays a "similar" role as the viscosity in turbulence that leads to the energy "dissipation" in collisionless dark matter flow, which initiates and sustains a continuous energy cascade across different scales. However, unlike turbulence, the reader should note that the term "dissipation" is used here to stand for the energy decrease. No energy is truly dissipated into radiation during structure formation.

Based on these similarities, we hypothesize the existence of a universal range and associated "cascade" physics to facilitate the energy "dissipation" in the collisionless dark matter flow (SG-CFD). However, because of the collisionless nature and long-range gravity, we also expect the cascade in the SG-CFD to be unique and distinct from turbulence. Previous work has already studied universal scaling laws, the halo mass function, and density profiles \citep{Xu:2023-Dark-matter-halo-mass-functions-and, Xu:2023-Universal-scaling-laws-and-density-slope}. This paper will focus on the origin of universality, further developing and formulating these fundamental ideas with the help of cosmological simulations. Despite the path already explored by many pioneers, these ideas provide a new understanding of the abundance and internal structure of haloes.  


The remainder of the paper is organized as follows: Section \ref{sec:2-2-2} introduces and formulates the universal transition range. Section \ref{sec:3} presents the mass and energy cascade in the halo mass space that leads to the halo random walk, halo mass functions, and universal scaling laws in the mass space. Section \ref{sec:3-3} presents the energy cascade in individual haloes and the particle random walk that leads to halo density profiles and universal scaling laws in haloes, followed by Section \ref{sec:5} discussing the constraints on dark matter particle mass. The appendix presents the evolution of the cosmic energy that is globally "dissipated" to enable the cascade in SG-CFD, and provides more details on the mass and energy cascade. 

\section{Universal transition spectrum with \MakeLowercase{n}=-1}
\label{sec:2-2-2}
The universal range between two critical scales $k_{NL1}$ and $k_{NL2}$ in Fig. \ref{fig:999} represents a transition between the linear and nonlinear regimes, with a universal spectrum index and an expanding range or ratio $k_{NL2}/k_{NL1}$. To find that spectrum index, we write a self-similar dimensionless spectrum $\Delta^2_X(k,a)$ around the characteristic scale $k_h^*$, which is inside that range. Here $k_h^*\sim 1/r_{hc}^*$ is the comoving scale associated with haloes of characteristic mass $m_h^*$. We can separate the time and k-dependence for the spectrum around $k_h^*$ as
\begin{equation} 
\label{eq:2-2-2-1} 
\Delta^2_X(k,a)=\frac{P(k,a)k^3}{2\pi^2}=\Delta^2_X(k^*_h,a)\left(kr^*_{hc}\right)^c.  
\end{equation} 
where $c$ is the index to be determined. Here, the comoving halo size $r_{hc}^*$ is related to the comoving density of the halo $\rho_{hc}^*$ as
\begin{equation} 
\label{eq:2-2-2-2} 
\rho_{hc}^*=\frac{m_h^*}{\frac{4}{3}\pi{r_{hc}^*}^3}=\delta_c\bar\rho_0 \quad \textrm{and} \quad H_0^2 = \frac{8}{3}\pi G\bar\rho_0,
\end{equation} 
where $\bar\rho_0$ is the mean density of matter at $z$=0. The density ratio $\delta_c=18\pi^2$ can be obtained from the spherical collapse model for the matter-dominant era. Substituting \eqref{eq:2-2-2-2} into \eqref{eq:2-2-2-1} leads to
\begin{equation} 
\label{eq:2-2-2-3} 
\Delta^2_X(k,a) = \Delta^2_X(k^*_h,a)\left(\frac{2Gm_h^*}{\delta_cH_0^2}\right)^{c/3}k^c.  
\end{equation} 
The time dependence of the spectrum comes from both $\Delta^2_X(k^*_h,a)$ and the characteristic mass $m_h^*(a)$. 

\begin{figure}
\includegraphics*[width=\columnwidth]{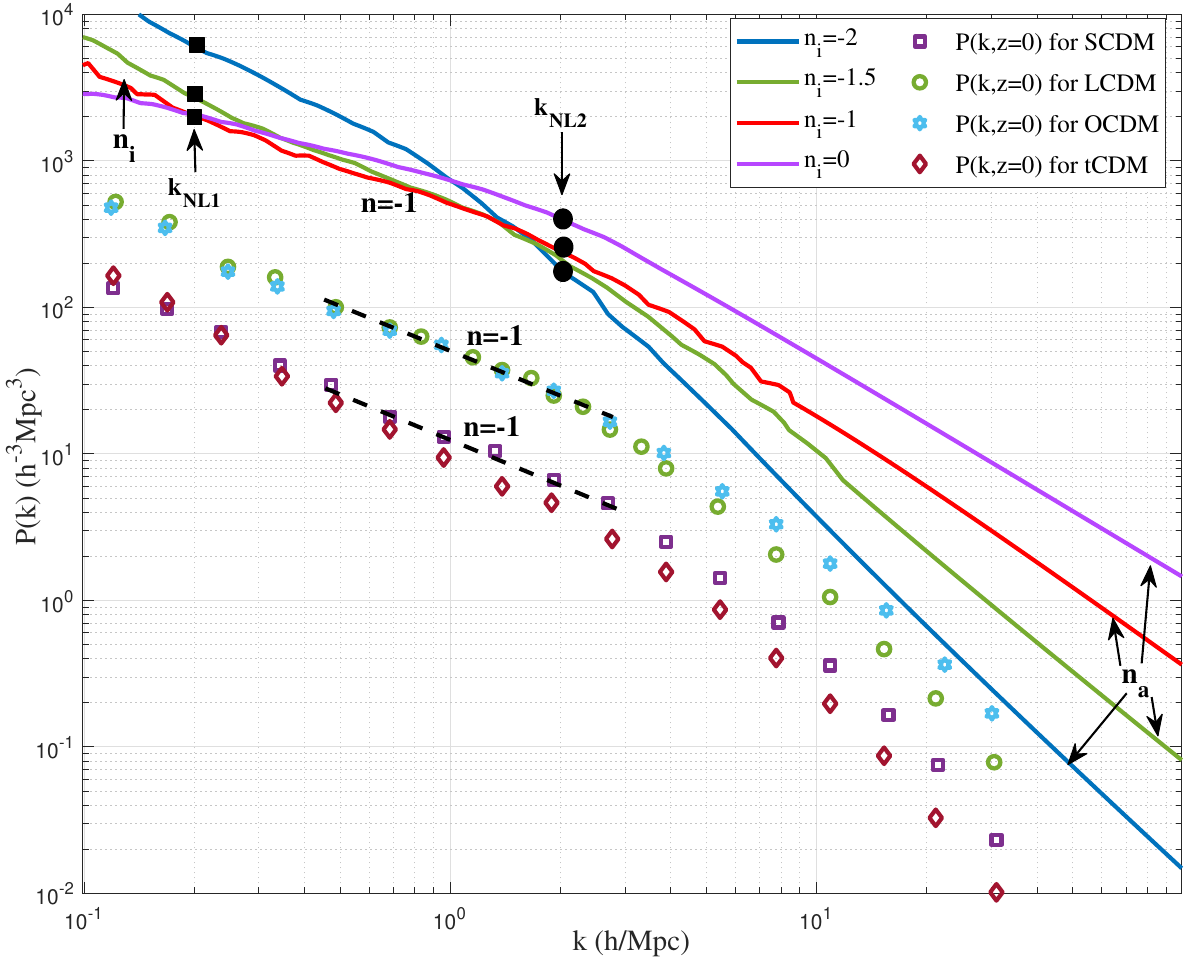}
\caption{The power spectrum $P(k,z)$ at $z$=0 from different Virgo simulations. Symbols plot the power spectrum with various cosmologies, exhibiting a universal transition range $P(k)\propto k^{-1}$ on scales $k_{NL1}<k<k_{NL2}$, where the gravitational dynamics dominate over the effect of the initial spectrum. Solid lines present the power spectrum from a power-law initial spectrum $P_{ini}(k)\propto k^{n_{i}}$, with $n_{i}$= -2, -1.5, -1, and 0. In the fully nonlinear range on scales $k>k_{NL2}$, the spectrum $P(k)$ retains the memory of the initial spectrum and develops an asymptotic slope $n_{a}$ as a function of the initial spectrum index $n_{i}$ (See Fig. \ref{fig:777}).}
\label{fig:888}
\end{figure}

When the universal spectrum is fully developed, we can relate the dimensionless spectrum to the two-point correlation function $\xi(r_{hc}^*,a)$ on the same scale. Since nonlinear structures are always formed on the scale $r_{hc}^*$, the correlation function can be further related to linear overdensity $\delta_{vir}^{lin}\approx 1.69$. Therefore, we write
\begin{equation} 
\label{eq:2-2-3-5} 
\Delta^2_X(k_h^*,a)\sim \xi(r_{hc}^*,a) \sim (\delta_{vir}^{lin})^2a^2.
\end{equation}
In the spherical collapse model, the critical density ratio $\delta_c$ is related to the linear density $\delta_{vir}^{lin}$ as  \citep{Gunn:1972-Infall-of-Matter-into-Clusters,Xu:2022-Postulating-dark-matter-partic} 
\begin{equation} 
\label{eq:2-2-3-6} 
\delta_c=32\xi_{ta} \quad \textrm{and} \quad \delta_{vir}^{lin} = \frac{3}{5}2^{2/3}(\xi_{ta})^{1/3},
\end{equation}
where $\xi_{ta}$ is the density ratio at the turn-around moment.Therefore, the linear overdensity and the dimensionless spectrum at $k_h^*$ read
\begin{equation} 
\label{eq:2-2-3-7} 
\delta_{vir}^{lin}=\frac{3}{10}\delta_c^{1/3} \quad \textrm{and} \quad \Delta^2_X(k_h^*,a)\sim \delta_c^{2/3} a^2.
\end{equation}
Substituting into Eq. \eqref{eq:2-2-2-3} leads to the fully developed spectrum
\begin{equation} 
\label{eq:2-2-3-8} 
\Delta^2_X(k,a) \propto (\delta_c)^{\frac{2-c}{3}}a^2 \left(\frac{Gm_h^*}{H_0^2}\right)^{c/3}k^c.  
\end{equation} 
Since gravitational dynamics and expansion are the only dominant physics in the transition region around scale $m_h^*$, relevant constants for the universal spectrum should include and only include the gravitational constant $G$, the characteristic mass $m_h^*$, and the Hubble parameter $H$. This requires $c$=2. The spectrum is independent of $\delta_c$ and $a^2$ is absorbed into $H$ so that
\begin{equation} 
\label{eq:2-2-3-9} 
\Delta^2_X(k,a) \propto \left({Gm_h^*}/{H^2}\right)^{2/3}k^2.  
\end{equation} 

Note that this argument is independent of the cosmological parameters or the initial spectrum, so that the spectrum $P(k)\propto \Delta^2(k)/k^3\propto k^{-1}$ should be universal. In addition, the mass cascade theory also supports a $k^{-1}$ spectrum, as we will discuss in Section \ref{sec:3}. The final expression can be conveniently written as (from Eq. \eqref{eq:2-2-2-3})
\begin{equation} 
\label{eq:2-2-2-4} 
\Delta^2_X(k,a)= \Delta^2_X(k_h^*,a) \left(\frac{2Gm_h^*H_0}{\delta_c(100\textrm{km/s})^3}\right)^{2/3}\left(k\frac{Mpc}{h}\right)^2. 
\end{equation} 

\begin{figure}
\includegraphics*[width=\columnwidth]{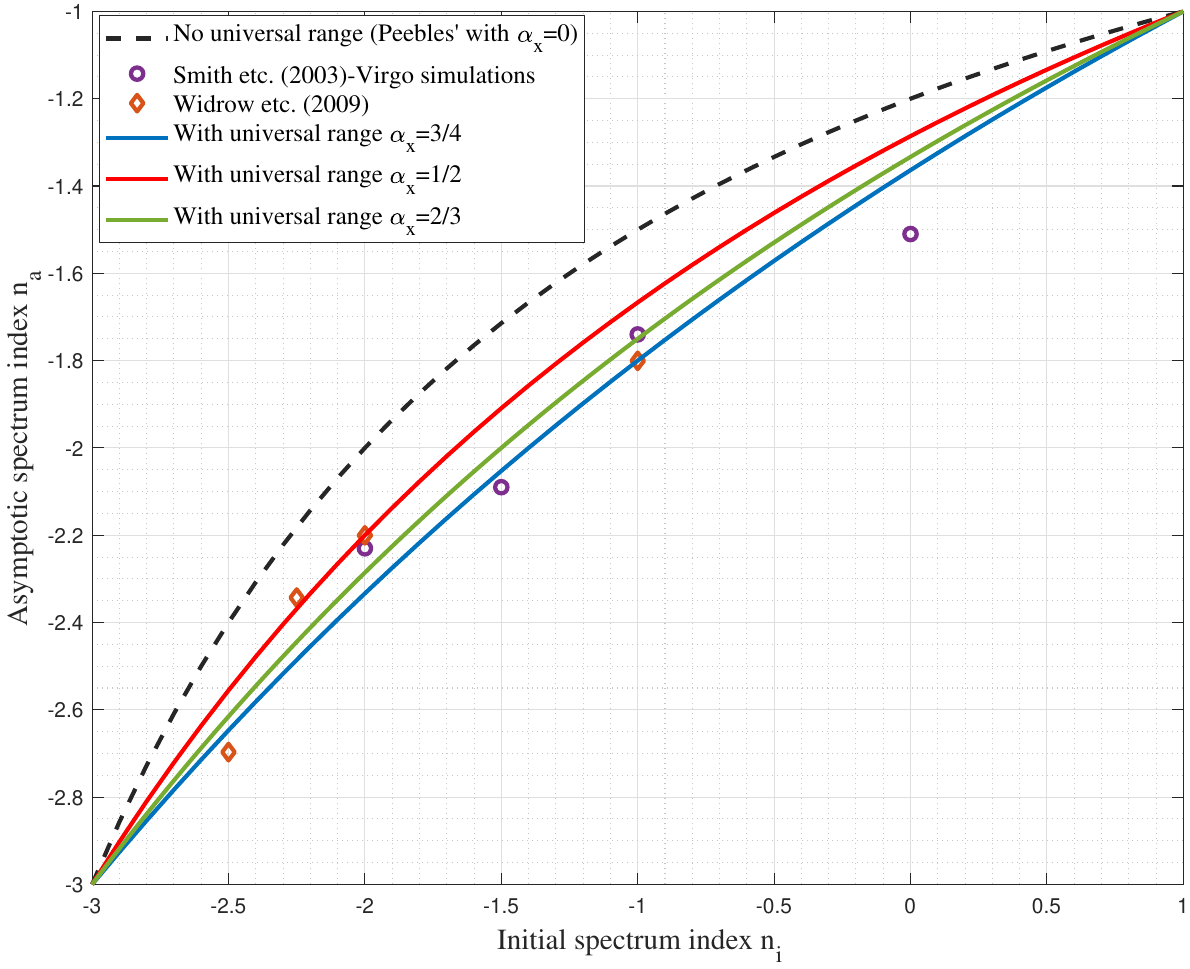}
\caption{The variation of asymptotic slope $n_a$ of nonlinear spectrum $P(k)$ with the slope $n_{i}$ of initial power-law spectrum. Symbols plot the simulation results. The dashed line plots Peebles' prediction without involving a universe transition range ($k_{NL1}=k_{NL2}$ and $\alpha_x=0$). The discrepancy with simulation data suggests a missing piece in that model. We propose that the existence of a universal range also impacts the asymptotic slope $n_a$, leading to a better agreement with simulations (Eq. \eqref{eq:2-2-2-9}). Solid lines plots the prediction with a ratio $k_{NL2}/k_{NL1}=a^{\alpha_x}$, where $\alpha_x=1/2$ is preferred. }
\label{fig:777}
\end{figure}

Figure \ref{fig:888} presents the power spectrum $P(k)$ at $z=0$ from different Virgo simulations \citep{Frenk:2000-Public-Release-of-N-body-simul}. Symbols plot the spectrum from various cosmologies, including the "standard" CDM with $\Omega_m=1$, $h=0.5$, and $\sigma_8=0.61$ (SCDM); "Lambda" CDM with $\Omega_m=0.3$, $\Lambda_0=0.7$, $h=0.7$, and $\sigma_8=0.9$ (LCDM); "Open" CDM with $\Omega_m=0.3$, $h=0.21$, and $\sigma_8=0.85$ (OCDM); and "Tau" CDM with $\Omega_m=1$, $h=0.5$, and $\sigma_8=0.61$ (tCDM). For clarity, the data were shifted (lowered) by one order of magnitude. A transition range can be clearly identified with a universal spectrum index of -1 that is independent of the cosmology model.

The effects of the initial power spectrum can best be studied by systematically varying a power-law initial spectrum. A more negative index $n$ often leads to less frequent halo merging and longer memory of the initial spectrum. Figure \ref{fig:888} also plots the simulated full spectrum with a power-law initial spectrum $P_{i}(k)\propto k^{n_{i}}$ (solid lines), with index $n_{i}$= -2, -1.5, -1 and 0, respectively \citep{Smith:2003-Stable-clustering--the-halo-mo}. The resultant spectrum $P(k)$ in the nonlinear range retains the memory of the initial spectrum and develops an asymptotic slope $n_{a}$ as a function of the initial index $n_{i}$. 

Figure \ref{fig:777} presents the asymptotic slope $n_a$ as a function of the initial index $n_i$. Symbols plot the simulation data from Smith et al. \citep{Smith:2003-Stable-clustering--the-halo-mo} and Widrow et al. \citep{Widrow:2009-Power-spectrum-for-the-small-scale-Universe}. Under the "stable clustering" hypothesis, Peebles shows that if the initial power spectrum was a pure power-law with an index $n_{i}$ in a matter dominant universe, the asymptotic slope $n_{a}$ of the nonlinear spectrum would be directly related to the initial index through the relation $n_{a} = -6/(5+n_{i})$ \citep{Smith:2003-Stable-clustering--the-halo-mo} (dashed black line).  However, the discrepancy between Peebles' prediction and simulation data suggests that an expanding universal range is required.

We first divide the entire spectrum $\Delta^2(k)$ into linear range, universal transition, and nonlinear range. 
\begin{equation} 
\label{eq:2-2-2-5} 
\begin{split}
&\Delta^2_L(k,a) \propto a^2k^{3+n_{i}} \quad \textrm{for}\quad k\le k_{NL1},\\
&\Delta^2_X(k,a) \propto a^bk^2 \quad\textrm{for} \quad k_{NL1}\le k \le k_{NL2},\\
&\Delta^2_{NL}(k,a)\propto a^{-n_{a}}k^{3+n_{a}} \quad\textrm{for} \quad k\ge k_{NL2},    
\end{split}     
\end{equation} 
where the spectrum $\Delta^2_{NL}$ in the nonlinear range satisfies the stable clustering hypothesis \citep{Smith:2003-Stable-clustering--the-halo-mo}. The spectrum $\Delta^2_{NL}\propto a^3$ in the nonlinear range is due to fully bound structures on the physical scale with the average density of the universe $\propto a^{-3}$. For the universal range, $b$ is a parameter for time evolution that depends on the initial conditions, while the spectrum $\Delta^2_X\propto k^2$ is universal and independent of the cosmology and the initial spectrum. Since $k_{NL1}$ is determined by $\Delta^2_L( k_{NL1},a)\equiv1$, we have the scale $k_{NL1}\propto a^{-2/(3+n_i)}$.

The spectrum should be continuous at two critical scales $k_{NL1}$ and $k_{NL2}$, leading to
\begin{equation} 
\label{eq:2-2-2-6} 
\begin{split}
&\Delta^2_L(k_{NL1},a) = \Delta^2_X(k_{NL1},a),\\
&\Delta^2_X(k_{NL2},a) = \Delta^2_{NL}(k_{NL2},a).\\
\end{split}     
\end{equation} 
Solving this continuity condition leads to the relation
\begin{equation} 
\label{eq:2-2-2-7} 
\begin{split}
&(k_{NL1})^{2}(k_{NL2})^{1+n_{a}}\propto a^{n_{a}} \quad \textrm{and} \quad b=4/(3+n_i).\\
\end{split}     
\end{equation} 
Next, we introduce a ratio parameter $\alpha_x>0$ to represent an expansion of the universal range
\begin{equation} 
\label{eq:2-2-2-8} 
\begin{split}
&k_{NL2}\propto a^{\alpha_x}k_{NL1}.\\
\end{split}     
\end{equation} 
Combining two equations \eqref{eq:2-2-2-7} and \eqref{eq:2-2-2-8} leads to the solution
\begin{equation} 
\label{eq:2-2-2-9} 
\begin{split}
n_{a}=\frac{\alpha_x\left(3+n_{i}\right)-6}{\left(1-\alpha_x\right)\left(3+n_{i}\right)+2}.\\
\end{split}     
\end{equation}
The solid lines in Fig. \ref{fig:777} plot three different ratio parameters $\alpha_x$=1/2, 2/3, and 3/4. Peebles' prediction corresponds to the special case with $\alpha_x=0$ or $k_{NL1}=k_{NL2}$ (dashed line). Compared with simulation data, an expanding universal transition range with a ratio parameter $\alpha_x=1/2$ is preferred. This ratio can be related to the characteristic halo mass $m_h^*\propto a^{3/2}$ (or a moving size $r_{hc}^*\propto a^{1/2}$) from the theory of the mass cascade (Eq. \eqref{eq:3-9-2}). The increase in characteristic halo size $r_{hc}$ reflects the expanding universal range and ratio $\alpha_x$.

This section discusses the existence of a universal transition range and its effects on the asymptotic spectrum index $n_a$. That transition range admits a unique and universal power-law spectrum $P(k)\propto k^{-1}$ that expands with time as $\propto a^{1/2}$. In the next, a cascade theory is developed to interpret the formation and evolution of such a universal spectrum, which also confirms the same spectrum index of $-1$ (Eq. \eqref{eq:3-6} in Section \ref{sec:3-3-3}). Similarly, associated with the mass and energy cascade, universal scaling laws for halo mass functions and density profiles also exist near the scale $m_h^*$ (Sections \ref{sec:3-3-3} and \ref{sec:4-4}).

\section{Mass cascade and halo mass function}
\label{sec:3}
The universal spectrum suggests a statistically steady state established in a nonequilibrium system that emerges from some "cascade" process, largely losing the memory of initial conditions. In this section, we introduce the notion of a mass and energy "cascade" across dark matter haloes spanning a range of masses. Because the dark component is effectively collisionless, it cannot sustain a classical turbulent cascade mediated by viscosity and vorticity. We therefore employ “cascade” in a generalized sense to denote the redistribution of mass and energy across halo-mass scales that drives the system toward its asymptotic equilibrium, rather than a Kolmogorov-type viscous turbulent process. A salient aspect of this evolution is the emergence of a statistically stationary state characterized by a scale-invariant flux of mass and energy through the halo mass space. 

\subsection{Halo random walk and mass function}
\label{sec:3-4}


The distribution of haloes depends both on the initial conditions and on gravitational dynamics. The Press-Schechter formalism suggests a mass function $\propto {m_h}^{\frac{n-3}{6}}$ dependent only on the spectrum index $n$ (Eq. \eqref{eq:3-9-2-2}). Using Eq. \eqref{eq:2-2-2-9} for the asymptotic spectrum index $n_a$, the asymptotic slope $\lambda_a$ of the mass function $m_h^{-\lambda_a}$ can be related to the initial power-law spectrum index $n_i$ as
\begin{equation} 
\label{eq:3-25-2} 
\begin{split}
\lambda_{a}=\frac{3-n_{a}}{6}=\frac{12+(3-4\alpha_x)\left(3+n_{i}\right)}{12+6\left(1-\alpha_x\right)\left(3+n_{i}\right)}.\\
\end{split}     
\end{equation}
In hierarchical (bottom-up) structure formation, low-mass haloes collapse at high redshift from peaks of the primordial density field. Their distributions are therefore largely determined by the properties of these peaks and retain a strong imprint of the primordial power spectrum. In contrast, high-mass haloes assemble later through repeated mergers and smooth accretion. This prolonged and violent assembly history progressively erases information about the initial conditions, driving the system toward a quasi-universal state. Consequently, the distribution for haloes around characteristic mass $m_h^*$ is expected to be universal, governed primarily by gravitational dynamics, and less sensitive to the primordial spectrum. 

\begin{figure}
\includegraphics*[width=\columnwidth]{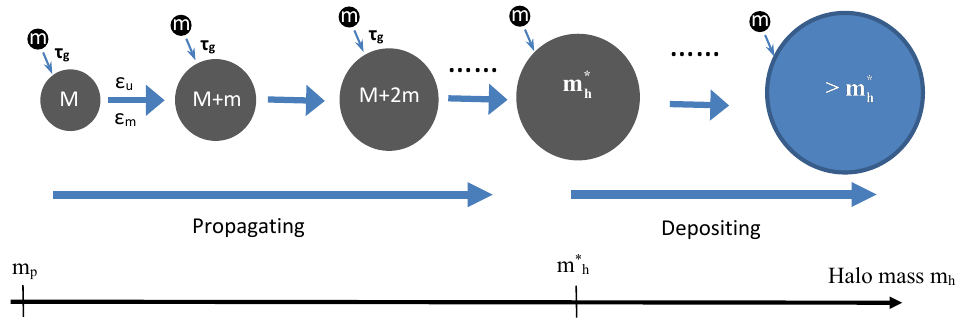}
\caption{Schematic plot for the halo random walk and mass and energy cascade in halo mass space. Haloes of mass $M$ merging with a single merger (free particles of mass $m$) leads to a mass flux to a larger scale $M+m$ (smooth accretion), that is, the halo of mass $M$ walking into the next mass scale $M+m$ after merging. For a given halo of mass $m_h$, a merge occurs with an average waiting time $\tau_g(m_h,z)$. 
Because haloes have finite mass and kinetic energy, continuous merging facilitates an inverse mass and kinetic energy cascade from small to large scales and a direct cascade of potential energy from large to small scales. 
The scale-independent mass and energy flux ($\varepsilon_m$ and $\varepsilon_u$ are independent of mass scale $m_h$) are expected in the propagation range ($m=m_p<m_h<m_h^*$), a typical feature of a statistical steady state in a nonequilibrium system. The rate of cascade becomes scale-dependent in the deposition range ($m_h>m_h^*$) (see Figs. \ref{fig:3-2}, \ref{fig:3-3} for $\varepsilon_m$ and Fig. \ref{fig:3-7} for $\varepsilon_u$).} 
\label{fig:3-1}
\end{figure}

To clarify this behavior, we consider an idealized initial condition in which the halo population is sharply concentrated at $m_h$=0 in mass space, representing the initial simultaneous collapse of an extremely large number of small and high-density proto-haloes at a high redshift and at various locations throughout the universe. In this limit, the influence of the primordial spectrum is confined to a small region near $m_h=0$, whereas subsequent evolution proceeds subject to self-gravity in an expanding dark matter background. We adopted an effective mean-field description to describe structure formation via smooth mass accretion from the background (Fig. \ref{fig:3-1}). The evolution of the halo distribution in mass space is modeled as a continuous-time random walk with a mass-dependent waiting time. In the gravity-dominated regime, this waiting time is controlled by the halo potential $\tau_{g}(m_h,z)\propto \Phi(m_h)^{-1}\propto m_h^{-\lambda}$ (Eq. \eqref{eq:4-3}). The larger the haloes, the deeper the halo potential, the hotter the local temperature, the shorter the residence times, and the faster the mass accretion. When gravity dominates the merging rate, a universal $\lambda=2/3$ can be obtained, as shown in Eq. \eqref{eq:3-6}. If the primordial spectrum dominates, the exponent $\lambda$ is expected to be related to the index of the initial spectrum, or $\lambda=(3-n)/6$, as shown in Eq. \eqref{eq:3-9-2-3}.

Figure \ref{fig:3-1} describes the random walk of haloes in mass space. The merging of a halo (mass $M$) with a single mass merger $m$ (smooth accretion, depending on the mass resolution in the N-body simulations) leads to a continuous mass and energy flux to a larger mass scale $M+m$, that is, the halo of mass $M$ walking into the next mass scale $M+m$ after merging. For a given halo mass $m_h$, the merge event occurs with an average waiting time $\tau_g(m_h,z)$. Therefore, haloes continuously migrate in mass space from one scale ($m_h$) to the neighboring scale ($m_h+m$), enabled by the merging with single mergers (smooth accretion). A diffusion-like particle random walk results in an evolving particle distribution, as described by the Fokker-Planck equation. Similarly, the random walk of haloes in mass space leads to a probability distribution to find a halo at a given mass. The distribution of haloes with respect to the halo mass, i.e., the halo mass function, can be naturally obtained from the corresponding Fokker-Planck equation. This approach for the mass function does not rely on any collapse model (either spherical or ellipsoidal).

This approach is mathematically convenient, fundamentally different from the Press-Schechter formalism, and reveals important aspects of the halo mass function. The evolving halo mass function continuously maximizes the system entropy contained in the distribution of the halo mass, just as the evolving particle distribution maximizes the entropy contained in the particle distribution. Mathematically, the random walk of haloes in mass space can be described by a stochastic Langevin equation \citep{Xu:2023-Dark-matter-halo-mass-functions-and}
\begin{equation}
\label{eq:3-10} 
\frac{\partial m_{h} \left(t\right)}{\partial t} =\sqrt{2D_{p} \left(m_{h} \right)} \varsigma \left(t\right)\propto \frac{m_p}{\tau_g}\propto \frac{m_p}{m_h^{-\lambda}}.        
\end{equation} 
Unlike the standard particle random walk for diffusion, the waiting time $\tau_g$ for the halo random walk is scale-dependent. For a power-law waiting time $\tau_{g} \propto {m_{h}^{-\lambda}}$, the scale-dependent diffusivity reads
\begin{equation} 
\label{eq:3-11} 
D_{p} \left(m_{h} \right)=D_{p0}(t) m_{h}^{2\lambda}.         
\end{equation} 
Here, $D_{p0}(t)$ is a proportional constant for the diffusivity $D_{p}$. The white Gaussian noise $\varsigma (t)$ satisfies the covariance $\langle \varsigma (t)\varsigma (t^{'})\rangle =\delta (t-t^{'})$ with zero mean $\langle \varsigma (t)\rangle =0$. 

Next, in Stratonovich's interpretation \citep{Stratonovich:1966-A-new-representation-for-stoch,Metzler:2014-Anomalous-diffusion-models-and}, the Langevin equation (Eq. \eqref{eq:3-10}) yields the Fokker-Planck equation for the halo mass function $f_M(m_h,t)$:
\begin{equation} 
\label{eq:3-12} 
\frac{\partial f_{M}(m_h,t) }{\partial t}=D_{p0} \frac{\partial }{\partial m_{h} } \left[m_{h}^{\lambda } \frac{\partial }{\partial m_{h} } \left(m_{h}^{\lambda } f_{M}(m_h,t) \right)\right].
\end{equation} 

Finally, the halo mass function $f_M(m_h,t)$ can be analytically solved from Eq. \eqref{eq:3-12}, which is a stretched Gaussian with an exponential cut-off at large $m_{h}$ and a power-law at small $m_{h}$,
\begin{equation}
\label{eq:3-13} 
f_{M} \left(m_{h},t\right)=\frac{m_{h}^{-\lambda} }{\sqrt{\pi D_{p0} t} } \exp \left[-\frac{m_{h}^{2-2\lambda } }{4\left(1-\lambda \right)^{2} D_{p0} t} \right].      
\end{equation} 
The characteristic mass $m_h^*(t)$ can be determined numerically from N-body simulations \citep{Xu:2023-Dark-matter-halo-mass-functions-and} (Fig. \ref{fig:3-4}). In principle, it is related to the mean square displacement in the mass space as
\begin{equation} 
\label{eq:3-14} 
\begin{split}
&\left\langle m_{h}^{2} \right\rangle =\int _{0}^{\infty }f_{M} \left(m_{h} ,t\right) m_{h}^{2} dm_{h}\\
&=\frac{1}{\sqrt{\pi } } \Gamma \left(\frac{3-\lambda }{2-2\lambda } \right)\left[4\left(1-\lambda \right)^2 D_{p0} t\right]^{\frac{1}{1-\lambda } } \equiv \gamma _{0} m_{h}^{*2}.
\end{split}
\end{equation} 
where $\gamma _{0} $ is a proportional constant. With $\lambda<1$ and the exponent ${1/\left(1-\lambda \right)} \ge 1$ in Eq. \eqref{eq:3-14}, the random walk of haloes in mass space is essentially a super-diffusion. With $m_h^*$ as the characteristic mass scale, the halo mass function can be rewritten as
\begin{equation}
\label{eq:3-15} 
f_{M} \left(m_{h} ,t\right)=\frac{\left(1-\lambda \right)}{\sqrt{\pi \eta _{0} } m_h^*} \left(\frac{m_{h}}{m_{h}^*} \right)^{-\lambda } \exp \left[-\frac{1}{4\eta _{0} } \left(\frac{m_{h} }{m_{h}^{*} } \right)^{2-2\lambda } \right],     
\end{equation} 
where the dimensionless constant
\begin{equation} 
\label{eq:3-16} 
\eta _{0} =\frac{1}{4} \left[\frac{\gamma _{0} \sqrt{\pi } }{\Gamma \left({\left(3-\lambda \right)/\left(2-2\lambda \right)} \right)} \right]^{1-\lambda } .     
\end{equation} 
From this derivation, the parameter $\lambda$ for waiting time $\tau_g\propto m_h^{-\lambda}$ is essentially the slope of the halo mass function. By introducing a dimensionless variable $\nu =(m_h/m_h^*)^{2/3}$, the mass function can be expressed in terms of $\nu$
\begin{equation} 
\label{eq:3-17} 
f_{D\lambda} (\nu)=\frac{p}{2\sqrt{\pi\eta_0}}{\nu}^{\frac{p}{2}-1} \exp \left(-\frac{{\nu}^p }{4\eta _{0} } \right),       
\end{equation} 
where the parameter $p=3\left(1-\lambda \right)$ is on the order of unity. 

The Press-Schechter (PS) mass function shares a similar form, but involves a spectrum index $n$
\begin{equation}
\begin{split}
f_{PS} \left(m_{h} \right)=\frac{3+n}{3\sqrt{2\pi}}\frac{1}{M^*}\left(\frac{m_h}{M^{*}}\right)^{-\frac{3-n}{6} }\exp \left[-\frac{1}{2} \left(\frac{m_{h}}{M^{*}}\right)^{\frac{3+n}{3}} \right].
\end{split}
\label{eq:3-9-2-2}
\end{equation}
When two mass functions are compared (Eqs. \eqref{eq:3-15} and \eqref{eq:3-9-2-2}), the parameter $\lambda$ is related to the spectrum index $n$
\begin{equation}
\lambda = \frac{3-n}{6}. 
\label{eq:3-9-2-3}
\end{equation}
Clearly, our mass function reduces to the PS mass function if $\lambda=(3-n)/6$ and $\eta _{0}={1/2}$ or $\gamma_0=15$. However, without considering the universal transition range, the mass scale $M^*$ in the Press-Schechter mass function is simply linked to the critical scale $k_{NL1}$ with $M^*\propto a^{\frac{6}{n+3}}$ in linear theory. While our characteristic mass $m_h^*$ initially tracks $M^*$, it becomes smaller than $M^*$, and scales as $m_h^*\propto a^{3/2}$ as the universal transition range develops and broadens (Fig. \ref{fig:999}). 

The parameter $\lambda$ is the slope of the mass function, which also quantifies the waiting time or the merging rate. A more negative $n$ leads to a larger $\lambda$ and less frequent merging. For the CDM spectrum, $\lambda$ could increase and approach 1 with decreasing mass scales (or increasing wavenumber k). Due to the dominance of initial power spectrum and gravitational dynamics on different scales, the simplest general scenario is to have two different "effective" $\lambda$ (that is, a double-$\lambda$ model) for two different ranges. The fully nonlinear region on small scales ($m_h\ll m_h^*$) retains a strong memory of initial conditions, whereas the universal transition range ($m_h\sim m_h^*$) is governed by gravitational dynamics. The halo mass function in Eq. \eqref{eq:3-15} can be naturally generalized to include two $\lambda$ parameters, with $\lambda_{1}$ for the effects of initial spectrum and $\lambda_{2}$ for gravity-dominated merging \citep{Xu:2023-Dark-matter-halo-mass-functions-and}. Based on this idea, a double-$\lambda$ mass function can be naturally constructed from Eq. \eqref{eq:3-15},
\begin{equation} 
\label{eq:3-18} 
\begin{split}
f_{M} \left(m_{h} ,z\right)&=\left(2\sqrt{\eta _{0} } \right)^{-q} \frac{2\left(1-\lambda _{1} \right)}{q\Gamma \left({q/2} \right)}\\ &\cdot \frac{1}{m_{h}^{*}} \left(\frac{m_{h}^{*}}{m_{h}}\right)^{\lambda _{1}}  \exp \left[-\frac{1}{4\eta _{0} } \left(\frac{m_{h}}{m_{h}^{*}} \right)^{2-2\lambda _{2} } \right], 
\end{split}
\end{equation} 
where $q={\left(1-\lambda _{1} \right)/\left(1-\lambda _{2} \right)} $ is the ratio between two $\lambda $ values to satisfy the normalization. The \textit{k}th order moment of the double-$\lambda $ mass function is
\begin{equation} 
\label{eq:3-19}
\begin{split}
\int _{0}^{\infty } &f_{M} \left(m_{h} ,z\right)\left(m_{h} \right)^{k} dm_{h} \\
&=\frac{{q/2} }{\Gamma \left(1+{q/2} \right)} \left(4\eta _{0} \right)^{\frac{k}{\left(2-2\lambda _{2} \right)} } \Gamma \left(\frac{q}{2} +\frac{k}{2-2\lambda _{2} } \right)\left(m_{h}^{*} \right)^{k}.
\end{split}
\end{equation} 
With a dimensionless variable defined as $v=\left({m_{h} /m_{h}^{*} } \right)^{2-2\lambda _{2} } $, the dimensionless double-$\lambda $ mass function reads
\begin{equation} 
\label{eq:3-20} 
f_{D\lambda } \left(\nu \right)=\frac{\left(2\sqrt{\eta _{0} } \right)^{-q} }{\Gamma \left({q/2} \right)} \nu ^{{q/2} -1} \exp \left(-\frac{\nu }{4\eta _{0} } \right).       
\end{equation} 

\begin{figure}
\includegraphics*[width=\columnwidth]{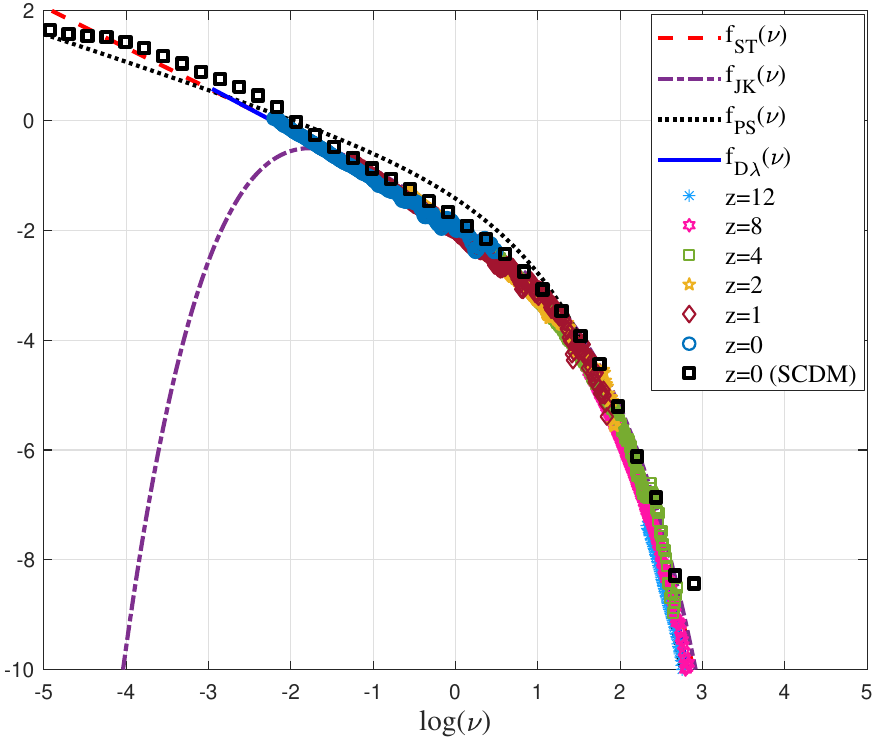}
\caption{Comparison between different mass functions (log(f($\nu$))) in Appendix \ref{sec:A3} and cosmological simulations. The black square represents the mass function from the Virgo SCDM simulation at \textit{z}=0. The color symbols represent the mass function from the Illustris simulation at different redshifts $z$. The PS mass function underestimates the mass in large haloes. The fitted JK mass function matches the simulation for a given range of halo size, but not the entire range. The double-$\lambda$ mass function (Eq. \eqref{eq:3-20}) matches both the simulation and ST mass function for the entire range.}
\label{fig:3-6}
\end{figure}

Figure \ref{fig:3-6} plots different analytical dimensionless mass functions (log(f($\nu$)) (Appendix \ref{sec:A3}) compared to the mass function obtained from the Virgo simulation (solid blue). The PS mass function overestimates the mass in small haloes and underestimates the mass in large haloes. The JK mass function matches the simulation for large-mass haloes with a large deviation for small-mass haloes. The new double-$\lambda$ mass function (Eq. \eqref{eq:3-20}) with the best fit of $\eta _{0} =0.76$ and $q=0.556$ matches both the simulation and the ST mass function for the entire range. With $\lambda _{2} ={2/3}$ in the universal transition range and $q=0.556$, we estimate $\lambda _{1} =0.815$, an effective $\lambda$ that reflects the average waiting time for a given CDM spectrum. 

\subsection{Universal scaling laws in transition region}
\label{sec:3-3-3}
The mass cascade in the universal range dominant in gravity enables universal scaling laws. We will first quantify the parameter $\tau_g$ in that universal range. By definition, $\tau_g$ quantifies the mass accretion rate, 
\begin{equation}
\begin{split}
&\frac{dm_h}{dt} = \frac{m_p}{\tau_g},
\end{split}
\label{eq:3-5-2}
\end{equation}
where $m_p$ denotes the characteristic mass increase per accretion event during smooth accretion (e.g., the mass associated with a single merger or a dark matter particle). This relation implies that, on average, the halo gains a mass $m_p$ over a time scale $\tau_g$. In practice, major mergers between haloes of comparable mass are infrequent, particularly during the early matter-dominated epoch when the halo number density is low. Equation \eqref{eq:3-5-2} should therefore be interpreted as a coarse-grained effective accretion rate that captures smooth growth and the cumulative contribution of minor mergers and continuous inflow, rather than the accretion of major mergers. The waiting time $\tau_g\propto m_h^{-\lambda}$ depends on the halo mass through the parameter $\lambda$. Intuitively, the waiting time depends on the initial distribution of matter. The more negative $n$ leads to a higher large-scale power and more massive haloes, a lower small-scale power, less frequent merging, thus a large $\lambda$ and a longer waiting time $\tau_g$ for low-mass haloes. For a general CDM power spectrum, $\lambda$ varies with scale. However, a unique value of $\lambda$ can be derived for haloes around $m_h^*$ in the universal transition range.

In the transition range, since gravitational dynamics is dominant, the waiting time is inversely proportional to the halo potential $\tau_g \propto \Phi_h^{-1}$ \citep{Xu:2023-Dark-matter-halo-mass-functions-and}. This is also explained by the random walk of particles in haloes in Section \ref{sec:3-3-2} (Eq. \eqref{eq:4-3}). Therefore, we can write the following for the haloes of $m_h\sim m_h^*$
\begin{equation}
\begin{split}
&\tau_g \propto \Phi_h^{-1} \propto (Gm_h/r_h)^{-1}, \\
&\rho_h=\frac{m_h}{\frac{4}{3}\pi {r_h}^3}=\Delta_c \rho_{DM} \propto a^{-3},
\end{split}
\label{eq:3-5}
\end{equation}
where $r_h$ is the size of the halo. The second equation relates the average halo density to the mean density $\rho_{DM}$, where $\Delta_c=18\pi^2$ is the critical density ratio. Solving these equations leads to:
\begin{equation}
\begin{split}
&\tau_g\propto m_h^{-2/3}a  \quad \textrm{and} \quad r_h\propto m_h^{1/3}a.
\end{split}
\label{eq:3-6}
\end{equation}
A unique value of $\lambda=2/3$ can be predicted for haloes around $m_h^*$ in the universal transition range. Using the relation between $\lambda$ and $n$ (Eq. \eqref{eq:3-9-2-3}), $\lambda=2/3$ corresponds to a universal spectrum index $n=-1$ for the gravitational dominant transition range, as shown in Fig. \ref{fig:999}. 

Substituting $\tau_g\propto m_h^{-2/3}a$ into Eq. \eqref{eq:3-5-2}, with $\lambda=2/3$, the scaling relations for haloes with characteristic mass $m_h^*$ read 
\begin{equation}
r_h^*\propto t, \quad m_h^*\propto t \quad \textrm{and} \quad \tau_g^* \propto (m_h^*)^{-2/3}a\propto t^0. 
\label{eq:3-9-2}
\end{equation}
Therefore, in the matter-dominant era, characteristic haloes accrete mass at a constant rate with a constant waiting time $\tau_g^*$ such that both mass $m_h^*$ and size $r_h^*$ are $\propto t \propto a^{3/2}$. Since the characteristic mass $m_h^*\propto a^{3/2}$, the comoving size $r_{hc}^*\propto a^{1/2}$ leads to an expanding universal range with an estimate of $\alpha_x=1/2$ in Eq. \eqref{fig:777}. Using solution \eqref{eq:3-15}, in the universal transition range, the halo mass function $f_M$ follows a simple power law (with $\lambda=2/3$ and $m_h^*\propto a^{3/2}$),
\begin{equation}
\begin{split}
&f_M(m_h,z) \propto {m_h^*}^{\lambda-1} {m_h}^{-\lambda}\propto m_h^{-2/3} a^{-1/2}
\end{split}
\label{eq:3-8-2}
\end{equation}

\begin{figure}
\includegraphics*[width=\columnwidth]{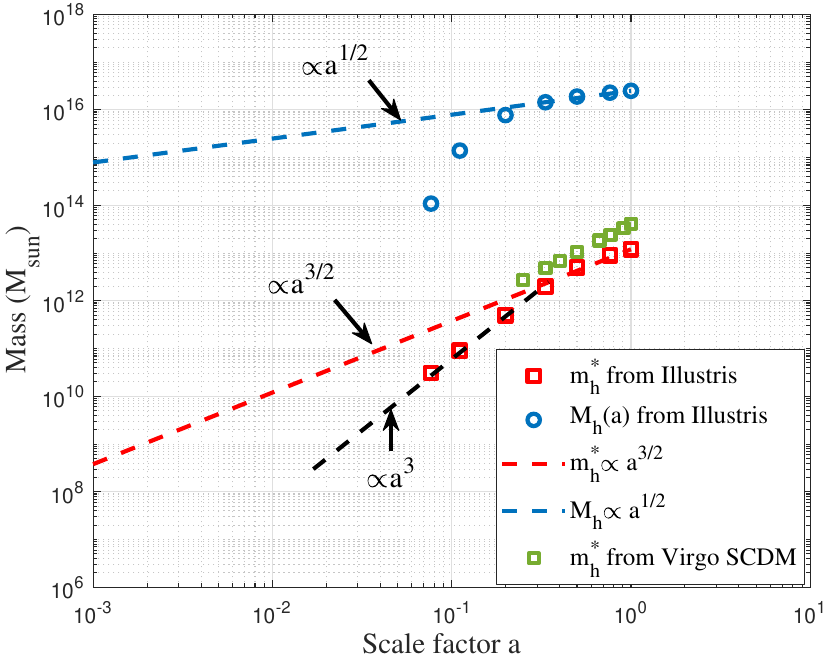}
\caption{The evolution of the characteristic halo mass $m_h^*(t)$ (red dashed) and total halo mass $M_h(t)$ (blue dashed) in the matter era, both can be numerically determined from N-body simulations \citep{Xu:2023-Dark-matter-halo-mass-functions-and}. Results from Illustris and Virgo simulations are presented as symbols, which are in agreement with the prediction of Eqs. \eqref{eq:3-9-2} and \eqref{eq:3-8}, i.e. $m_h^*\propto a^{3/2}\propto t$ and $M_h\propto a^{1/2}\propto t^{1/3}$. At high redshift, before the universal spectrum being fully developed, $m_h^*\propto a^3$ follows the mass scale $M^*$ at $k_{NL1}$.}   
\label{fig:3-4}
\end{figure}

In this section, we identify the waiting time $\tau_g\propto m_h^{-2/3}a$ with $\lambda=2/3$ in the gravitationally dominant universal range, confirm the universal power spectrum $n$=-1, and derive the scaling relations for the evolution of $m_h^*$. Figure \ref{fig:3-4} presents the time variation of $m_h^*$ from the Illustris and Virgo simulations. After the system reaches a statistically stable state with a fully developed universal spectrum, the characteristic mass is reached $m_h^*\propto a^{3/2}$. To derive the evolution of total mass $M_h(t)$ in all haloes, i.e., $M_h\propto a^{1/2}$, we need to discuss the scale-independent rate $\varepsilon_m$ of the mass cascade in the next section.

\subsection{The scale-to-scale mass cascade in mass space}
\label{sec:3-2}
In this section, we present a quantitative description of the scale-to-scale mass transfer in the halo mass space, i.e. a mass cascade process. As shown in Fig. \ref{fig:3-1}, the halo random walk in mass space generates mass and energy flux across scales. On average, it takes a waiting time $\tau_g(m_h,z)$ for a given halo to merge and go to the next scale. For a group of $n_h$ haloes of the same mass $m_h$, the average waiting time for a merge event of any halo in that group can be found as $\tau_h(m_h,z)=\tau_g/n_h$. A simple explanation is provided here. 

Let us randomly pick one halo in that group to merge in every round and repeat until the given halo is picked. The probability of picking that given halo in the first round is $1/n_h$, and the corresponding waiting time for that specific halo is $\tau_h$. The probability of picking that given halo in the second round is $(n_h-1)/n_h*1/n_h$ (product of the probability not selected in the first round and the probability selected in the second round). The corresponding waiting time is $2\tau_h$. Therefore, the waiting time $\tau_g$ for a given halo is
\begin{equation} 
\label{eq:3-1} 
\begin{split}
\tau_{g} =\sum _{k=1}^{\infty }\frac{k\tau _{h} }{n_{h} } \left(1-\frac{1}{n_h} \right)^{k-1} = \frac{\tau _{h} }{n_{h} } +...=n_{h} \tau _{h}.
\end{split}
\end{equation} 

Based on this description, the rate of mass cascade (or mass flux) in halo mass space can be conveniently defined as
\begin{equation}
\begin{split}
&\Pi_m(m_h,z) = -\frac{m_h}{\tau_h}=-\frac{n_hm_h}{\tau_g}=-\frac{m_g}{\tau_g}, \\ 
&\varepsilon_m(z) = \Pi_m(m_h,z) \quad \textrm{for} \quad m_h\le m_h^*. 
\end{split}
\label{eq:3-2}
\end{equation}
The mass $m_h$ of the halo is transferred to a larger mass scale during a period of $\tau_h$. Here, $m_g=n_hm_h$ is the total mass for all $n_h$ haloes of the same mass $m_h$ (or the mass of the halo group), such that $\tau_g$ is the time required to transfer the mass of the entire group $m_g$ to a larger mass scale. The negative sign represents the inverse cascade from small to large scales, in contrast to the direct cascade. 

Here, we introduce an important hypothesis implied by Eq. \eqref{eq:3-2}: the rate of the mass cascade is scale independent on scales $m_h<m_h^*$; that is, $\varepsilon_m$ is independent of the mass scale $m_h$ in that range. The mass and energy cascade drives the nonequilibrium system to establish a statistically steady state to continuously release the energy of the system and maximize entropy \citep{Xu:2023-Maximum-entropy-distributions-of-dark-matter}. When the dark matter flow reaches that statistically steady state, the rates of mass and energy cascade must be scale independent. If this is not the case, there would be a net accumulation of mass and energy on some intermediate-mass scale below $m_{h}^{*}$. 

To validate this hypothesis by N-body simulations, we need to calculate the scale-to-scale transfer in simulations. We first introduce the mass flux ($\Pi_m$) across haloes of different sizes
\begin{equation} 
\label{eq:3-3} 
\begin{split}
\Pi _{m} \left(m_{h} ,t\right) &=-\int _{m_{h} }^{\infty }\frac{\partial }{\partial t} \left[M_{h} \left(t\right)f_{M} \left(m,m_{h}^{*} \right)\right] dm, \\
&=-\frac{\partial }{\partial t} \left[\int _{m_{h} }^{\infty } M_{h} \left(t\right)f_{M} \left(m,m_{h}^{*} \right) dm \right]=-\frac{\partial \Lambda_m}{\partial t},
\end{split}
\end{equation} 
where $M_{h}$ is the total mass in all haloes of all sizes. Here, $f_{M}$ is the halo mass function. The cumulative mass function $\Lambda_m(m_h,t)$ for the total mass in all haloes greater than $m_h$ reads
\begin{equation} 
\label{eq:3-4} 
\Lambda_m(m_h,t) = \int _{m_{h}}^{\infty } M_{h} \left(t\right)f_{M} \left(m,m_{h}^{*} \right) dm.  
\end{equation} 
The total mass in all haloes can be obtained by setting $m_h\rightarrow 0$ in Eq. \eqref{eq:3-4}, i.e., $M_h(t)=\Lambda_m(m_h\rightarrow 0,t)$. The time derivative of $\Lambda_m$ describes the mass flux from all haloes below the scale $m_h$ to all haloes above the scale $m_h$, i.e., the rate of mass cascade $\Pi_m$ across the scale $m_h$. 

Equation \eqref{eq:3-3} describes the total mass flux induced by the binary merging involving three haloes (two merging haloes and one remnant halo), with at least one merging halo being below $m_h$ and the remnant halo above $m_h$. Under the assumption of smooth mass accretion, the dominant contribution to this total mass flux is the merging of $m$ and $m_h$ to produce a halo of mass $m+m_h$, such that Eq. \eqref{eq:3-3} is equivalent to Eq. \eqref{eq:3-2}, especially in the early universe when halo number density is low. Appendix \ref{sec:A1.1} provides more discussion on this locality feature for halo interactions in mass space.

\begin{figure}
\includegraphics*[width=\columnwidth]{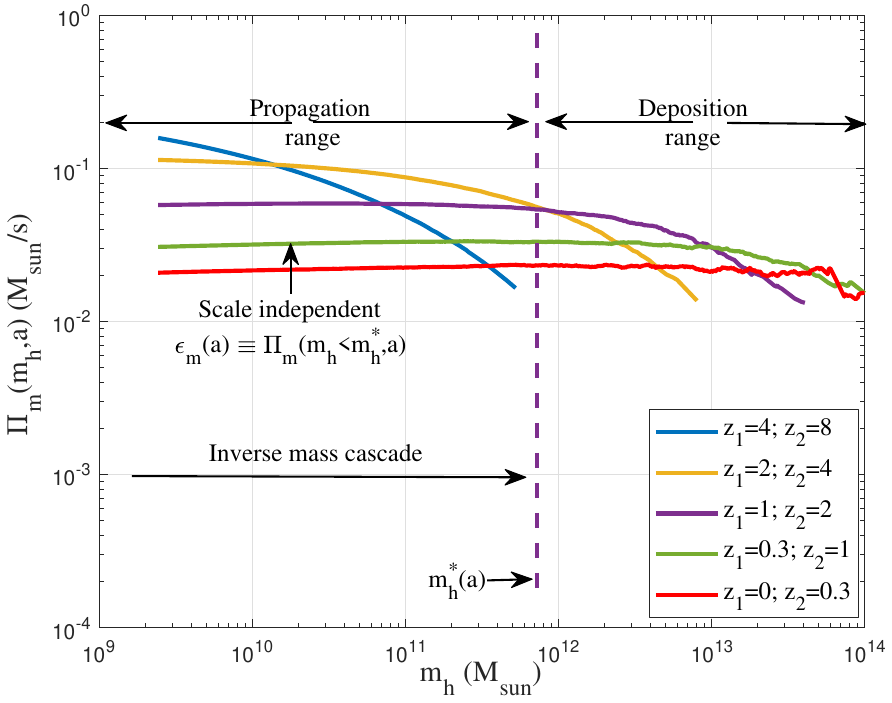}
\caption{The variation of the mass flux $-\Pi_m(m_h,z)$ from the Illustris simulation ($<0$ for "inverse"), calculated from the cumulative mass function $\Lambda_m$ at two different redshifts $z_1$ and $z_2$ (Eq. \eqref{eq:3-3}). After reaching a statistically steady state at around $z=4$, a propagation range is formed for the scales below a characteristic mass, i.e., $m_h<m_h^*(a)$. A scale-independent rate of the mass cascade $\varepsilon_m(a)\equiv \Pi_m(m_h, a)$ can be identified in that range, which decreases with time ($\varepsilon_m(a)\propto a^{-1}$) and is around -0.02$M_{\odot}/s$ at $z=0$. Haloes in the propagation range pass their mass to larger haloes, while the group mass $m_g$ (total mass of all haloes with the same mass $m_h$) remains constant (see Fig. \ref{fig:3-5}). The energy cascade is shown in Fig. \ref{fig:3-7}.} 
\label{fig:3-2}
\end{figure}

In N-body simulations, we use the difference of $\Lambda_m$ at two different redshifts $z_1$ (or $t_1$) and $z_2$ (or $t_2$) to compute the time derivative and hence the mass flux $\Pi_m$.
Figure \ref{fig:3-2} presents the variation of the mass flux $\Pi_m(m_h,z)$ from the Illustris simulation. The mass flux $\Pi_{m} \left(m,a\right)$ can be computed from the cumulative mass $\Lambda_m\left(m_{h} ,t\right)$ at two different redshifts (see Eq. \eqref{eq:3-3}). We propose a scale-independent rate of cascade $\varepsilon_m(z)$ for $m_h<m_h^*$ in the propagation range. The simulation results confirm this concept. After reaching a statistically steady state around $z=4$, a propagation range is formed with a scale-independent $\varepsilon_m(a)$ for $m_h$ below a critical mass scale $m_h^*$. That rate $\varepsilon_m(a)\propto a^{-1}$ (predicted in Eq. \eqref{eq:3-8}), that is, decreases with time and is about -0.02$M_{\odot}/s$ at $z=0$ from the Illustris simulation. 

\begin{figure}
\includegraphics*[width=\columnwidth]{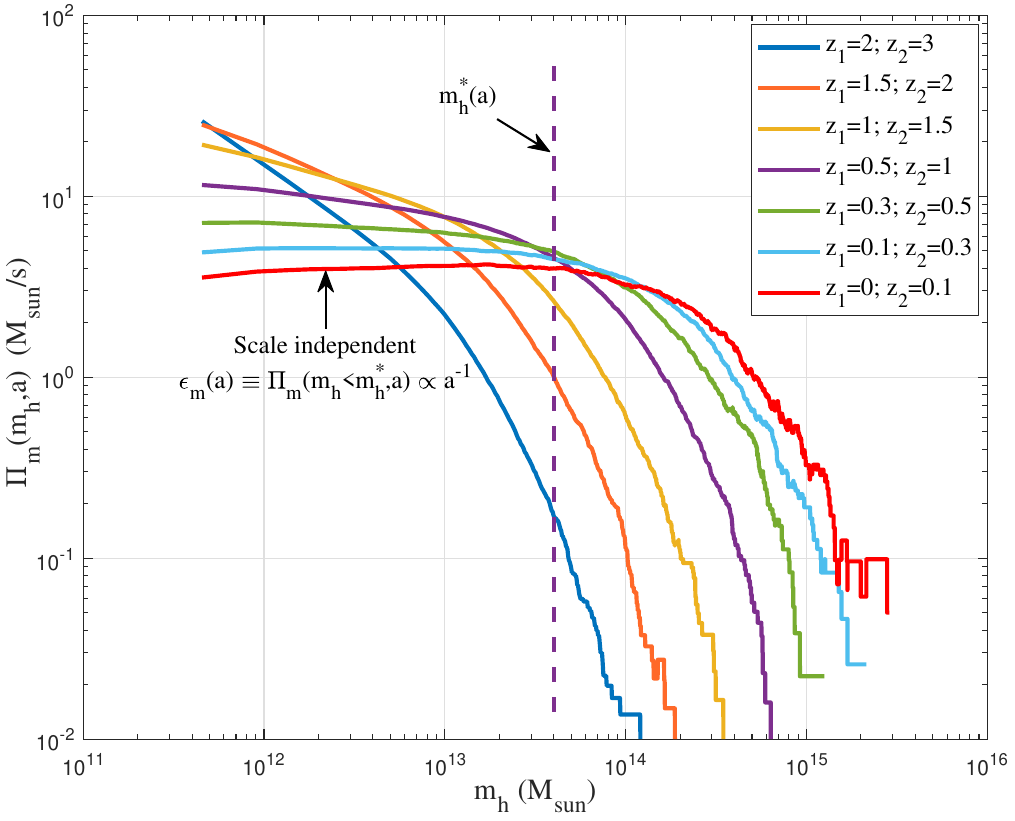}
\caption{The variation of mass flux $-\Pi _{m} \left(m_{h} ,z\right)$ from the Virgo simulation using halo group mass $m_{g} \left(a\right)$ at two different redshifts \textit{z} (Eq. \eqref{eq:3-3}). A scale-independent mass flux $\varepsilon _{m} \left(a\right)$ can be found for haloes smaller than the characteristic mass $m_{h}<m_{h}^{*}$. The negative mass flux $\varepsilon _{m}(z)<0$ indicates an inverse mass cascade from small to large scales. A mass propagation range with a scale-independent $\varepsilon_{m}$ is formed at around $z=0.3$, much later than the Illustris simulation in Fig. \ref{fig:3-2} due to a lower mass resolution.}
\label{fig:3-3}
\end{figure}

Figure \ref{fig:3-3} presents the evolution of the mass flux function $\Pi_{m} \left(m,z\right)$ from the Virgo simulation. Similarly, a scale-independent mass flux $\varepsilon_{m}(z)$ can be clearly identified for the halo mass $m_{h}<m_{h}^{*} $. The propagation range with scale-independent mass flux $\varepsilon_{m}$ is formed around $z=0.3$ and gradually expands in mass space. The time to reach the statistical state with a scale-independent rate of cascade is dependent on the simulation resolution. The higher mass resolution in the Illustris simulation leads to an earlier formation of the propagation range. The simulation with a higher resolution establishes the statistically steady state at a higher redshift. Both simulations confirm a scale-independent rate of mass cascade.

\begin{figure}
\includegraphics*[width=\columnwidth]{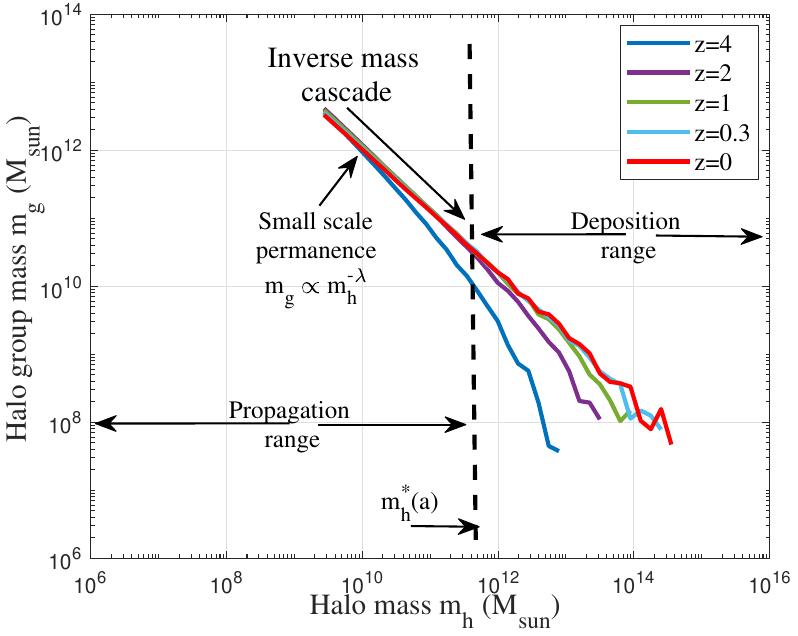}
\caption{The variation of halo group mass $m_g(m_h)$ for all haloes of the same mass $m_h$ at different redshifts $z$ from Illustris-1-Dark simulation. The figure demonstrates the small-scale permanence of the group mass $m_g$. Once the statistically steady state is established ($z\le4$), the rate of the inverse mass cascade $\varepsilon_m$ becomes scale independent. The mass flux into a halo group balances the mass flux out of the same group such that the group mass $m_g$ at different redshifts $z$ collapses into a time-independent power law $m_g\propto m_h^{-\lambda}$ on small mass scales. This is the so-called "propagation range," where the mass cascaded from the smallest (DM particle) scale is propagated to larger scales, that is, a constant group mass $m_g$. The cascaded mass is eventually consumed to grow haloes greater than $m_h^*$ (the "deposition range"). The propagation range gradually extends to large scales ($m_h^*(a)$ increases over time) due to the continuous inverse mass cascade. The slope $\lambda\approx$1.9 is related to the waiting time, which can be dependent on the initial spectrum.}
\label{fig:3-5}
\end{figure}

An important and direct consequence of a scale-independent rate of the cascade $\varepsilon_m(a)$ is the total mass of a halo group $m_g$ invariant in time. Taking the derivative of Eq. \eqref{eq:3-3} leads to
\begin{equation} 
\label{eq:3-3-3} 
\begin{split}
&\frac{\partial \Pi _{m}}{\partial m_h} = -\frac{\partial}{\partial t}\left(\frac{\partial \Lambda_m}{\partial m_h}\right)=-\frac{\partial m_g}{\partial t}=0,\\
&m_g(m_h) = M_h(t)f_M(m_h,z)m_p = n_h m_h.
\end{split}
\end{equation} 
Here, $n_h$ is the number of haloes with the same mass $m_h$. The mass of the entire halo group $m_g$ is independent of the redshift $z$ due to the scale-independent mass cascade. That is, the mass flux into a halo group is equal to the mass flux out of the same group, so the mass of the group $m_g$ is a function of $m_h$ only and is constant over time. This is the so-called small-scale permanence that is independent of the cosmological model and initial conditions. As long as a statistical steady state is established in dark matter flow, the halo group mass $m_g$ at different redshifts $z$ always collapses into a redshift-independent $m_g(m_h)$. Figure \ref{fig:3-5} demonstrates this concept from Illustris simulations. The mass of the halo group $m_g$ at different $z$ collapses into a simple power law $m_g(m_h)\propto m_h^{-\lambda}$, with an effective $\lambda\approx$1.9 determined by the initial power spectrum of Illustris.

In addition, with mass injected from the smallest scale $m_p$ at a scale-independent rate of $\varepsilon_m$, the rate of the mass cascade also equals the rate of change in total mass $M_h(t)$ in all haloes. From the definitions of the rate of mass cascade $\varepsilon_m$ and the halo group mass $m_g$ (Eq. \eqref{eq:3-2}), we have the following.
\begin{equation}
\begin{split}
&\frac{d}{dt}M_h(t)=-\varepsilon_m(z) = \frac{m_g(m_h)}{\tau_g(m_h,a)} \propto \frac{m_g(m_h)}{m_h^{-2/3}a}. \\
\end{split}
\label{eq:3-7}
\end{equation}
Combining Eq. \eqref{eq:3-7} with Eq. \eqref{eq:3-6} for $\tau_g\propto m_h^{-2/3}a$, additional scaling laws can be identified when the system reaches the statistically steady state with a scale-independent rate $\varepsilon_m$. In the universal transition range around mass $m_h^*$, when small-scale permanence is satisfied (time-independent $m_g$), 
\begin{equation}
\begin{split}
&m_g(m_h) = -\varepsilon_m(z)\tau_g \propto -\varepsilon_m(z) a {m_h}^{-2/3}, \\
&\textrm{leads to}\\
&\varepsilon_m(z)\propto a^{-1}, \quad M_h\propto a^{1/2}, \quad m_g\propto m_h^{-2/3}, \quad n_h\propto m_h^{-5/3}.
\end{split}
\label{eq:3-8}
\end{equation}
The rate of mass cascade $\varepsilon_m$ decreases with time as $\propto a^{-1}$, while the total halo mass $M_h(t)$ increases with time as $a^{1/2}$ (shown in Fig. \ref{fig:3-4}). More generally, the mass of the halo group $m_g(m_h)$ is independent of the redshift regardless of the initial conditions (small-scale permanence). In particular,  the group mass $m_g^*$ and the total number of haloes $n_h^*$ for the characteristic haloes of mass $m_h^*$ are read (using Eqs. \eqref{eq:3-3-3} and \eqref{eq:3-8-2})
\begin{equation}
\begin{split}
&m_g^* \propto M_h(t)(m_h^*)^{-1}m_p \propto a^{-1}, \\
&n_h^* \propto {M_h(t)m_p}/{{m_h^*}^2} \propto a^{-5/2}.
\end{split}
\label{eq:3-8-3}
\end{equation}
Therefore, the total halo mass $M_h$ (or the total mass of dark matter) is known as long as we know the particle mass $m_p$ and the number $n_h^*$ for characteristic haloes. If $n_h^*$=1 at $z$=0, $M_h\approx {m_h^*}^2/m_p$.

\subsection{The scale-to-scale energy cascade in mass space}
\label{sec:3-5}
In parallel with the mass cascade, there always exists a concurrent flux of energy across haloes of different characteristic scales. This is a unique feature of SG-CFD, as there is only an energy cascade in the incompressible viscous fluid. On cosmological scales, the total energy is not conserved, as implied by Noether’s theorem, because the expanding background lacks time-translation symmetry. As shown in Section \ref{sec:2-2}, the cosmic energy monotonically decreases with time in an expanding background. This behavior is analogous to dissipation in fluid turbulence, but here it arises from the expanding background rather than from viscous processes, and it does not entail radiation production. However, effective “dissipation” is not an intrinsic process operating on a global scale. Instead, energy is injected near the scale of the most massive haloes ($m_h^*$) and subsequently transferred to smaller scales, where it is ultimately “dissipated” by certain small-scale mechanisms. To make this mechanism explicit, we first demonstrate the intrinsic energy loss associated with each halo merging event, which provides a small-scale channel for the “dissipation” of the global cosmic energy budget (Fig. \ref{fig:S1-3-2}). Because merging events continuously remove energy, a sustained energy flux through halo mass space—i.e., an energy cascade—is required to balance the ongoing “dissipation.”
 
We consider binary merging for two haloes of mass $M_1$ and $M_2$, and size $r_{h1}$ and $r_{h2}$, respectively. The two haloes are initially well separated. Following merging, the remnant halo has a mass of $M_3=M_1+M_2$ and a size of $r_{h3}$. Assuming that all three haloes share a common bulk velocity $v_h$ that is approximately independent of halo mass \citep{Xu:2021-Inverse-and-direct-cascade-of-}, the kinetic energy associated with their translational motion is conserved, i.e., $M_1v_h^2+M_2v_h^2=(M_1+M_2)v_h^2$. We next evaluate the change in the internal (halo) energy due to merging, excluding the large-scale interaction energy between particles belonging to different haloes. By the virial theorem for a potential with an effective exponent $n_e$ ($PE\propto r^{n_e}$, see Section \ref{sec:2-2}), the internal kinetic and potential energies satisfy KE = $n_e/2$ PE. Therefore, the change in total internal energy associated with the merging is
\begin{equation}
\begin{split}
\Delta E_1 = -\left(1+\frac{n_e}{2}\right)\frac{GM_3^2}{r_{h3}}+\left(1+\frac{n_e}{2}\right)\left[\frac{G{M_1}^2}{r_{h1}}+\frac{G{M_2}^2}{r_{h2}}\right].
\end{split}
\label{eq:3-2-s1}
\end{equation}
The density of haloes is related to the mean density $\bar\rho_{DM}$ as
\begin{equation}
\begin{split}
\frac{M_1}{r_{h1}^3}=\frac{M_2}{r_{h_2}^3}=\frac{M_3}{r_{h_3}^3}=\frac{4}{3}\pi \Delta_c \bar{\rho}_{DM},
\end{split}
\label{eq:3-2-s2}
\end{equation}
where $\Delta_c=18\pi^2$ is the critical density ratio from the spherical collapse model. Substituting Eq. \eqref{eq:3-2-s2} into Eq. \eqref{eq:3-2-s1}, the change in total energy due to the halo merging reads
\begin{equation}
\small
\begin{split}
&\Delta E_1 = \left(1+\frac{n_e}{2}\right)\left(\frac{4}{3}\pi\Delta_c\bar\rho_{DM}\right)^{\frac{1}{3}}G{M_1}^{\frac{5}{3}}\left(1+\gamma^{\frac{5}{3}}-(1+\gamma)^{\frac{5}{3}}\right),
\end{split}
\label{eq:3-2-s3}
\end{equation}
where $\gamma=M_2/M_1$ is the mass ratio between two merging haloes. Clearly, the change in total energy due to the merging is always negative or $\Delta E_1<0$. Therefore, the structure formation must proceed in a "bottom-up" fashion via halo merging to continuously release system energy, not in a "top-down" fashion that requires an energy penalty. The halo merging provides a viable mechanism to "dissipate" the global-scale cosmic energy. For two merging haloes with a large mass ratio of $\gamma=M_2/M_1\ll 1$,  
\begin{equation}
\begin{split}
\Delta E_1 &= -\frac{5}{3}\left(1+\frac{n_e}{2}\right)G\left(\frac{4}{3}\pi\Delta_c\bar\rho_{DM}\right)^{\frac{1}{3}}{M_1}^{\frac{2}{3}}M_2\\
         &= -\frac{5}{3}\left(1+\frac{n_e}{2}\right)\frac{GM_1}{r_{h1}}M_2<0,
\end{split}
\label{eq:3-2-s4}
\end{equation}
where the change in energy is proportional to $M_1^{2/3}M_2$. In the smooth mass accretion limit, $M_2$ is just the mass of a single merger in Fig. \ref{fig:3-1}. The global-scale energy is continuously "dissipated" by every elementary merging event.

Next, we will consider scale-to-scale energy transfer. In halo mass space, the kinetic energy is cascaded from small to large mass scales at a rate of $\varepsilon_u$ (in the same direction as the mass cascade). In contrast, the potential energy cascades in the opposite direction at a rate of $-1.4\varepsilon_u$ (Fig. \ref{fig:S1-3-2}) to satisfy the virial theorem with $n_e=-10/7$ (Section \ref{sec:2-2}). Therefore, the total energy is cascaded to small scales at a rate of $-0.4\varepsilon_u$ and "dissipated" at the same rate. Similarly to the mass flux in Eq. \eqref{eq:3-2}, the flux of the specific kinetic energy across haloes reads (also see Eq. \eqref{eq:3-25})
\begin{equation}
\begin{split}
&\Pi_{pv}(m_h,z) = -\frac{m_h}{\tau_h}\frac{\overline {K_{pv}}}{\Lambda_m}=\Pi_m \frac{\overline {K_{pv}}}{\Lambda_m}, \\ 
&\varepsilon_u = \Pi_{pv}(m_h,z) \quad \textrm{for} \quad m_h\le m_h^*, 
\end{split}
\label{eq:3-2-2}
\end{equation}
where $\overline {K_{pv}}$ is the mean specific kinetic energy in all haloes greater than $m_h$ (defined in Eq. \eqref{eq:3-23}), while $\Lambda_m$ is the total mass in all haloes greater than $m_h$ (defined in Eq. \eqref{eq:3-4}). For every waiting time of $\tau_h$, the energy cascade leads to an increase in the specific kinetic energy of $m_h{\overline {K_{pv}}}/{\Lambda_m}$ on all scales greater than $m_h$.

To better describe the energy cascade, we first decompose the kinetic energy of the halo particles into two parts of different nature. In N-body simulations, every halo particle can be characterized by a mass $m_{p}$ and a velocity vector $\boldsymbol{\mathrm{v}}_{\boldsymbol{\mathrm{p}}}$. The particle velocity $\boldsymbol{\mathrm{v}}_{p}$ can be decomposed as \citep{Xu:2023-Maximum-entropy-distributions-of-dark-matter}
\begin{equation} 
\label{eq:3-21} 
\boldsymbol{\mathrm{v}}_{p} =\boldsymbol{\mathrm{v}}_{h} +\boldsymbol{\mathrm{v}}_p',           
\end{equation} 
namely, the mean velocity of the halo, $\boldsymbol{\mathrm{v}}_{h}=\langle \boldsymbol{\mathrm{v}}_{p} \rangle_h$ , and the fluctuation of the velocity, $\boldsymbol{\mathrm{v}}_{p}^{'} $. Here, $\left\langle \right\rangle _{h} $ represents the average of all particles in the same halo and $\boldsymbol{\mathrm{v}}_{h}$ represents the velocity of that halo. Consequently, the total kinetic energy $K_p$ of a given halo particle can be divided into $K_p = K_{ph}+K_{pv}$. Here, $K_{ph}=\boldsymbol{\mathrm{v}}_{h}^2/2$ (halo kinetic energy) is the contribution from the motion of entire haloes $\boldsymbol{\mathrm{v}}_{h}$ due to the inter-halo interaction of that particle with all other particles in different haloes and all out-of-halo particles. This part of the kinetic energy is related to interactions on large scales in the linear regime. 

The other part, $K_{pv}={\boldsymbol{\mathrm{v}}_p'}^2/2$ (virial kinetic energy), is the contribution of the velocity fluctuation $\boldsymbol{\mathrm{v}}_p'$ due to the intra-halo interaction of that particle with all other particles in the same halo. This part of the kinetic energy is from halo virialization and is due to interactions on a shorter distance and smaller scales in the non-linear regime. Similarly to the energy cascade associated with nonlinear interactions in turbulence, the energy cascade in dark matter flow focused on the cascade of the virial kinetic energy $K_{pv}$ due to nonlinear interactions on small scales.

Similarly to the cumulative mass function $\Lambda_m$ in Eq. \eqref{eq:3-4}, the cumulative kinetic energies ($\Lambda_{ph}$ and $\Lambda_{pv}$) represent the total kinetic energies $K_{ph}$ and $K_{pv}$ in all haloes greater than $m_h$, such that
\begin{equation} 
\label{eq:3-22} 
\begin{split}
&\Lambda_{ph}(m_h,t) = \int _{m_{h}}^{\infty } M_{h} \left(t\right)f_{M} \left(m,m_{h}^{*} \right) K_{ph} dm,  \\
&\Lambda_{pv}(m_h,t) = \int _{m_{h}}^{\infty } M_{h} \left(t\right)f_{M} \left(m,m_{h}^{*} \right) K_{pv} dm, \\
&K_{ph} = \frac{3}{2}\sigma_h^{2}\left(m,t\right) \quad \textrm{and} \quad K_{pv} = \frac{3}{2}\sigma_v^{2}\left(m,t\right). 
\end{split}
\end{equation} 
Here, $\sigma_h^{2}$ and $\sigma_v^{2}$ are the dispersion of the one-dimensional velocity for the halo velocity $\boldsymbol{\mathrm{v}}_{h}$ and the velocity fluctuation $\boldsymbol{\mathrm{v}}_{p}^{'}$.

Next, we will use the cumulative kinetic energy and cumulative mass $\Lambda_m$ to calculate the mean specific halo kinetic energy (energy per unit mass) $\overline {K_{ph}}$ and the virial kinetic energy $\overline {K_{pv}}$ in all haloes above any mass scale $m_h$, that is,
\begin{equation} 
\label{eq:3-23} 
\overline {K_{ph}}(m_h,t)=\frac{\Lambda_{ph}}{\Lambda_{m}} \quad \textrm{and} \quad \overline {K_{pv}}(m_h,t)=\frac{\Lambda_{pv}}{\Lambda_{m}}.
\end{equation}

We will focus on the energy cascade of the specific virial kinetic energy $K_{pv}$ due to nonlinear interactions on small scales. For the inverse mass cascade in Eq. \eqref{eq:3-3}, the change in total halo mass above the scale $m_h$, that is, the cumulative mass function $\Lambda_m(m_h, a)$, comes entirely from the mass cascade or the interactions between all haloes below the scale $m_h$ and all haloes above $m_h$. Similarly, the change in the (specific) virial kinetic energy $\overline {K_{pv}}$ for all haloes above the scale $m_h$ comes entirely from the energy cascade due to interactions between haloes below and above the scale $m_h$. 
Therefore, similar to the mass cascade $\Pi_m$ in Eq. \eqref{eq:3-3}, the rate of cascade for the virial kinetic energy $K_{pv}$ reads
\begin{equation} 
\label{eq:3-24} 
\begin{split}
\Pi_{pv}(m_h,t) &=-\frac{\partial }{\partial t} \left( \overline {K_{pv}} \right) = -\frac{\partial }{\partial t} \left(\frac{\Lambda_{pv}}{\Lambda_m} \right) \\
&=-\frac{\partial }{\partial t}\int _{m_{h} }^{\infty } \frac{M_{h} \left(t\right)f_{M} \left(m,m_{h}^{*} \right) K_{pv}}{\int _{m_{h} }^{\infty } M_{h} \left(t\right)f_{M} \left(m,m_{h}^{*} \right) dm} dm,
\end{split}
\end{equation}
where $\overline {K_{pv}}$ is defined in Eq. \eqref{eq:3-23}, i.e. the specific virial kinetic energy in all haloes greater than $m_h$. Similarly to Eq. \eqref{eq:3-3}, Eq. \eqref{eq:3-24} describes the rate of transfer of specific virial kinetic energy ($K_{pv}$) from haloes below the scale $m_h$ to haloes above the scale $m_h$ at a rate of $\Pi_{pv}$. 

\begin{figure}
\includegraphics*[width=\columnwidth]{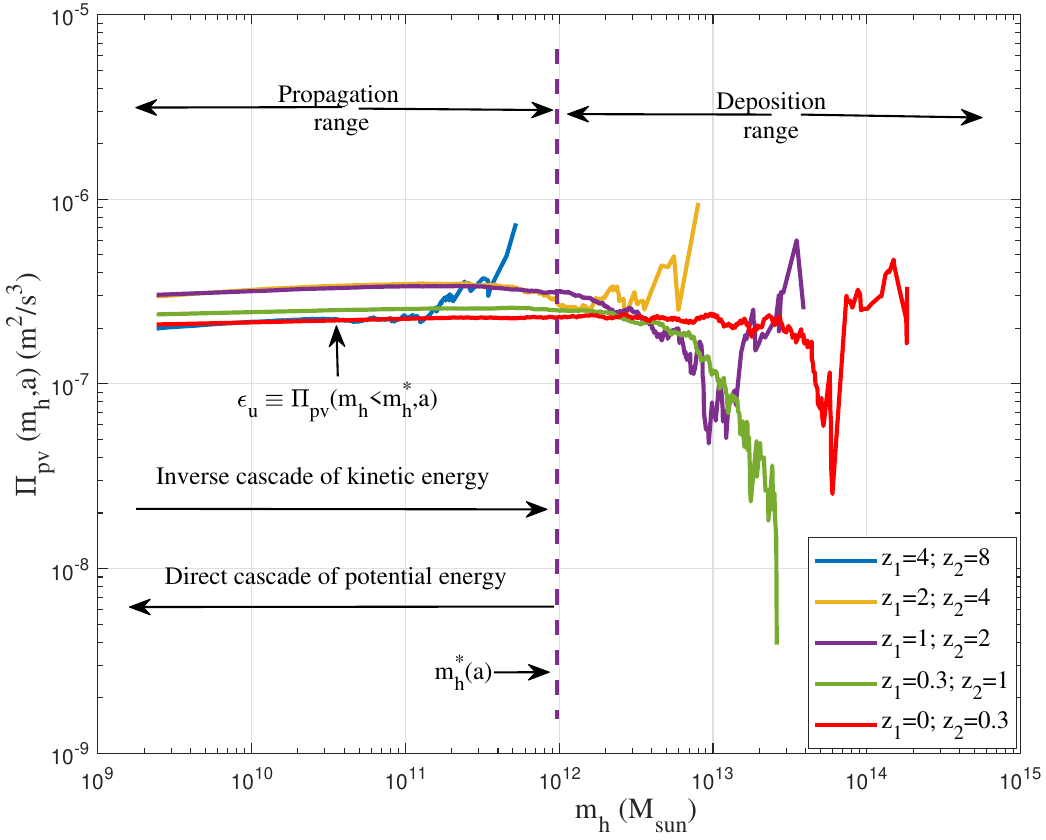}
\caption{The variation of the rate of energy cascade $\Pi_{pv}(m_h,a)$ (Eq. \eqref{eq:3-24}) with halo mass $m_h$ at different redshifts $z$ from Illustris-1-Dark simulation.  A scale-independent constant rate of $\varepsilon_u$ can be identified in the propagation range for an inverse cascade of virial kinetic energy ${K_{pv}}$. That rate is also relatively independent of time and is around $\varepsilon_u=-2.5\times 10^{-7}m^2/s^3$. According to the virial theorem, there also exists a simultaneous direct cascade of potential energy from large to small scales at -1.4$\varepsilon_u$.} 
\label{fig:3-7}
\end{figure}

Figure \ref{fig:3-7} plots the variation of $\Pi_{pv}$ with the halo mass $m_h$ and the redshifts $z$. The mean (specific) virial kinetic energy $\overline {K_{pv}}$ at two different redshifts $z_1$ and $z_2$ was used to calculate $\Pi_{pv}$ in this figure. Similarly to the mass cascade in Fig. \ref{fig:3-2}, if the statistical structures of the haloes are self-similar and scale-free for haloes smaller than $m_h^*$, the rate of the energy cascade $\varepsilon_u$ should also be independent of the scale $m_h$ for $m_p<m_h<m_h^*$. The simulation results confirm a scale- and time-independent rate of cascade $\varepsilon_u\approx -10^{-7}m^2/s^3$. Therefore, in the propagation range ($m_h<m_h^*$),
\begin{equation} 
\label{eq:3-25} 
\begin{split}
\varepsilon_u \equiv \Pi _{pv}(m_h,t) =-\frac{\partial \langle K_{pv}\rangle}{\partial t}\propto\varepsilon_m\frac{\langle K_{pv}\rangle}{M_h},
\end{split}
\end{equation} 
where the rate of energy cascade $\varepsilon_u$ can be directly related to the rate of the mass cascade $\varepsilon_m$ (see Eq. \eqref{eq:3-2-2}). Here, $\langle K_{pv}\rangle=\overline {K_{pv}}(0,t)$ is the mean specific kinetic energy for all halo particles. Similarly to Eq. \eqref{eq:3-7}, since $\varepsilon_u$ is scale independent, with continuous injection of virial kinetic energy $K_{pv}$ at a constant rate of $\varepsilon_u$ on the smallest mass scale, the rate of the energy cascade equals the rate of change of $\langle K_{pv}\rangle$ in all haloes. Therefore, we expect the total virial kinetic energy of all halo particles to be proportional to time $t$ or $\langle K_{pv}\rangle\propto t$. This is consistent with the evolution of cosmic energy in Eq. \eqref{eq:4-1-s}.

\section{Energy cascade and halo density profile}
\label{sec:3-3}

In this section, we analyze the cascade of mass and energy operating within individual dark matter haloes. Because the dark component is collisionless, it does not support turbulent cascades driven by viscous dissipation and vorticity; similarly, a halo that has reached full virial equilibrium cannot sustain a net transfer of mass or energy across spatial scales. Accordingly, we use the term "cascade" to denote the scale‑to‑scale redistribution of mass and energy that occurs in nonequilibrium haloes as they relax toward that limiting equilibrium, rather than to imply a viscous, Kolmogorov‑type turbulent cascade. Analogously to the cascade and halo random walk in halo mass space, we characterize the intra‑halo mass and energy fluxes that play a central role in shaping halo internal structure (i.e., the density profile) \citep{Xu:2023-Dark-matter-halo-mass-functions-and}, underpinning the universal scaling relations of dark‑matter haloes \citep{Xu:2023-Universal-scaling-laws-and-density-slope}, and informing constraints on the properties of dark matter particles \citep{Xu:2022-Postulating-dark-matter-partic}. The same theory may also be applied to the evolution of the galaxy, yielding scalings for the mass and size of the bulge \citep{Xu:2024-Cosmic-quenching-and-scaling-laws}.

\subsection{Particle random walk and distribution in haloes}
\label{sec:3-3-1}
Similarly to the halo mass function, the halo structure also depends on both initial conditions and gravitational dynamics. A more negative index or steeper spectrum usually leads to haloes with lower concentrations and less dense central cusps. Using Eq. \eqref{eq:2-2-2-9} for asymptotic spectrum index $n_a$, a simple relation between the asymptotic density $r^{-\gamma_a}$ and the initial spectrum index $n_i$ can be obtained as
\begin{equation} 
\label{eq:3-25-3} 
\begin{split}
\gamma_{a}=3+n_{a}=\frac{(3-2\alpha_x)\left(3+n_{i}\right)}{\left(1-\alpha_x\right)\left(3+n_{i}\right)+2}.\\
\end{split}     
\end{equation}
Here, we assume that the slope of the halo density is equal to the slope of the two-point correlation function $\xi(r)\propto r^{-(3+n_a)}$. For the ratio parameter $\alpha_x=0$, this result reduces to the standard suggested result in \cite{Syer:1998-Dark-halo-mergers-and-the-formation}.

Due to the inside-out nature of halo assembly, central regions collapse at high redshift from peaks in the primordial density field. Their structure is therefore largely set by the properties of the initial high-density peak and retains a stronger memory of the shape of the primordial power spectrum than do outer regions. By contrast, the outskirts are assembled later through mergers and smooth accretion; these processes drive violent relaxation, phase mixing, and tidal stripping, which progressively erase the imprint of the initial spectrum and render the dynamics near a characteristic scale $r_s$ predominantly universal. Consequently, the density profile near $r_s$ is expected to be approximately universal, governed primarily by gravitational dynamics, and only weakly sensitive to the initial spectrum. 

To elucidate this behavior, consider an idealized initial condition in which the proto-halo is highly concentrated, with its mass localized at $r$=0, representing the collapse of an extreme high-redshift peak. In this limit, the influence of the primordial spectrum is confined to a central boundary, whereas the subsequent evolution proceeds subject to self-gravity in an expanding background. This initial setting resembles the spherical collapse model (SCM). However, SCM enforces a uniform (top-hat) density and thus cannot capture the density variation. We instead adopt an effective mean-field kinetic description without considering gravity between individual particles. In contrast, dark matter particles move within a self-consistent gravitational potential field determined by the instantaneous particle distribution, permitting analytical halo density profiles.

The evolution of the particle distribution is modeled as a three-dimensional continuous-time random walk with a radius-dependent residence (waiting) time. In the gravity-dominated regime, this residence time is controlled by the local potential $\tau_{gr}(r)\propto \Phi_r(r)^{-1}\propto r^{-\gamma}$ (Eq. \eqref{eq:4-3}), which admits an interpretation in terms of an effective local temperature: deeper potentials correspond to higher effective temperatures, larger particle kinetic energies, and shorter residence times. Around scale $r_s$, migration is dominated by gravity, yielding $\gamma=2/3$ (Eq. \eqref{eq:4-4-4-2}). When the primordial spectrum dominates, the exponent $\gamma$ is expected to be related to the effective spectral index of the initial power spectrum (Eq. \eqref{eq:3-25-3}). In a smooth dark-matter background, stochastic particle random walk expands the halo size and facilitates accretion. With smooth mass accretion, mass growth increases the total mass of the halo while leaving its internal structural profile essentially unchanged.

Next, we need to derive the particle distribution due to the random walk of particles. The 3D particle random walk can be described by a Langevin equation for the particle position $\boldsymbol{X}_{t}$, 
\begin{equation} 
\label{eq:4-4-1} 
\frac{d\boldsymbol{X}_{t} }{dt} = \sqrt{2 D_P(\boldsymbol{X}_t)} \boldsymbol{\xi}\left(t\right)a^{-1/2},       
\end{equation} 
where the factor $a^{-1/2}$ reflects the Langevin equation in an expanding background with redshifting velocity fluctuations as $v^2\propto T\propto a^{-1}$. Without loss of generality, a power law $\tau_{gr}(r)\propto r^{-\gamma}$ can be assumed, i.e., a position-dependent waiting time. Here, $r\equiv |\boldsymbol{X_t}|$ is the distance from the particle to the center of the halo. From this, the position-dependent diffusivity reads 
\begin{equation} 
\label{eq:4-4-2} 
D_{P}(\boldsymbol{X}_t)=D_{0}(t) r^{2\gamma},         
\end{equation} 
where $D_{0}(t)$ is a proportional constant. The closer to the halo center (smaller $r$), the smaller the diffusivity or the longer the waiting time, and the higher the density of the particles. 
By introducing a time $\tau$ proportional to the scale factor $\tau=(3/2)at_0$ ($t_0$ is the present epoch), $a^{-1/2}$ can be absorbed, and the original Eq. \eqref{eq:4-4-1} reads 
\begin{equation} 
\label{eq:4-4-1-2} 
\frac{d\boldsymbol{X}_{\tau} }{d\tau} = \sqrt{2 D_P(\boldsymbol{X}_\tau)} \boldsymbol{\xi}\left(\tau\right).       
\end{equation} 

For the stochastic integral, the Ito interpretation yields the limit of a discrete random process as the length of the discrete intervals tends towards zero. In It$\hat{\textrm{o}}$ convention \citep{Sokolov:2010-Ito-Stratonovich-Hänggi}, the 3D Fokker-Planck equation corresponding to Langevin Eq. \eqref{eq:4-4-1-2} can be directly obtained for particle distribution function $P_{r} \left(\boldsymbol{X},\tau\right)$ ($i=1,2,3$ for Cartesian coordinates),
\begin{equation} 
\label{eq:4-4-3} 
\begin{split}
\frac{\partial P_{r} \left(\boldsymbol{X},\tau\right)}{\partial \tau} = D_0\frac{\partial }{\partial X_i}\left[ \frac{\partial }{\partial X_i}\left(r^{2\gamma}P_{r} \left(\boldsymbol{X},\tau\right)\right) \right]. 
\end{split}
\end{equation} 
The corresponding solution of \eqref{eq:4-4-3} in spherical coordinates reads 
\begin{equation}
\label{eq:4-4-4} 
P_{r} \left(r,\tau\right) = \frac{(2-2\gamma)^{\frac{\gamma-2}{1-\gamma}} r^{-2\gamma}}{4\pi\left(D_0\tau\right)^{\frac{3-2\gamma}{2-2\gamma}}\Gamma\left(\frac{3-2\gamma}{2-2\gamma}\right)}\exp \left(-\frac{r^{2-2\gamma}}{4(1-\gamma)^2D_0\tau}\right).        
\end{equation} 
Since the distribution function $P_r(r,\tau)$ is equivalent to the halo density, we find that the halo density $\rho_r\propto r^{-2\gamma}$ for the waiting time $\tau_{gr}(r)\propto r^{-\gamma}$. From this insight, assuming that $\gamma$ is unknown, we can predict $\gamma=2/3$ for gravity-dominant halo evolution.

Since the waiting time $\tau_{gr} \propto \Phi(r)^{-1} \propto r^{-\gamma}$, the halo density should scale as $\rho_r \propto r^{-2\gamma}$ from Eq. \eqref{eq:4-4-4}. The halo mass enclosed in $r$ scales as $m_r \propto \rho_r r^3 \propto r^{3-2\gamma}$. The local potential at $r$ should scale as $\Phi(r) \propto Gm_r/r \propto r^{3-2\gamma-1}$. The waiting time of the particle at $r$ satisfies $\tau_{gr}\propto \Phi^{-1}\propto r^{2\gamma-2}$. Combined with $\tau_{gr} \propto r^{-\gamma}$, we have 
\begin{equation}
\label{eq:4-4-4-2} 
2\gamma-2=-\gamma \quad \textrm{such that}\quad \gamma=2/3. 
\end{equation} 
The density slope should be $2\gamma=4/3$. Therefore, the halo density scales as $\rho_r\propto r^{-4/3}$ and the waiting time $\tau_{gr}\propto r^{-2/3}$. This interesting result demonstrates that in the absence of the effect of the initial spectrum, gravity should drive the haloes toward a limiting slope of $-4/3$. For a realistic CDM spectrum, although the core structure can be affected by the initial spectrum, the universal slope of $-4/3$ should be true for large halos of characteristic mass $m_h^*$ or for haloes on scales close to $r_s$. It should be noted that the random walk theory for halo formation and evolution confirms a -4/3 law ($\rho_r\propto r^{-4/3}$) that can also be predicted by the energy cascade in haloes with a scale-independent rate (Eq. \eqref{eq:4-3-2}). The predictions of the -4/3 scaling law are tested against the simulations in Figs. \ref{fig:4-5} and \ref{fig:4-4}. 

Similarly to the halo mass function (Eq. \eqref{eq:3-18}), the exponent $\gamma$ can be different in two different scale ranges, the inner halo below the scale radius $r_s$ that is affected by the initial spectrum and the outer region around scale $r_s$ dominated by gravitational dynamics. Using two different $\gamma$, that is, $\gamma_1$ and $\gamma_2$ for two different ranges, the double-$\gamma$ distribution function can be obtained based on the single-$\gamma$ distribution in Eq. \eqref{eq:4-4-4},
\begin{equation}
\label{eq:4-4-5} 
P_{r} \left(r\right) = \frac{(2-2\gamma_2)^{\frac{2\gamma_1-2-\gamma_2}{1-\gamma_2}} r^{-2\gamma_1} }{4\pi\left(D_0\tau\right)^{\frac{3-2\gamma_1}{2-2\gamma_2}}\Gamma\left(\frac{3-2\gamma_1}{2-2\gamma_2}\right)} \exp \left(-\frac{r^{2-2\gamma_2}}{4(1-\gamma_2)^2D_0\tau}\right).         
\end{equation} 
Introducing the conventional scale radius $r_s(t)$ where the logarithmic slope of $P_{r}(r,t)$ is equal to -2, we should have
\begin{equation}
\label{eq:4-4-6} 
4(1-\gamma_2)^2D_0\tau = \frac{2-2\gamma_2}{2-2\gamma_1}r_s^{2-2\gamma_2}.         
\end{equation} 
Here, the characteristic scale $r_s\propto a^{1/(2-2\gamma_2)}$ also corresponds to the logarithmic slope of the dimensionless spectrum $\Delta^2(k)$ that is 2, or where the power spectrum $P(k)\propto k^{-1}$, i.e., in the universal transition range in Fig. \ref{fig:999}. For gravitational dominant dynamics,$\gamma_1=2/3$ and the scale radius $r_s\propto a^{3/2}\propto t$.

Substituting Eq. \eqref{eq:4-4-6} into Eq. \eqref{eq:4-4-5} and introducing a dimensionless spatial-temporal variable $x=r/r_s(t)$, the distribution reads
\begin{equation}
\label{eq:4-4-7} 
P_{r} \left(x\right) = \frac{(1-\gamma_2) x^{-2\gamma_1}}{2\pi\Gamma\left(\frac{3-2\gamma_1}{2-2\gamma_2}\right)\left(\frac{1-\gamma_2}{1-\gamma_1}\right)^{\frac{3-2\gamma_1}{2-2\gamma_2}}}\exp \left(-\frac{1-\gamma_1}{1-\gamma_2}x^{2-2\gamma_2}\right).         
\end{equation} 
Finally, the two-parameter particle distribution reads
\begin{equation}
\label{eq:4-4-8} 
P_{r} \left(x\right) = \frac{\alpha \beta^{-(\frac{1}{\alpha}+\frac{1}{\beta})}}{4\pi\Gamma\left(\frac{1}{\alpha}+\frac{1}{\beta}\right)} x^{\frac{\alpha}{\beta}-2} \exp\left(-\frac{x^{\alpha}}{\beta}\right),      
\end{equation} 
where two dimensionless parameters $\alpha$ and $\beta$ are
\begin{equation}
\label{eq:4-4-9-2} 
\alpha = 2-2\gamma_2 \quad \textrm{and} \quad \beta = \frac{1-\gamma_2}{1-\gamma_1}.
\end{equation} 
Both parameters $\alpha$ and $\beta$ can be related to $\gamma$ for the scale dependence of waiting time $\tau_{gr}(r)\propto r^{-\gamma}$. The time variation of the distribution function is absorbed into the scale radius $r_s(t)$. The double-$\gamma$ distribution (Eq. \eqref{eq:4-4-8}) reduces to the Einasto profile with $\alpha=2\beta$ or $\gamma_1=0$. The cumulative distribution in spherical coordinates reads
\begin{equation}
\label{eq:4-4-10} 
\int_{0}^{x} P_{r} \left(y\right) 4\pi {y}^2 d{y} = 1-\frac{\Gamma\left(\frac{1}{\alpha}+\frac{1}{\beta},\frac{x^{\alpha}}
{\beta}\right)}{\Gamma\left(\frac{1}{\alpha}+\frac{1}{\beta}\right)},    
\end{equation} 
where $\Gamma(x,y)$ is an upper incomplete gamma function. 

\subsection{Halo density profiles and universal scaling laws}
\label{sec:4-4}
The halo density profile can be inferred from the particle radial distribution function $P_r$. The well-known core–cusp problem highlights the tension between the cuspy inner profiles produced by cosmological cold dark matter (CDM)–only simulations and the approximately constant-density cores indicated by observations. In CDM simulations, the inner density typically follows $\rho\propto r^s$, with reported logarithmic slopes s in the range -1.0 to -1.5 (see Introduction Section \ref{sec:1}). There is neither consensus on a single value of s nor a robust first-principles theory that uniquely predicts it. This non-universality reflects the joint influence of the primordial power spectrum and nonlinear gravitational dynamics on halo structure: the inner halo retains greater memory of the initial spectrum and thus exhibits CDM-dependent variation in $s$, whereas the region near the scale radius $r_s$ is more strongly shaped by gravity and correspondingly displays a more universal profile.

A double-$\gamma$ density profile can accommodate both effects through two distinct parameters. We interpret $\gamma_1$ as an effective descriptor of the inner structure, governed by the primordial power spectrum, whereas $\gamma_2$ encapsulates the influence of gravitational dynamics. The general expression reads
\begin{equation}
\label{eq:4-4-15} 
\rho_r(r,t)= \rho_s(m_h,t) \left(\frac{r}{r_s}\right)^{\frac{\alpha}{\beta}-2} \exp\left(\frac{1}{\beta}\left(1-{\left(\frac{r}{r_s}\right)^{\alpha}}\right)\right),
\end{equation}
where $\rho_s(t)$ is the density at scale radius $r_s(t)$. This four-parameter double-$\gamma$ density profile ($\rho_s$, $r_s$, $\alpha$, and $\beta$ in Eq. \eqref{eq:4-4-15}) reduces to the three-parameter Einasto profile (($\rho_s$, $r_s$, and $\alpha$) with $\alpha=2\beta$. 

For massive haloes near the characteristic mass $m_h^*$, the structure across most spatial scales is governed primarily by gravitational dynamics and the associated radial mass and energy flux\citep{Xu:2023-Universal-scaling-laws-and-density-slope, Xu:2023-Dark-matter-halo-mass-functions-and}, producing a universal logarithmic density slope of -4/3. The same behavior also holds in the vicinity of the scale $r_s$ for haloes of arbitrary mass $m_h$. Under these conditions, the double-$\gamma$ density profile is obtained with $\alpha/\beta=2/3$ in Eq. \eqref{eq:4-4-15}, or $\gamma_1=2/3$, and it reads:
\begin{equation}
\label{eq:4-4-16} 
\begin{split}
&\rho_r(r,m_h,z) = \beta_r\varepsilon^{\frac{2}{3}}G^{-1}r_s^{-\frac{4}{3}}\left(\frac{r}{r_s}\right)^{-\frac{4}{3}}\exp\left[-\frac{1}{\beta}\left(\frac{r}{r_s}\right)^{\frac{2\beta}{3}}\right],\\
&\varepsilon(m_h,z) = \left({m_h}/{m_h^*(z)}\right)^{{2}/{3}}\varepsilon_u,
\end{split}
\end{equation}
where $\beta_r\approx 1$ is an amplitude parameter, $\beta$ is a shape parameter in Eq. \eqref{eq:4-4-9-2}. Here, we introduce a parameter $\varepsilon$ (m$^2$s$^{-3}$) to replace the density $\rho_s(t)$ in Eq. \eqref{eq:4-4-15} to respect the -4/3 law on scales near $r_s$. This -4/3 law for density should only involve the gravitational constant $G$ that reflects the dynamics governed by gravity (see the derivation of scaling laws in Eq. \eqref{eq:4-3-2}): 
\begin{equation}
\label{eq:4-4-16-2} 
\begin{split}
&\rho_s(m_h,t)=\beta_r\exp(-1/\beta)\varepsilon^{2/3}G^{-1}r_s^{-4/3}.
\end{split}
\end{equation}
The parameter $\varepsilon(m_h,z)$ represents the rate of the energy flow in haloes of mass $m_h$ (defined in Eq. \eqref{eq:4-3-1}), while $\varepsilon_u=\varepsilon(m_h^*,z)\approx 10^{-7}m^2/s^3$ is the rate of the energy flow in haloes with characteristic mass $m_h^*(z)$. This is also the rate of "dissipation" for global-scale cosmic energy (Section \ref{sec:2-2}). Both parameters and the associated mass and energy flow in nonequilibrium haloes control the halo structure near $r_s$. Its physical meaning is discussed in the next subsection \ref{sec:3-3-2}. The values are estimated from N-body simulations (Figs. \ref{fig:4-2} and \ref{fig:4-3}).

In particular, for small $r$, the inner density for gravitational dynamics dominant haloes (Eq . \eqref{eq:4-4-16}) is reduced to the -4/3 law:
\begin{equation}
\label{eq:4-4-17} 
\rho_r(r,m_h,z) = \beta_r \varepsilon^{\frac{2}{3}}G^{-1}r^{-\frac{4}{3}}\propto m_h^{\frac{4}{9}}a^{-\frac{2}{3}}r^{-\frac{4}{3}} \quad \textrm{for} \quad r\rightarrow 0.
\end{equation}
The redshift dependence is absorbed into the parameter $\varepsilon$. For haloes of characteristic mass $m_h^*(z)\propto a^{3/2}$, there exists a small-scale permanence for the halo density, i.e., the density profiles at different redshifts $z$ converge to a time-unvarying scaling,
\begin{equation}
\label{eq:4-4-18} 
\rho_r(r,m_h^*,z) \equiv \rho_r(r) =\beta_r \varepsilon_u^{2/3}G^{-1}r^{-4/3} \quad \textrm{for} \quad r\rightarrow 0.
\end{equation}
The scaling law  shows the density of the halo $\rho_r(r)\propto m_h^{4/9}a^{-2/3}$ for haloes of mass $m_h$. Clearly, halo mass accretion leads to an increase in $m_h$ and density $\rho_r$, while expansion leads to a decrease in $\rho_r$ over time. The two competing effects are equal for haloes with characteristic mass $m_h^*\propto a^{3/2}$ such that the inner density $\rho_r$ for haloes with characteristic mass $m_h^*$ remains constant over time. This is the so-called small-scale permanence for the halo density. The density of haloes of mass $m_h^*$ at different redshifts collapses into the same -4/3 law. For old and small haloes with a mass less than $m_h^*$, the mass accretion is slower with a smaller $\varepsilon$, leading to a decrease in density $\rho_r$ over cosmic time $t$, as shown in Eq. \eqref{eq:4-4-17}.  

\begin{figure}
\includegraphics*[width=\columnwidth]{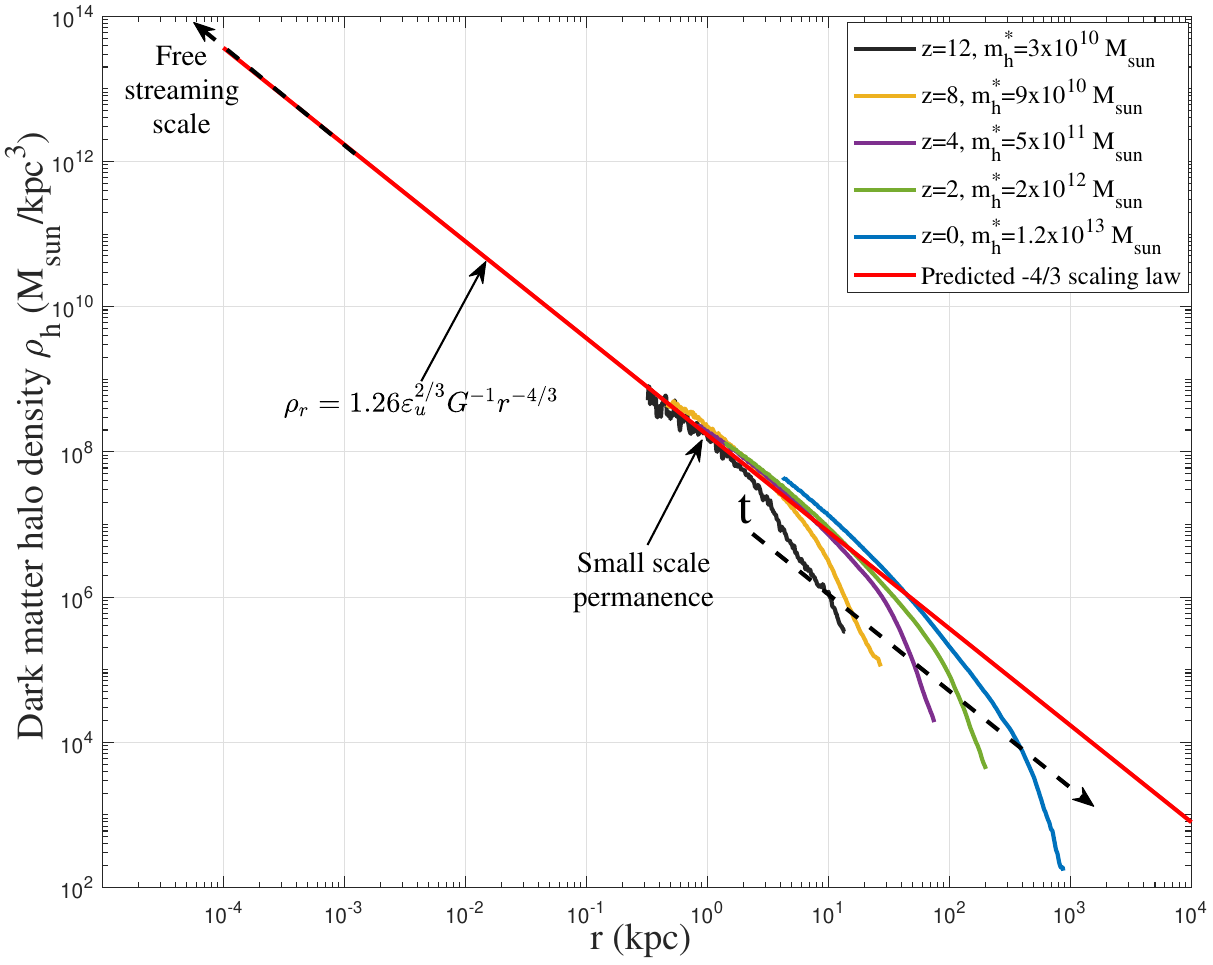}
\caption{The redshift evolution of halo density profiles for haloes with a characteristic mass $m_h^*(z)$. Density profiles for haloes with other masses (from the galaxy cluster to the Earth size) are presented in Fig. \ref{fig:4-4}. This figure demonstrates the small-scale permanence, i.e., the density profiles for haloes of mass $m_h^*(z)$ at different redshifts $z$ all collapse onto the predicted time-unvarying scaling (solid red line from Eq. \eqref{eq:4-4-18}). Due to the time and scale-independent rate of the cascade $\varepsilon_u$, the -4/3 scaling should extend to smaller scales (or earlier time) until reaching the smallest scale (free streaming scale or the formation time of the smallest structures) that depends on the nature of dark matter.} 
\label{fig:4-5}
\end{figure}

To validate the predicted halo density and scaling laws, we present the N-body simulation results. Figure \ref{fig:4-5} demonstrates the small-scale permanence for the halo density profile, similarly to the small-scale permanence for the halo group mass $m_g$ (Fig. \ref{fig:3-5}). The figure plots the time evolution of the average density profiles for all haloes with characteristic masses $m_h^*(z)$ from Illustris simulations at different redshifts. All haloes have a limiting constant density slope of -4/3. The density profiles of all dark matter haloes of mass $m_h^*$ converge to the predicted time-unvarying scaling (solid red line from Eq. \eqref{eq:4-4-18}) on small scales, i.e., the small-scale permanence. The -4/3 scaling law in Fig. \ref{fig:4-5} should extend to smaller and smaller scales until the smallest scales. That scale can be the free streaming scale, which is dependent on the nature and mass of dark matter particles. The properties of dark matter particles may be obtained by extending these scaling laws to the smallest scale \citep{Xu:2022-Postulating-dark-matter-partic}.

\begin{figure}
\includegraphics*[width=\columnwidth]{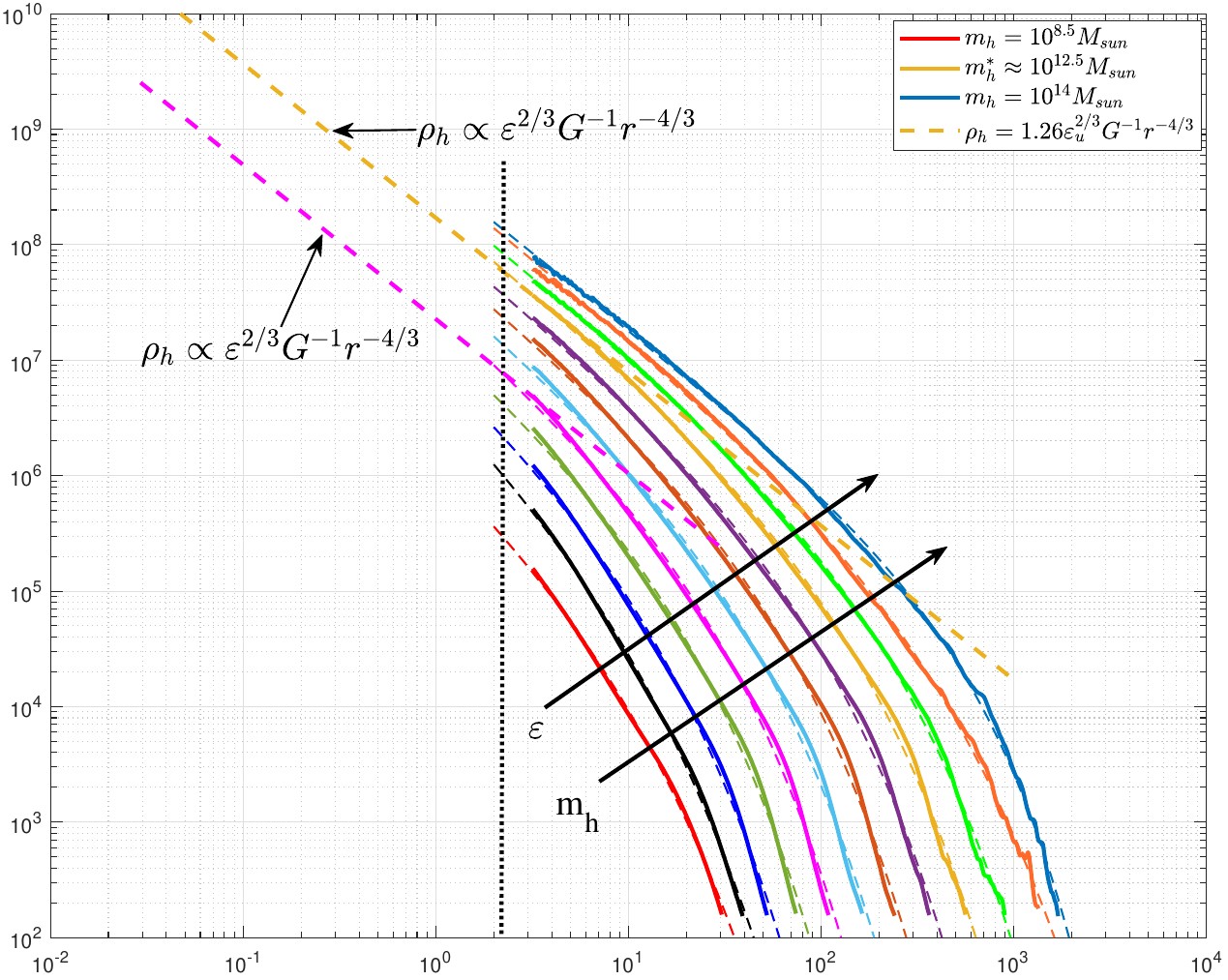}
\caption{The density profiles for haloes of different masses at $z=0$. The solid lines present the average density profiles for all haloes with a mass between $10^{\pm 0.1}m_h$ from the Illustris simulation for galactic haloes of mass $m_h=[10^{8.5} 10^{14}]M_{\odot}$. The thin dashed lines present the double-$\lambda$ density profile from Eq. \eqref{eq:4-4-16}. The thick dashed (straight) lines represent the -4/3 scaling law for inner halo density with $\varepsilon\propto m_h^{2/3}$ (Eq. \eqref{eq:4-2-6}), and for haloes of characteristic mass $m_h^*(z=0)$. The dotted lines indicate the softening length. The figure demonstrates the -4/3 law for the inner density of haloes of different masses over six orders of magnitude. The density profiles of haloes with mass $m_h^*(z)$ at different redshifts $z$ are presented in Fig. \ref{fig:4-5}.} 
\label{fig:4-4}
\end{figure}

Figure \ref{fig:4-4} presents the density profiles from Illustris simulations for (composite) haloes of different masses $m_h$ at $z=0$. The solid lines show the average halo density profiles for all haloes with a mass between $10^{\pm 0.1}m_h$ from the Illustris simulation, where the halo mass $m_h$ is between $10^8M_{\odot}$ and $10^{14}M_{\odot}$. 
The thick dashed (straight) line represents the -4/3 scaling laws for haloes of mass $m_h$ that involve the rate of cascade $\varepsilon\propto m_h^{2/3}$ (Eq. \eqref{eq:4-2-6}). Here, $\varepsilon=\varepsilon_u$ for haloes of characteristic mass $m_h^*$ (Eq. \eqref{eq:4-2-6}). The figure demonstrates the -4/3 law that we obtained from the energy cascade in haloes. That scaling was in agreement with the density profiles of haloes of more than six orders of magnitude. The thin dashed lines in both Figs. \ref{fig:4-4} and \ref{fig:4-5} present the double-$\gamma$ density profiles from Eq. \eqref{eq:4-4-16} that agree well with the simulation data. 

Universal scaling laws are well known to be associated with the energy cascade in fluid turbulence \citep{Richardson:1922-Weather-Prediction-by-Numerica, Kolmogoroff:1941-Dissipation-of-energy-in-the-l}. Next, we focus on the scaling laws in spherical haloes governed by gravitational dynamics, which can be derived from the particle distribution. Because of the slope $\rho_r\propto r^{-4/3}$ (Eq .\eqref{eq:4-4-4}), we have the total mass $m_r\propto \rho_r r^3\propto r^{5/3}$ enclosed within r, the characteristic velocity $v_r^2\propto Gm_r/r\propto r^{2/3}$, and the time $t_r\propto r/v_r\propto r^{2/3}$ on scale r. Next, a simple definition of $\varepsilon$ that is independent of scale r reads,
\begin{equation}
\begin{split}
\varepsilon = \frac{v_r^2}{2t_r} = \frac{v_r^3}{3r}.
\end{split}
\label{eq:4-3-1}
\end{equation}
The microscopic interpretation of $\varepsilon$ is associated with the energy transfer in haloes in the next section (Eq. \eqref{eq:2-2-9}).

The rate of the energy cascade $\varepsilon$ is independent of $r$ on a certain range of scales. The 2/3 law can be easily obtained for kinetic energy $v_r^2\propto(-\varepsilon r)^{2/3}$ from Eq. \eqref{eq:4-3-1}. With the virial theorem ($v_r^2\propto Gm_r/r$), the 5/3 scaling law can be recovered for the mass enclosed within $r$, i.e., $m_r\propto r^{5/3}$. We can write the scaling laws for mass $m_r$, density $\rho_r$, velocity $v_r$, time $t_r$, and kinetic energy $v_r^2$, and mass flux $\dot m_r$ on scale $r$, all determined by three and only three quantities $\varepsilon$, $G$, and scale $r$ reflecting gravity-dominant dynamics:
\begin{equation} 
\label{eq:4-3-2} 
\begin{split}
&m_r = \alpha_r \varepsilon^{2/3}G^{-1}r^{5/3} \textrm{,} \quad \rho_r = \beta_r \varepsilon^{2/3}G^{-1}r^{-4/3}, \\
&v_r^2=\beta_h^{-2/3}(\varepsilon r)^{2/3} \textrm{,} \quad t_r =0.5 \beta_h^{-2/3} \varepsilon^{-1/3}r^{2/3}, \\
&v_r^2=\beta_h^{-2/3}\alpha_r^{-2/5}(\varepsilon G m_r)^{2/5}, \quad \dot m_r=4\pi \beta_h^{-1/3}\beta_r\varepsilon G^{-1}r
\end{split}
\end{equation} 
where $\alpha_r$ and $\beta_r$ are two numerical factors that can be determined by fitting the data (Fig. \ref{fig:4-5}). The evolution of these quantities is absorbed into $\varepsilon(m_h,z)$. The relations between numerical factors are read as
\begin{equation} 
\label{eq:4-3-2-1} 
\begin{split}
\beta_h=\frac{1}{3},\quad \gamma_r\alpha_r=\beta_h^{-2/3},\quad \textrm{and}\quad \beta_r=\frac{5}{4\pi}\alpha_r\beta_h
\end{split}
\end{equation} 

Another approach to derive these scaling laws using the dimensional argument is relatively simpler but more heuristic. When a statistically steady state is established due to gravitational dynamics, the fast motion on small scales does not feel the slow motion on large scales directly, except through the scale-independent rate of the energy cascade $\varepsilon$. The flow fields on these scales are statistically similar so that all relevant physical quantities can be determined by and only by three quantities: $\varepsilon$, $G$, and scale $r$. Therefore, the scaling laws for any quantity $Q=\varepsilon_u^xG^yr^z$ can be determined from the dimensional analysis to obtain the exponents $x$, $y$, and $z$. This approach leads to the same scaling laws as in Eq. \eqref{eq:4-3-2}.

\subsection{Particle random walk and energy cascade}
\label{sec:3-3-2}
Fully virialized haloes cannot sustain mass or energy transport in the limit of vanishing energy flux ($\varepsilon$=0); Therefore, $\varepsilon$=0 represents the formal equilibrium state toward which haloes asymptotically relax but never truly attain. In this section, we examine the mass and energy fluxes in nonequilibrium haloes that are evolving toward that limiting equilibrium. These fluxes are driven by the stochastic random walk of individual particles within the halo potential, which leads to a transfer of mass and energy across scales. This microscopic diffusion of particles is analogous, on the internal halo scale, to the random walk of haloes in the mass space that underlies the mass and energy cascade in the mass space (Section \ref{sec:3}). 

\begin{figure}
\includegraphics*[width=\columnwidth]{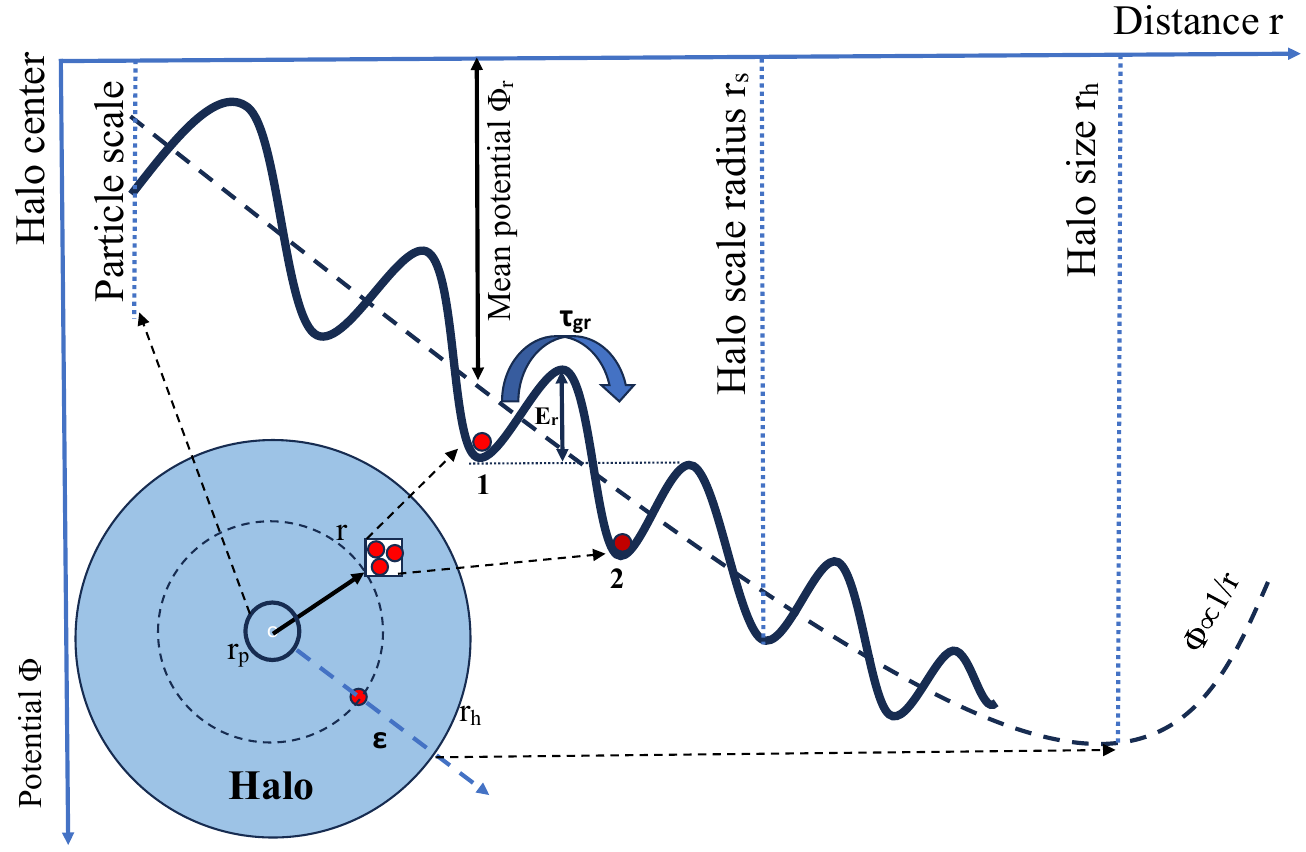}
\caption{Schematic plot for the random walk of dark matter particles in spherical haloes. Particles conduct a random walk in a fluctuating gravitational potential field $\Phi=\Phi_r+E_v$, where $\Phi_r$ is the mean potential and $E_r$ is the potential fluctuation. There is an energy flux associated with every elementary jump of a particle from small to large $r$ (position "1" to "2"), where the particle kinetic energy $v_r^2/2$ is cascaded to a larger scale $r$ after every migration. For a given particle, migration occurs with an average waiting time $\tau_{gr}(r,m_h,z)\propto \Phi_r^{-1}$ (Eq. \eqref{eq:4-3}). Because particles have finite kinetic and potential energy, continuous migration facilitates an inverse cascade of kinetic energy from small to large $r$ at a rate of $\varepsilon$ and a direct cascade of potential energy from large to small $r$ at a rate of -1.4$\varepsilon$.}
\label{fig:4-1}
\end{figure}

Figure \ref{fig:4-1} presents a schematic diagram for the random walk of particles in haloes with a fluctuating gravitational potential, a mean field representation for halo structure formation (Section \ref{sec:3-3-1}). Due to the discrete particle nature of haloes, the gravitational potential $\Phi$ in the haloes can never be smooth, as shown by N-body simulations. The total potential on scale $r$ can be decomposed into the mean and fluctuations. Here, $\Phi(r)=\Phi_r(r)+E_r(r)$, where $\Phi_r$ is the mean potential at scale $r$, and $E_r$ is the fluctuation of the potential at $r$. The mean potential, potential fluctuation, and density at scale $r$ read
\begin{equation} 
\label{eq:4-1} 
\begin{split}
\Phi_r(r) = -\frac{Gm_r}{r}, \quad E_r(r) = \frac{Gm_p}{d_r} \quad \text{and} \quad \rho_r(r) = \frac{m_p}{d_r^3},
\end{split}
\end{equation} 
where $m_r$ is the mass enclosed by $r$, $d_r$ is the mean spacing between particles on scale $r$, and $m_p$ is the mass of a single particle. The 

Next, we focus on the average waiting time for the random walk of particles in a fluctuating potential, for example, jumping from position "1" to "2" in Fig. \ref{fig:4-1}. The local temperature $T_r$ and the probability $P_r$ of a successful attempt read
\begin{equation}
\begin{split}
& k_BT_r = m_p v_r^2= m_p\gamma_r\frac{Gm_r}{r} \quad \textrm{and} \quad P_r = \exp\left(-\frac{m_pE_r}{k_BT_r}\right),\\
\end{split}
\label{eq:4-2}
\end{equation}
where $\gamma_r$ is a numerical factor. The typical velocity $v_r$ on scale $r$ is related to the local potential $\Phi_r$ by the Virial theorem. Hereafter, we adopt the subscript 'r' to represent quantities on scale $r$, the subscript 'p' for these quantities on the smallest scale in haloes that is determined by the particle mass and dependent on the resolution of the N-body simulations, the subscript 's' for these quantities at scale radius $r_s$, the subscript 'h' for these quantities on the scale of halo size $r_h$, and the superscript '*' for quantities in haloes of characteristic mass $m_h^*$.  

Since potential fluctuations are usually much smaller than the mean potential ($m_pE_r\ll k_BT_r$), the probability of a successful attempt and the waiting time $\tau_{gr}$ for particle random walk read 
\begin{equation}
\begin{split}
& P_r \approx 1-\left(\frac{m_pE_r}{k_BT_r}\right), \\
& \tau_{gr}(r) = \frac{1}{\Gamma_r}(1-P_r)= \frac{1}{\Gamma_rd_r}\frac{m_pr}{\gamma_r m_r}= \frac{1}{\Gamma_rd_r}\frac{Gm_p}{\gamma_r|\Phi_r|},
\end{split}
\label{eq:4-3}
\end{equation}
where $\tau_{gr}(r)$ is the waiting time for the random walk at location $r$ that is inversely proportional to the local potential $\Phi(r)$, as we assumed in Section \ref{sec:3-3-1}. Here, $\Gamma_r(r)$ is the attempt frequency at $r$ (similarly to the attempt frequency in solid diffusion), i.e., the number of attempts per unit of time for a particle to jump to a neighboring position. The waiting time $\tau_{gr}$ can be related to the waiting time $\tau_g$ for the random walk of haloes in mass space (Eq. \eqref{eq:3-5}), i.e. $\tau_g\equiv\tau_{gh}=\tau_{gr}(r=r_h)$. 

If we follow the growth of haloes of characteristic mass $m_h^*$ over time, we can write
\begin{equation}
\begin{split}
& \frac{dm_h^*}{dt} =\frac{m_p}{\tau_g^*}, \\
& \tau_g^* = \tau_{gr}(r=r_h^*) = \frac{m_p}{\Gamma_rd_r}\frac{r_h^*}{m_h^*}.
\end{split}
\label{eq:4-4}
\end{equation}
These two equations lead to 
\begin{equation}
\begin{split}
& u_h^*=\Gamma_rd_r =\frac{dm_h^*}{dt}\frac{r_h^*}{m_h^*}=\frac{r_h^*}{t}=\frac{dr_h^*}{dt},
\end{split}
\label{eq:4-5}
\end{equation}
where $m_h^*$ and $r_h^*$ are the mass and size of the haloes and $u_h^* = \Gamma_rd_r$ represents the speed of growth of the haloes. Here, we use the fact that $m_h^*\propto t$, $r_h^*\propto t$, and $\tau_g^*\propto t^0$ (Eq. \eqref{eq:3-9-2}). For a typical characteristic mass $m_h^*\approx 3\times 10^{13}M_{\odot}$, the size $r_h^*\approx 1$Mpc at $z=0$, and the age of the universe of 13.7 billion years, the velocity $u_h^*$ is around 80km/s. Since the speed of halo growth $u_h^*$ remains constant while the density of characteristic haloes $\rho_h^*\propto a^{-3}$ decreases with cosmic time, we estimate that $u_h^*$ is independent of the halo density and should also be roughly independent of the scale $r$. The larger the mean particle distance $d_r$, the smaller the attempt frequency $\Gamma_r$ (also see Eq. \eqref{eq:5-9-4}). Therefore, the waiting time of the particles is inversely proportional to the local potential, that is, $\tau_{gr}\propto \Phi_r^{-1}$ from Eq. \eqref{eq:4-3}. Similarly, the waiting time for haloes in mass space is inversely proportional to the halo potential, i.e., $\tau_{g}\propto \Phi_h^{-1}$ in Eq. \eqref{eq:3-5}.

Now, we focus on the change in particle energy associated with that elementary jump during infinitesimal time $dt_r$
\begin{equation} 
\label{eq:2-2-8} 
\begin{split}
&\Delta E_2 = \frac{1}{2}\left(1+\frac{2}{n_e}\right) m_p dv_r^2,
\end{split}
\end{equation}
where $n_e$ is the effective potential index ($n_e$=-10/7 on global scales in Section \ref{sec:2-2}). Here, $dv_r^2$ is the infinitesimal difference in the specific kinetic energy between two positions "1" and "2" in Fig. \ref{fig:4-1}. We assume that the particle is in virial equilibrium before and after every jump. With $n_e>-2$, $\Delta E_2<0$ means that the particle energy decreases after this elementary jump. Similarly to the merging of haloes (Eq. \eqref{eq:3-2-s1}), the migration of particles also leads to the energy "dissipation". Both merging of haloes and migration of particles provide small-scale channels that contribute to the global-scale energy "dissipation" (Fig. \ref{fig:S1-3-2}). Now we introduce a key parameter $\varepsilon$ to represent the rate of change in the specific energy of the particle
\begin{equation} 
\label{eq:2-2-9} 
\begin{split}
\varepsilon = \frac{\Delta{E_2}}{m_p dt_r} = \frac{1}{2}\left(1+\frac{2}{n_e}\right)\frac{dv_r^2}{dt_r},
\end{split}
\end{equation}
where $\varepsilon<0$ represents the direction of energy transfer from small to large scales (inverse cascade). This is the microscopic origin of the definition of the parameter $\varepsilon$ in Eq. \eqref{eq:4-3-1}.

Next, we need to identify the microscopic counterpart of terms $dv_r^2$ and $dt_r$ in Eq. \eqref{eq:2-2-9}. Considering a thin spherical shell of thickness $s_r$ located at $r$ that contains $N_r$ particles. We will identify the waiting time $\tau_{hr}(r)$ for a successful jump for any one of these $N_r$ particles in that spherical shell. In analogy to Eq. \eqref{eq:3-1} for halo random walk in mass space, we have $\tau_{gr}(r)=N_r\tau_{hr}(r)$ and the waiting time
\begin{equation}
\begin{split}
\tau_{hr}(r)=\frac{\tau_{gr}}{N_r}=\frac{\tau_{gr}}{4\pi r^2\rho_r s_r/m_p}=\frac{m_p^2}{\Gamma_rd_r}\frac{1}{4\pi r\rho_r m_r s_r},
\end{split}
\label{eq:4-6}
\end{equation}
where $s_r$ is the thickness of the spherical shell or the jump length of the random walk. Similarly to Eq. \eqref{eq:3-2} for random walk in mass space, the corresponding rate of mass flow due to the outward migration of particles should read as follows: 
\begin{equation}
\begin{split}
\dot m_r=\frac{m_p}{\tau_{hr}}=4\pi u_h^*{r\rho_r s_r}\frac{m_r}{m_p},
\end{split}
\label{eq:4-7}
\end{equation}
i.e., a mass of $m_p$ is transferred to larger scales during the time $\tau_{hr}$ by migration. The waiting time $\tau_{hr}$ is the microscopical quantity equivalent to infinitesimal time $dt_r$, as well as the quantity $dv_r^2$
\begin{equation}
\begin{split}
dt_r = \tau_{hr} \quad \textrm{and} \quad {dv_r^2}=\frac{2}{5}{dm_r}\frac{v_r^2}{m_r}=\frac{2}{5}\frac{m_pv_r^2}{m_r}.
\end{split}
\label{eq:4-8}
\end{equation}
Here, the particle mass $m_p$ is the infinitesimal mass $dm_r$, and we use the scaling law $v_r^2\propto m_r^{2/5}$ in Eq. \eqref{eq:4-3-2}. Inserting $dt_r$ and $dv_r^2$ in Eq. \eqref{eq:2-2-9}, we can express the rate of the energy cascade $\varepsilon$ as:
\begin{equation}
\begin{split}
\varepsilon = \frac{1}{5}\left(1+\frac{2}{n_e}\right)\frac{m_pv_r^2}{m_r\tau_{hr}}=\frac{1}{5}\left(1+\frac{2}{n_e}\right)\frac{Gm_p}{r\tau_{hr}}.
\end{split}
\label{eq:4-8-2}
\end{equation}
With a particle on scale $r$ migrating from position "1" to "2" (Fig. \ref{fig:4-1}) during the waiting time $\tau_{hr}$, the specific energy of $m_pv_r^2/m_r$ is transferred across scale $r$. Therefore, the parameter $\varepsilon$ represents the energy flux associated with the migration across scale $r$. 

The mass flux is also related to the characteristic velocity $v_r$, the typical velocity at $r$ (Eq. \eqref{eq:4-2}), so that the mass flux also reads
\begin{equation}
\begin{split}
\dot m_r=4\pi r^2\rho_r v_r.
\end{split}
\label{eq:4-9}
\end{equation}
By equating the two mass fluxes in Eqs. \eqref{eq:4-7} and \eqref{eq:4-9}, the thickness $s_r$ and waiting time $\tau_{hr}$ read
\begin{equation}
\begin{split}
s_r=v_r\tau_{gr}=\frac{m_p}{u_h^*}\frac{rv_r}{m_r}\quad \textrm{and} \quad \tau_{hr}=\frac{m_p}{\dot m_r}=\frac{m_p}{4\pi r^2\rho_r v_r}.
\end{split}
\label{eq:4-10}
\end{equation}
Other relevant quantities include the mean particle distance $d_r$, the attempt frequency $\Gamma_r$, the waiting time $\tau_{gr}$ and $\tau_{hr}$, and the thickness of the shell $s_r$. These quantities depend on the particle mass $m_p$ and the growth speed of haloes of characteristic mass $m_h^*$, i.e., $u_h^*=\Gamma_rd_r\approx$80 km/s. From Eqs. \eqref{eq:4-3} to \eqref{eq:4-10}, on scales $r\le r_s$,
\begin{equation} 
\label{eq:4-3-3} 
\begin{split}
&d_r =\beta_r^{-\frac{1}{3}} m_p^{\frac{1}{3}}\varepsilon^{-\frac{2}{9}}G^{\frac{1}{3}}r^{\frac{4}{9}} \textrm{,} \quad \Gamma_r = \beta_r^{\frac{1}{3}}u_h^* m_p^{-\frac{1}{3}} \varepsilon^{\frac{2}{9}}G^{-\frac{1}{3}}r^{-\frac{4}{9}}, \\
&\tau_{gr} = \frac{m_p}{\alpha_r\gamma_ru_h^*}\varepsilon^{-\frac{2}{3}}G r^{-\frac{2}{3}} \textrm{,} \quad \tau_{hr} = \frac{3}{5}\beta_h\gamma_r m_p \varepsilon^{-1}Gr^{-1}, \\
&N_r = \frac{5{u_h^*}^{-1}}{3\alpha_r\beta_h\gamma_r^2}\varepsilon^{\frac{1}{3}}r^{\frac{1}{3}},\quad s_r = \frac{{u_h^*}^{-1}}{\alpha_r\beta_h^{1/3}} m_p \varepsilon^{-\frac{1}{3}}Gr^{-\frac{1}{3}}.
\end{split}
\end{equation} 
Since these quantities involve particle mass $m_p$, information on these quantities may provide useful insights into the particle mass and properties of dark matter (Section \ref{sec:5}). Specifically, for haloes of characteristic mass $m_h^*$, the rate of energy cascade $\varepsilon =\varepsilon_u\approx$-$10^{-7}$m$^2$/s$^3$. On scale $r=r_s^*\propto a^{3/2}$, with subscript 's' standing for the scale radius, relevant quantities satisfy
\begin{equation} 
\label{eq:4-3-3-2} 
\begin{split}
&\rho_s^*\propto a^{-2}\textrm{,}\quad d_s^* \propto a^{2/3} \textrm{,} \quad \Gamma_s^* \propto a^{-2/3}, \\
&\tau_{gs}^* \propto a^{-1}\textrm{,} \quad \tau_{hs}^* \propto a^{-3/2}\textrm{,} \quad N_s^*\propto a^{1/2},\quad s_s^*\propto a^{-1/2}.
\end{split}
\end{equation}
From Eqs. \eqref{eq:2-2-9} and \eqref{eq:4-8-2}, the rate of energy cascade can be related to the energy "dissipated" due to the merging of haloes ($\Delta E_1$) and the random walk of particles in haloes ($\Delta E_2$):
\begin{equation}
\begin{split}
\varepsilon \propto \frac{\Delta E_2}{m_p \tau_{hr}} \propto \frac{\Delta E_1}{m_h \tau_{hr}(r=r_h)}.
\end{split}
\label{eq:4-8-3}
\end{equation}

\subsection{The scale-to-scale energy cascade in haloes}
\label{sec:4-2}
This section quantifies the energy cascade in haloes from N-body simulations. The same theory can also be applied to the energy flow in galaxy bulges and the associated scaling laws for bulge mass, size, dynamics, and the evolution of supermassive black holes \citep{Xu:2024-Cosmic-quenching-and-scaling-laws}. In analogy to the formulation of the energy cascade in halo mass space (Section \ref{sec:3-5}), we examine the energy cascade in all haloes of a given mass $m_h$. We start by introducing cumulative functions along the radial direction $r$. The cumulative mass function $\Lambda^h_m(m_h,r,z)$ (similar to $\Lambda_m$ in Eq. \eqref{eq:3-4}) represents the total mass above $r$ that is averaged for all haloes of the same mass $m_h$
\begin{equation} 
\label{eq:4-2-1} 
\Lambda^h_m(m_h,r,z) = \int_{r}^{\infty} \rho_r \left(m_h,r',z) \right)4\pi r'^2 dr',
\end{equation} 
where $\rho_r$ is the average mass density for all haloes of mass $m_h$. 

Next, similarly to the formulation in Section \ref{sec:3-5}, we also decompose the velocity of the halo particles $\boldsymbol{\mathrm{v}}_{p}$ into the mean halo velocity, $\boldsymbol{\mathrm{v}}_{h}=\langle \boldsymbol{\mathrm{v}}_{p} \rangle_h$, and the velocity fluctuation, $\boldsymbol{\mathrm{v}}_{p}^{'} $, that is, $\boldsymbol{\mathrm{v}}_{p} =\boldsymbol{\mathrm{v}}_{h} +\boldsymbol{\mathrm{v}}_p'$ (Eq. \eqref{eq:3-21}). Here, $\boldsymbol{\mathrm{v}}_{h}$ represents the velocity of that halo, the average velocity of all particles in the same halo. Consequently, the total kinetic energy $K_p$ of a given halo particle can be divided into $K_p = K_{ph}+K_{pv}$. The virial kinetic energy, $K_{pv}={\boldsymbol{\mathrm{v}}_p'}^2/2$, is the contribution from the velocity fluctuation due to the intra-halo interactions on small scales that are in the nonlinear regime. Only this part of the kinetic energy is relevant for the energy cascade in haloes. We introduce a cumulative function $\Lambda^h_{pv}$ for $K_{pv}$
\begin{equation} 
\label{eq:4-2-2} 
\begin{split}
&\Lambda^h_{pv}(m_h,r,z) = \int_{r}^{\infty} K_{pv} \rho_r \left(m_h,r',z) \right)4\pi r'^2 dr'. 
\end{split}
\end{equation} 

Next, similarly to the energy cascade in the mass space (Eq. \eqref{eq:3-23}), we introduce the specific kinetic energy on scale $r$ for all haloes of the same mass $m_h$ that reads
\begin{equation} 
\label{eq:4-2-3} 
\begin{split}
\overline {K^h_{pv}}(m_h,r,z) = \frac{\Lambda^h_{pv}}{\Lambda^h_{m}}=\frac{\int_{r}^{\infty} K_{pv} \rho_h \left(m_h,r',z) \right)4\pi r'^2 dr'}{\int_{r}^{\infty} \rho_h \left(m_h,r',z) \right)4\pi r'^2 dr'}.
\end{split}
\end{equation} 
Here, $\overline {K^h_{pv}}$ is the specific energy (energy per unit mass) contained on scales above $r$. With $\overline {K^h_{pv}}$ increasing with time, the energy flux $\Pi^h_{pv}$ along the halo radial direction is defined as (similarly to Eq. \eqref{eq:3-24})   
\begin{equation} 
\label{eq:4-2-4} 
\begin{split}
\Pi^h_{pv}(m_h,r,z) &=-\frac{\partial }{\partial t} \left( \overline {K^h_{pv}} \right) = -\frac{\partial }{\partial t} \left(\frac{\Lambda^h_{pv}}{\Lambda^h_m} \right).
\end{split}
\end{equation}

\begin{figure}
\includegraphics*[width=\columnwidth]{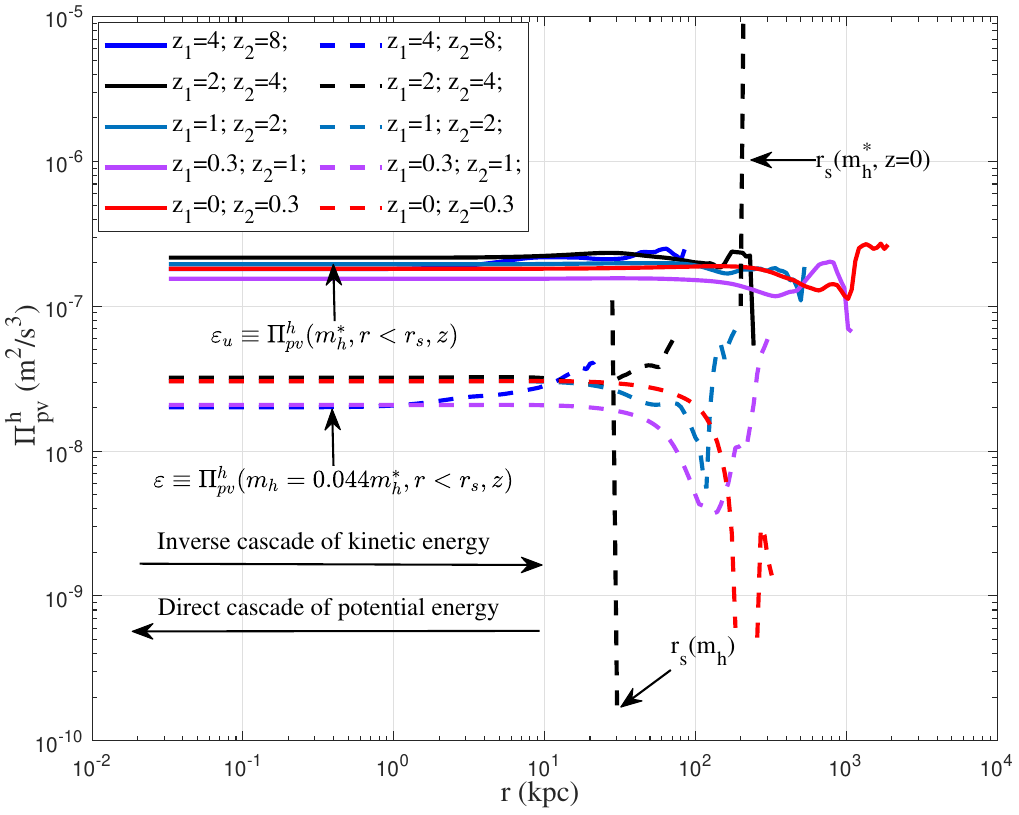}
\caption{The variation of the rate of energy cascade $\Pi^h_{pv}(m_h,r,z)$ (Eq. \eqref{eq:4-2-4}) with radial scale $r$ at different redshifts $z$ for haloes of characteristic mass $m^*_h(z)$ and $m_h=0.044m^*_h(z)$. A scale-independent (independent of $r$) constant rate of $\varepsilon\equiv \Pi^h_{pv}(r<r_s)$ can be identified for an inverse cascade of virial kinetic energy ${K_{pv}}$ from small to larger scales. There also exists a simultaneous direct cascade of potential energy from large to small scales at a rate of $-7/5\varepsilon$ \citep{Xu:2022-Postulating-dark-matter-partic}. For haloes with characteristic mass $m^*_h$, the rate of energy cascade $\varepsilon(m_h^*,z)\equiv \varepsilon_u\approx -10^{-7}m^2/s^3$.} 
\label{fig:4-2}
\end{figure}

In a certain range of scales $r<r_s$ (inner haloes), the characteristic time on small scales is very small compared to the time on large scales. The small-scale motion does not feel the large-scale motion directly, except through the energy flux $\varepsilon$ across scales. When a statistical equilibrium is established, similar to the energy cascade in the halo mass space, we expect the rate of energy flow $\varepsilon(m_h,z)$ to be independent of the scale $r$, i.e.
\begin{equation} 
\label{eq:4-2-5} 
\begin{split}
\varepsilon\left(m_h,z\right) \equiv \Pi^h _{pv}(m_h,r,z) =-\frac{\partial }{\partial t} \overline {K^h_{pv}}\quad \textrm{for}\quad r<r_s.
\end{split}
\end{equation} 
This can be tested by N-body simulations. 

Figure \ref{fig:4-2} plots the variation of the energy flux $\Pi_{pv}^h(m_h,z)$ using Eq. \eqref{eq:4-2-4} and the kinetic energy $\overline {K^h_{pv}}$ in Eq. \eqref{eq:4-2-3} at two different redshifts from the Illustris simulations. In this figure, the rate of energy flow $\varepsilon$ is clearly independent of the scale $r$ below a characteristic scale $r_s$, usually the scale radius of haloes. The key parameter $\varepsilon$ increases with the halo mass $m_h$ and the redshift $z$. For haloes with characteristic mass $m^*_h$, $\varepsilon(m_h^*,z)\equiv \varepsilon_u$, i.e., the rate of energy cascade in halo mass space (Fig. \ref{fig:3-7}). 

\begin{figure}
\includegraphics*[width=\columnwidth]{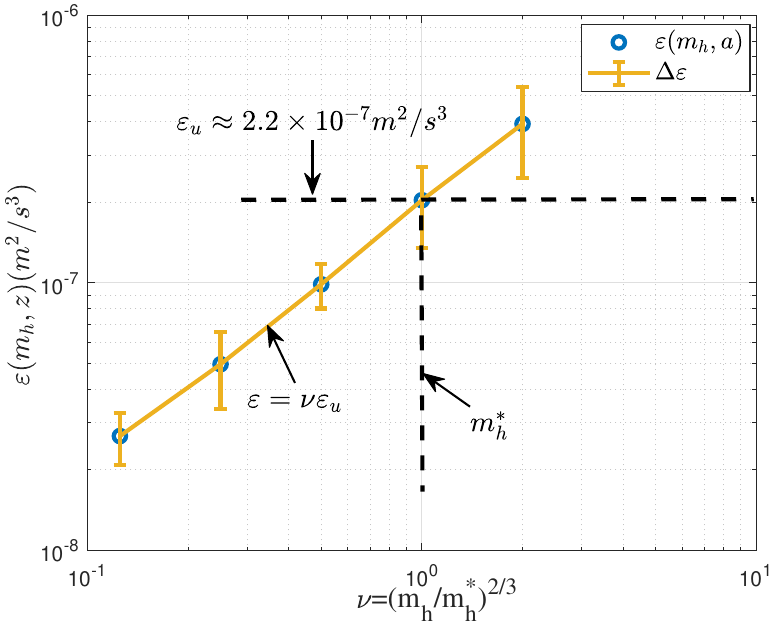}
\caption{The variation of the rate of energy cascade $\varepsilon(m_h,z)$ (Eq. \eqref{eq:4-2-5}) with halo mass $m_h$ at different redshifts $z$ in terms of a dimensionless parameter $\nu=(m_h/m_h^*)^{2/3}$. A scale-independent constant rate of $\varepsilon_u$ can be identified at $\nu=1$ or $m_h=m^*_h$. That rate is also relatively independent of time and is around -$10^{-7}m^2/s^3$ (also see Fig. \ref{fig:3-7}).} 
\label{fig:4-3}
\end{figure}

Similarly, we can calculate the rate of the energy cascade for haloes of different masses. Figure \ref{fig:4-3} plots the variation of $\varepsilon(m_h,z)$ with the halo mass $m_h$ and the redshift $z$. The figure shows that $\varepsilon \propto m_h^{2/3}$ and increases with redshift $z$, 
\begin{equation} 
\label{eq:4-2-6} 
\varepsilon(m_h,z) = \varepsilon_u \nu = \varepsilon_u \left({m_h}/{m_h^*}\right)^{2/3}\propto m_h^{2/3}a^{-1}, 
\end{equation} 
where $\varepsilon_u\equiv \varepsilon(m_h^*,z)$ is the rate of energy flow in haloes of characteristic mass $m_h^*$. The dimensionless parameter $\nu$ is defined as $\nu = ({m_h}/{m_h^*})^{2/3}$ \citep{Xu:2023-Dark-matter-halo-mass-functions-and}. The rate of the energy cascade varies with the mass and time as $\varepsilon\propto m_h^{2/3}a^{-1}$, while the haloes evolve towards the limiting equilibrium. We only have a vanishing energy cascade or $\varepsilon=0$ in fully virialized haloes with $a\rightarrow \infty$ (usually small and old haloes), a limiting state that can never be reached.

Finally, for haloes of any mass, the rate of energy cascade in that halo can be conveniently related to the energy "dissipated" by merging (Eq. \eqref{eq:3-2-s4}). For haloes of characteristic mass $m_h^*$ or any mass $m_h$, that rate is 
\begin{equation} 
\label{eq:4-2-6-2} 
\varepsilon_u\propto \frac{\Delta E_1}{m_{h0}^*\tau_g^*}\propto \frac{Gm_p}{r_{h0}^*\tau_g^*}\propto \frac{Gm_h^*}{r_h^*t_0} \quad \textrm{and} \quad \varepsilon(m_h,z) \propto \frac{Gm_h}{r_ht_0},
\end{equation} 
where $t_0$ is the age of the Universe, $m_{h0}^*\equiv m_h^*(z=0)$ and $r_{h0}^*\equiv r_h^*(z=0)$. From Eq. \eqref{eq:3-2-s4}, it can be directly related to the energy released $\Delta E_1$ during every merging. The continuous halo merging drives the system continuously, releasing energy and maximizing global entropy during the structure evolution.

\section{Constraints on dark matter particle mass}
\label{sec:5}
Previous sections examined the cascade of mass and energy both across halo mass scales and within individual halos, and discussed how these cascades may depend on the mass and microphysical properties of dark matter particles. In this section, we outline a few illustrative constraints that can, in principle, be derived from these ideas. The discussion that follows is necessarily preliminary; further work is required to quantify these constraints and test their robustness. Therefore, these remarks are presented as suggestions intended to stimulate community follow-up studies.

Two general ideas motivate the constraints. First, if gravity is the only interaction among dark matter particles, for gravitationally dominant haloes of characteristic mass $m_h^*$ with universal spectrum and scaling laws, a scale–independent cascade rate may persist down to the smallest physical scales (the free‑streaming cutoff illustrated in Fig. \ref{fig:4-5}). When the free‑streaming scale approaches the intrinsic length scale of the particle, the scaling relations derived above could provide direct insight into the properties of the particle \citep{Xu:2022-Postulating-dark-matter-partic, Xu:2023-Universal-scaling-laws-and-density-slope}. Second, several key physical quantities that govern the effective random walk and migration of particles (e.g., the waiting time) depend explicitly on the particle mass $m_p$ (Eq. \eqref{eq:4-3-3}). If independent bounds can be placed on these quantities, they can be translated into novel constraints on $m_p$. In the remainder of this section, we illustrate the second approach with several examples.

\begin{enumerate}
\item \noindent First, from the scaling for waiting time $\tau_{hr}$ in Eq. \eqref{eq:4-3-3}, we have the expression for particle mass $m_p$,
\begin{equation}
\begin{split}
m_p = \frac{5}{\gamma_r}\varepsilon G^{-1}r\tau_{hr}.
\end{split}
\label{eq:5-4-2}
\end{equation}
For haloes with characteristic mass $m_h^*$, the rate of energy flow $\varepsilon=\varepsilon_u\approx 10^{-7}m^2/s^3$, and the halo size $r_h^*\approx$1Mpc, we should have
\begin{equation}
\begin{split}
m_p = \frac{5}{\gamma_r}\varepsilon_u G^{-1}r_h^* \tau_h^*.
\end{split}
\label{eq:5-4-3}
\end{equation}
The waiting time $\tau_h^*\equiv \tau_{hr}(r=r_h^*)$ represents a characteristic timescale for a real physical process, i.e., the waiting time for the migration of particles across the halo boundary (Eq. \eqref{eq:4-6}). It is reasonable to assume that the waiting time $\tau_h^*\ge t_p$, i.e. is greater than the Planck time $t_p=5.4\times 10^{-44}$s, the smallest possible unit of time for any physical process. With this and $\gamma_r\approx 0.5$ (Eq. \eqref{eq:4-3-2-1}), the particle mass satisfies
\begin{equation}
\begin{split}
m_p \ge \frac{5}{\gamma_r}\varepsilon_u G^{-1}r_h^*t_p=2.5\times 10^{-18}kg\approx 10^{9}GeV.
\end{split}
\label{eq:5-4-4}
\end{equation}

\item \noindent Second, we consider the growth of haloes of characteristic mass $m_h^*\propto t$ and size $r_h^*\propto t$ (Eq. \eqref{eq:3-9-2}). Since the waiting time $\tau_g\propto m_h^{-2/3}$ (Eq. \eqref{eq:3-6}) in mass space and $\tau_{gr}\propto r^{-2/3}$ (Eq. \eqref{eq:4-3-3}) in individual haloes, haloes with a characteristic mass $m_h^*$ must have the fastest mass accretion and the shortest waiting time $\tau_g^*$. The mass accretion of these characteristic haloes reads 
\begin{equation}
\begin{split}
&\dot{m_h^*} = \frac{dm_h^*}{dt}=\frac{m_p}{\tau_g^*}=\frac{m_h^*(z=0)}{t_0},
\end{split}
\label{eq:5-1}
\end{equation}
where $\dot {m_h^*}$ is the rate of mass accretion. On average, characteristic haloes accrete one particle of mass $m_p$ during an average waiting time $\tau_g^*$. For a typical mass $m_h^*(z=0)\approx 3.2\times 10^{13}M_{\odot}$ at the current epoch \citep{Cooray:2002-Halo-models-of-large-scale-str} and $t_0\approx 4.3\times 10^{17}$ s (the age of the universe), the rate of mass accretion $\dot {m_h^*}\approx 1.5\times 10^{26}$ kg/s. The particle mass $m_p$ can be related to the waiting time as
\begin{equation}
\begin{split}
m_p = \dot {m_h^*}\tau_g^* = 1.5\times 10^{26}\frac{kg}{s}\tau_g^*.
\end{split}
\label{eq:5-2}
\end{equation}
In N-body simulations, $m_p$ is dependent on the mass resolution. For Illustris simulation with $m_p\approx 10^6M_{\odot}$ (Table \ref{tab:1}), the waiting time is on the order of $\tau_g^*=7\times 10^{10}$s. 

The waiting time $\tau_g^*\equiv \tau_{gr}(r=r_h^*)$ also represents a time scale for a real physical process, i.e., the waiting time for particle random walk in haloes. Therefore, it is also reasonable to assume that $\tau_g^*\ge t_p$. With this consideration, the particle mass should satisfy
\begin{equation}
\begin{split}
m_p = 1.5\times 10^{26}\frac{kg}{s}\tau_g^*\ge 1.5\times 10^{26}\frac{kg}{s} t_p =10^{-17}kg.
\end{split}
\label{eq:5-4}
\end{equation}
This provides a similar lower limit for particle mass $m_p$. 
\\
\begin{figure}
\includegraphics*[width=\columnwidth]{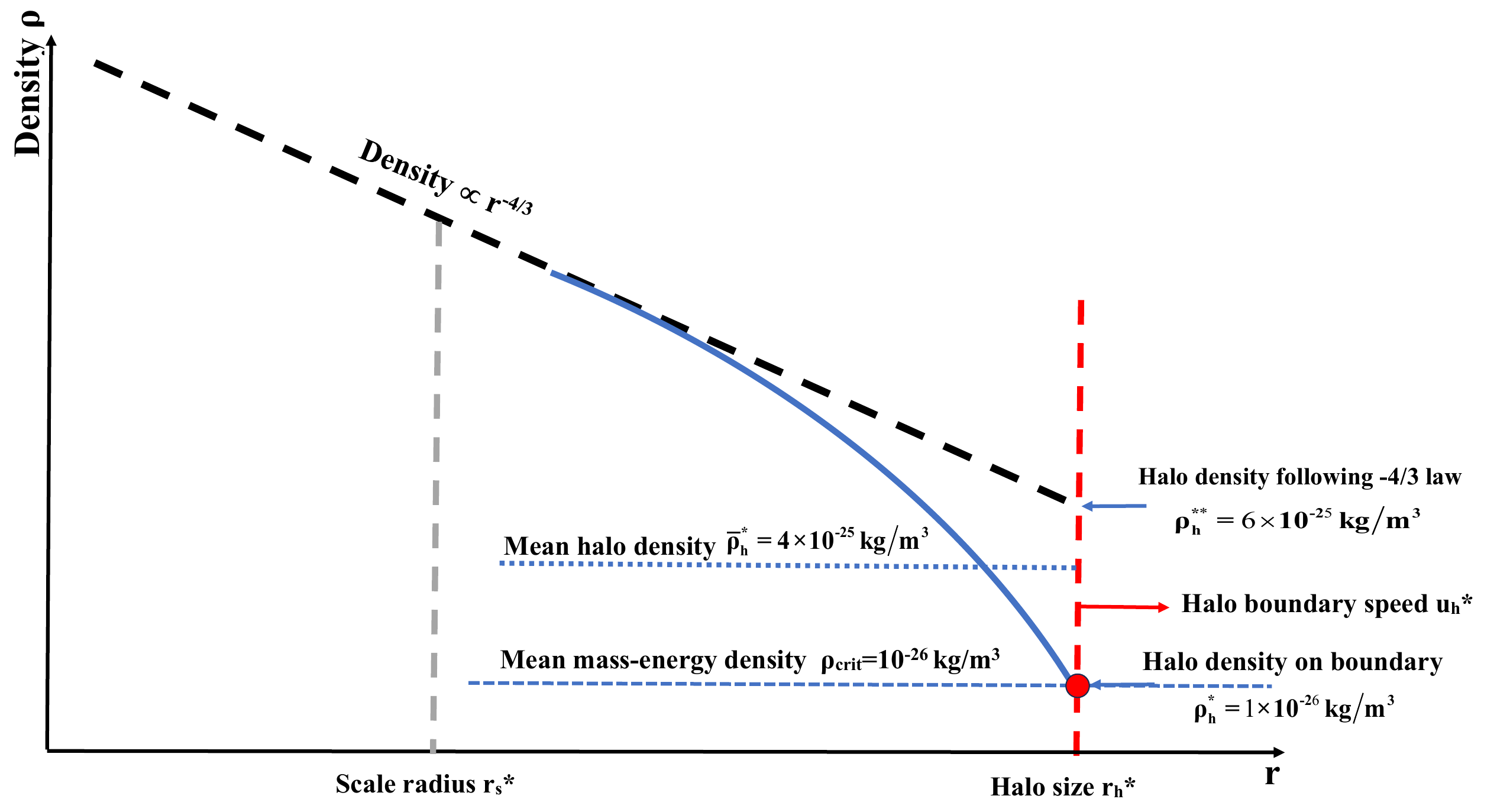}
\caption{The schematic plot of relevant densities next to the boundary of haloes at $z=0$. The mass-energy density of the universe is $\rho_{crit}=10^{-26}$kg/m$^3$. The man halo density $\bar\rho_h^*=4\times 10^{-25}$kg/m$^3$. Characteristic haloes have a mass of $m_h^*=3.2\times 10^{13}M_{\odot}$ and a halo size $r_h^*\approx$1Mpc. The halo density on the boundary $\rho_h^*$ is in the same order as the critical density $\rho_{crit}$. The halo boundary density $\rho_h^{**}$ obtained from the -4/3 scaling is around $6\times 10^{-25}$kg/m$^3$ (Eq. \eqref{eq:4-3-2}). Relevant quantities are also listed in Table \ref{tab:5-1}} 
\label{fig:5-1}
\end{figure}

\item \noindent Third, a more stringent constraint estimates that the waiting time $\tau_g^*\ge \Delta_ct_p$, where $\Delta_c\approx$ 200 is the contrast ratio of the halo density to the background dark matter density. To understand this, we start from equations for the growth of halo mass $m_h^*$ and size $r_h^*$,
\begin{equation}
\begin{split}
&\frac{dm_h^*}{dt}=\frac{m_h^*}{t}=\frac{m_p}{\tau_g^*} \quad \textrm{and} \quad \frac{dr_h^*}{dt}=\frac{r_h^*}{t}=u_h^*, \\
&\frac{m_h^*}{\frac{4}{3}\pi {r_h^*}^3}=\rho_{crit}\Omega_{DM}\Delta_c \quad \textrm{and} \quad \rho_{crit} = \frac{3H^2}{8\pi G},
\end{split}
\label{eq:5-5}
\end{equation}
where $H$ is the Hubble parameter, $\rho_{crit}\approx 10^{-26}$kg/m$^3$ is the critical mass-energy density of the present universe, and $\Omega_{DM}\approx 0.2449$ is the mass fraction of dark matter. For a typical halo mass of $m_h^*=3.2\times 10^{13}M_{\odot}$ and halo size $r_h^*=$1Mpc (listed in Table \ref{tab:5-1}, the speed of the halo growth is around $u_h^*\approx$ 80km/s (Eq. \eqref{eq:4-5}). Figure \ref{fig:5-1} presents a schematic plot of the relevant densities next to the halo boundary. From Eq. \eqref{eq:5-5}, we can have the relation between the speed of halo growth $u_h^*$ and the waiting time $\tau_g^*$,
\begin{equation}
\begin{split}
{u_h^*}^2 = \frac{Gm_p}{\frac{1}{2}H^2t^2\Omega_{DM}\Delta_c u_h^* \tau_g^* }. 
\end{split}
\label{eq:5-6}
\end{equation}

Without loss of generality, let the mass density on the halo boundary be $\rho_h^* = \rho_{crit}/\alpha$. The mass-energy density in haloes is determined mostly by the dark matter density since dark matter is the dominant component in haloes. The mass-energy density of the universe is $\rho_{crit}$, which includes all cosmic components, including dark matter, baryonic matter, and dark energy. For $\rho_h^*\ge \rho_{crit}$, we expect $\alpha \le 1$. The halo mass now reads (from Eq. \eqref{eq:5-5})
\begin{equation}
\begin{split}
m_h^*=\frac{4}{3}\pi {r_h^*}^3\alpha \rho_h^*\Omega_{DM}\Delta_c.
\end{split}
\label{eq:5-7}
\end{equation}
With Eq. \eqref{eq:5-5}, we can further express the rate of mass accretion in terms of the boundary density $\rho_h^*$,
\begin{equation}
\begin{split}
\frac{1}{3}\Omega_{DM}\Delta_c\alpha \rho_h^* 4\pi {r_h^*}^2 u_h^*=\frac{m_h^*}{t}=\frac{m_p}{\tau_g^*}=\dot{m_h^*}.
\end{split}
\label{eq:5-8}
\end{equation}

For a spherical shell on the halo boundary that contains one and only one particle ($\tau_{gr}=\tau_{hr}$ for $r=r_h$), the thickness of that spherical shell $s_h^*=s_r(r=r_h^*)$ reads (from Eq. \eqref{eq:5-8}) 
\begin{equation}
\begin{split}
&s_h^* = \frac{m_p}{\rho_h^*4\pi {r_h^*}^2} = \frac{1}{3}\Omega_{DM}\Delta_c \alpha u_h^* \tau_g^*,
\end{split}
\label{eq:5-9-1}
\end{equation}
where $s_h^*$ is approximately the jumping length for the migration of particles on the halo boundary. Using Eq. \eqref{eq:5-6}, the speed of halo growth is related to the jumping length as
\begin{equation}
\begin{split}
&{u_h^*}^2 = \frac{2\alpha}{3H^2t^2} \frac{Gm_p}{s_h^*}\quad \textrm{and}\quad m_p = \frac{3H^2t^2}{2\alpha}{u_h^*}^2 {s_h^*} G^{-1}.
\end{split}
\label{eq:5-9-2}
\end{equation}
For every waiting time $\tau_g^*$, the particle migrates at a distance of $s_h^*$. Similarly, since $s_h^*$ represents a physical length scale for particle migration, it is reasonable to assume that the distance $s_h^*\ge l_p$, where $l_p=1.6\times 10^{-35}$m is the Planck length, the smallest possible unit of length. With this constraint, the particle mass should satisfy
\begin{equation}
\begin{split}
& m_p \ge \frac{3H^2t^2}{2\alpha}{u_h^*}^2 G^{-1}l_p = 10^{-15}kg.
\end{split}
\label{eq:5-9-3}
\end{equation}

\item \noindent Especially, from Eq. \eqref{eq:5-9-2}, we estimate the speed of halo growth and the jumping frequency at scale $r$ as
\begin{equation}
\begin{split}
&{u_h^*}=\Gamma_rd_r \propto \sqrt{\frac{Gm_p}{l_p}}\quad\textrm{or}\quad \Gamma_r \propto \sqrt{\frac{Gm_p}{d_r^2l_p}}.
\end{split}
\label{eq:5-9-4}
\end{equation}
Since gravity is assumed to be the only force between dark matter particles, for particle migration in haloes, the smallest separation between two particles should be the Planck length scale $l_p$. The typical acceleration $Gm_p/d_r^2$ on scale $r$ changes its direction on the Planck length $l_p$, which determines the jumping frequency $\Gamma_r$.

Next, the relevant speed of that particle reads (using Eq. \eqref{eq:5-9-2})
\begin{equation}
\begin{split}
&\frac{s_h^*}{\tau_g^*} = \frac{1}{3}\Omega_{DM}\Delta_c \alpha u_h^* = \frac{1}{3}\Omega_{DM}\Delta_c \alpha \sqrt{\frac{2\alpha}{3H^2t^2}} \sqrt{\frac{Gm_p}{s_h^*}}.
\end{split}
\label{eq:5-10}
\end{equation}
Finally, normalizing the jumping length $s_h^*=\beta l_p$ by Planck length $l_p$ and waiting time $\tau_g^* = \gamma t_p$ by Planck time $t_p$, we obtain 
\begin{equation}
\begin{split}
&\left(\frac{\beta}{\alpha\gamma}\right)^{3/2} = \frac{1}{3}\Omega_{DM}\Delta_c \sqrt{\frac{2}{3H^2t^2}} \sqrt{\frac{G\dot{m_h^*}}{c^3}},
\end{split}
\label{eq:5-11}
\end{equation}
where $c=3\times 10^8$m/s is the speed of light. Plugging in numbers into Eq. \eqref{eq:5-11} and considering that $\alpha\le 1$ and $\beta\ge 1$ ($\dot {m_h^*}\approx 1.5\times 10^{26}$ kg/s, $\Omega_{DM}\approx$0.2449, and $Ht$=2/3), we should have
\begin{equation}
\begin{split}
\gamma\approx \Delta_c \frac{\beta}{\alpha} \quad \textrm{or} \quad \tau_g^*=\Delta_c t_p \frac{\beta}{\alpha} \ge \Delta_c t_p.
\end{split}
\label{eq:5-11-2}
\end{equation}

The density contrast ratio $\Delta_c$ should influence the waiting time $\tau_g^*$. The higher $\Delta_c$, the longer the waiting time $\tau_g^*$. An alternative interpretation of this can be based on the Eq. \eqref{eq:4-6} for waiting time,
\begin{equation}
\begin{split}
\tau_{hr} \propto (r \rho_r m_r s_r)^{-1}.
\end{split}
\label{eq:5-12}
\end{equation}
Let us consider a series of possible characteristic haloes of different size $r$, mass $m_r$, and density $\rho_r=\Delta_c \bar\rho_{DM}$, where $\Delta_c$ is the density ratio. Since characteristic haloes have the fastest mass accretion such that the shell thickness on the halo boundary $s_r=s_h^*=l_p$ should be the same for all of these haloes. According to the scaling law in Eq. \eqref{eq:5-12}, the waiting time on the halo boundary should follow
\begin{equation}
\begin{split}
\tau_g^*=\tau_{hr} \propto (r \rho_r m_r)^{-1} \propto (r^1 r^{-4/3} r^{5/3})^{-1} \propto \rho_r \propto \Delta_c \bar\rho_{DM}.
\end{split}
\label{eq:5-13}
\end{equation}
Therefore, the waiting time is proportional to the density ratio $\Delta_c$. In the limiting situation $\Delta_c=1$ or the halo has the same density as the background, the waiting time satisfies $\tau_g^*\ge t_p$ (Eq. \eqref{eq:5-11-2}).

For waiting time $\tau_g^*$ satisfying $\tau_g^*\ge \Delta_ct_p$, a more stringent constraint for particle mass reads (Eq. \eqref{eq:5-2})
\begin{equation}
\begin{split}
m_p = 1.5\times 10^{26}\frac{kg}{s}\tau_g^*\ge 1.6\times 10^{-15}kg \approx 10^{12} GeV.
\end{split}
\label{eq:5-14}
\end{equation}
\end{enumerate}

All these constraints (Eqs. \eqref{eq:5-4-4}, Eq. \eqref{eq:5-4}, \eqref{eq:5-9-3}, and \eqref{eq:5-14}) exclude the standard WIMPs and strongly suggest a heavy dark matter scenario, where superheavy right-handed neutrinos might be a good candidate. Detailed discussion on the possible properties of dark matter particles from the mass and energy cascade is presented in a different article \citep{Xu:2022-Postulating-dark-matter-partic}. Let us take the particle mass $m_p=10^{12}$GeV in Eq.\eqref{eq:5-14} as an example; all other quantities can be easily obtained from the scaling laws (Eq. \eqref{eq:4-3-2}) and equations in this section. Table \ref{tab:5-1} summarizes these values of relevant physical quantities on the smallest and largest scale of $r$ in haloes of characteristic mass $m_h^*$ at the current epoch. For $m_p=10^{12}$GeV, on the halo boundary, the waiting time $\tau_{gr}=\tau_{hr}=\Delta_c t_p$ and the jump length $s_h^*=l_p$.
 
  \begin{table}
    \begin{center}
    \caption{Relevant quantities in haloes of characteristic mass $m_h^*$ at $z$=0. The lower table contains quantities dependent on the particle mass $m_p$.}
    \label{tab:5-1}
    \begin{tabular}{lcccc} 
    \hline
    Quantity            & Symbol        & Scaling               &{Value at $r_p$}           & {Value at $r_h^*$}          \\
    \hline  \hline    
    Scale (m)           & {$r$}         &                       & $3\times 10^{-13}$       & $3.2\times 10^{22}$        \\
    Mass (kg)           & {$m_r$}       & $\propto r^{5/3}$     & $1.6\times 10^{-15}$     & $6.4\times 10^{43}$         \\
    Density ($kg/m^3$)  & {$\rho_r$}    & $\propto r^{-4/3}$    & $5.3\times 10^{22}$      & $1\times 10^{-26}$ \\
    Time (s)        & {$t_r$}   &  {$\propto r^{2/3}$}  & $1\times 10^{-6}$              & $4\times 10^{17}$  \\
    Velocity ($m/s$)  & {$v_r$} &  {$\propto r^{1/3}$} &  $4\times 10^{-7}$   & $2.6\times 10^{5}$  \\
    Mass flow ($kg/s$)  & {$\dot m_r$}  &  {$\propto r^{1}$} &  $1.6\times 10^{-9}$   & $1.6\times 10^{26}$  \\
    \hline \hline
    Mean distance (m)   & {$d_r$}       & $\propto r^{4/9}$    & $3\times 10^{-13}$         & $5.4\times 10^{3}$              \\
    Frequency ($1/s$)   & {$\Gamma_r$}  & $\propto r^{-4/9}$     & $2.8\times 10^{17}$      & $14.8$             \\
    Waiting time (s)    & {$\tau_{gr}$} & $\propto r^{-2/3}$    & $4\times 10^{-18}$        & $1.1\times 10^{-41}$   \\
    Waiting time (s)    & {$\tau_{hr}$} & {$\propto r^{-1}$}    &  $1\times 10^{-6}$        &  $1.1\times 10^{-41}$ \\
    \hline 
    \end{tabular}
  \end{center}
\end{table}

\section{Conclusions}
\label{sec:6}
In this paper, we propose that, between the fully linear and fully nonlinear regimes, there exists a universal transition range centered on a characteristic mass $m_h^*$, within which a quasi-universal spectrum with index $n$=-1 is established by gravity and becomes largely independent of the specifics of the initial conditions. We identify and develop the physical mechanism that enables this universality as a cascade of mass and energy across scales, and we derive the attendant universal scaling laws governing both the halo mass function and the halo density profiles.

We formulate the mass and energy cascade at two distinct but coupled levels. Globally, a cascade in halo mass space governs the redistribution of mass among haloes and thereby determines the halo mass function. Locally, within individual haloes, a cascade directed along the radius governs how the particles are arranged, and thus determines the halo density profile. On both levels, the cascade drives the system toward a statistically steady state that continuously releases energy and maximizes entropy. A defining property of this steady state is a scale-independent cascade rate: statistical halo structures are self-similar and scale-free, and there is no net accumulation of mass or energy at intermediate scales.

Because an N-body system in an expanding background has a decreasing total energy (Fig. \ref{fig:S1-3-2}), the global decrease or "dissipation" of energy facilitates the cascades: energy is injected around scale $m_h^*$, transferred from large to small haloes in mass space and from large to small radii within haloes, and is ultimately dissipated on small scales through both smooth halo merging and the motions of particles in haloes. The cascade establishes a statistically steady state with scale-independent rates. We find that kinetic energy is transferred inversely from small to large scales at a rate $\varepsilon(m_h,z)\propto m_h^{2/3}a^{-1}$, while potential energy undergoes a direct cascade from large to small scales at a rate of $1.4\varepsilon$ (Fig. \ref{fig:4-2}).

For the mass cascade in the halo mass space, the net mass transfer is upward in a bottom-up fashion, i.e., consistent with hierarchical structure formation. Two distinct regimes emerge, i.e., a propagation range with a scale-independent rate of mass transfer $\varepsilon_m$ below a characteristic mass $m_{h}^{*}$ and a deposition range to actively consume the mass cascaded from small scales to grow halo larger than $m_{h}^{*}$ (Fig. \ref{fig:3-2}). The inverse mass cascade leads to a random walk of haloes in the mass space with a position-dependent waiting time $\tau_g$. The distribution of haloes in the halo mass space, i.e. the halo mass function, is naturally given by the solution of the corresponding Fokker-Planck equation for halo random walk (double-$\lambda$ mass function in Eq. \eqref{eq:3-18}). This approach for the halo mass function is simple without resorting to a spherical or elliptical collapse model. The non-Gaussian features in the density field around scales $m_h^*$ are highly expected to be an important signature of the inverse mass cascade. Since haloes have finite kinetic and potential energy, there also exists an inverse cascade of kinetic energy at a rate of $\varepsilon_u$ from small to large mass scales and a direct cascade of potential energy at a rate of $1.4\varepsilon_u$ from large to small mass scales (Fig. \ref{fig:3-7}). Universal scaling laws can be developed with small-scale permanence for halo group mass $m_g$ (Fig.\ref{fig:3-5}). 

For the energy cascade in haloes, the kinetic energy is transferred outward, while the potential energy is transferred inward. The origin of halo structure can be described by the random walk of particles with a position-dependent waiting time $\tau_{gr}$. The distribution of particles in haloes, i.e., the halo density profile, can be analytically derived from the corresponding Fokker-Planck equation (double-$\gamma$ profile in Eq. \eqref{eq:4-4-7}). Universal scaling laws (Eqs. \eqref{eq:4-3-2}) can be identified with small-scale permanence for the halo density profile (Fig. \ref{fig:4-5}).

Since the waiting time and jumping length for particle random walk in haloes depend on the particle mass $m_p$ (Eq. \eqref{eq:4-3-3}), new constraints on the particle mass can be identified from the physical constraints on these quantities. Based on the assumption that the waiting time should be greater than the Planck time (the smallest unit of time) or the jumping length should be greater than the Planck length (the smallest unit of length), we propose a new constraint for particle mass $m_p\ge 10^{-15}$kg or $10^{12}$GeV (Eqs. \eqref{eq:5-4-4}, Eq. \eqref{eq:5-4}, \eqref{eq:5-9-3}, and \eqref{eq:5-14}). These constraints exclude the standard WIMPs and strongly suggest a heavy dark matter scenario (superheavy right-handed neutrinos, etc.). This constraint is also consistent with the particle mass obtained by extending the established scaling laws to the smallest free-streaming scale \citep{Xu:2022-Postulating-dark-matter-partic}.

\section*{Author contributions}
Z.J. Xu performed the conception, data analysis, and writing.

\section*{Funding}
This research was supported by the Laboratory Directed Research and Development at Pacific Northwest National Laboratory (PNNL). PNNL is a multiprogram national laboratory operated for the U.S. Department of Energy (DOE) by Battelle Memorial Institute under contract no. DE-AC05-76RL01830.

\section*{Data Availability}
Two datasets underlying this article, that is, halo-based and correlation-based statistics of dark matter flow, are available on Zenodo \citep{Xu:2022-Dark_matter-flow-dataset-part1,Xu:2022-Dark_matter-flow-dataset-part2}, along with the presentation 'A comparative study of dark matter flow \& hydrodynamic turbulence and its applications' \citep{Xu:2022-Dark_matter-flow-and-hydrodynamic-turbulence-presentation}. All data are also available on GitHub \citep{Xu:Dark_matter_flow_dataset_2022_all_files}.

\bibliographystyle{Papers}
\bibliography{Papers}

\appendix
\addtocontents{toc}{\protect\setcounter{tocdepth}{-1}}

\section{N-body simulations and data}
\label{sec:2}
The large-scale cosmological Illustris simulation (Illustris-1-Dark) \citep{NELSON:2015-The-illustris-simulation} was used to demonstrate and validate the concepts. Illustris is a suite of large-volume cosmological dark matter only and hydrodynamical simulations. The selected Illustris-1-Dark is a dark matter only simulation that has a cosmological volume of 106.5Mpc$^3$ and 1820$^3$ DM particles for a high mass resolution. Each DM particle has a mass around $m_p=7.6\times 10^6 M_{\odot}$. The gravitational softening length is around 1.4 kpc. The simulation has cosmological parameters of dark matter density $\Omega_{DM}=0.2726$, dark energy density $\Omega_{DE}=0.7274$ at $z=0$, and Hubble constant $h=0.704$. Some key parameters of N-body simulations are listed in Table \ref{tab:1}.

For cross-validation,  the cosmological \textit{N}-body simulations carried out by the Virgo consortium were also used. A detailed description can be found in \citep{Frenk:2000-Public-Release-of-N-body-simul,Jenkins:1998-Evolution-of-structure-in-cold}. The SCDM simulation of the Virgo consortium focuses on matter-dominant dark matter-only simulations with a standard CDM power spectrum (SCDM). The same set of simulation data has been widely used in several different studies, from clustering statistics \citep{Jenkins:1998-Evolution-of-structure-in-cold} to the formation of halo clusters in large-scale environments \citep{Colberg:1999-Linking-cluster-formation-to-l}, and testing models for halo abundance and mass functions \citep{Sheth:2001-Ellipsoidal-collapse-and-an-im}. This simulation has a lower mass resolution with particle mass $m_{p} =2.27\times 10^{11} {M_{\odot } /h} $. The simulation box is around 240 Mpc/h, where \textit{h} is the dimensionless Hubble constant in units of 100 km/s/Mpc. 

The friends-of-friends algorithm (FOF) was used to identify all haloes in each simulation that depend only on a dimensionless parameter \textit{b}, which defines the linking length $b\left({N/V} \right)^{{-1/3} } $, where $V$ is the volume of the simulation box. In this work, haloes were identified with a linking length parameter of $b=0.2$. Identifying all haloes of different sizes, all dark matter particles were divided into halo particles with a total mass $M_h$ and out-of-halo particles that do not belong to any halo. Therefore, $M_h$ is the total mass of all haloes. We focus on the evolution of mass and energy in haloes of different mass $m_h$. All haloes were grouped into halo groups of various sizes according to the halo mass $m_{h} $ (or $n_{p} $, the number of particles in the halo), where $m_{h} =n_{p}m_{p}$. The total mass for a halo group of mass $m_{h}$ is $m_{g} =m_{h} n_{h}$, where $n_{h} $ is the number of haloes in that group. Two different cosmological simulations were used to demonstrate the fundamental concepts. The same approach can be easily extended to other cosmological simulations. 

\begin{table}
\caption{Virgo (SCDM) and Illustris (Illustris-1-Dark) parameters}
\begin{tabular}{p{0.3in}p{0.15in}p{0.15in}p{0.4in}p{0.3in}p{0.5in}p{0.4in}} 
\hline 
Run & $\Omega_{0}$ & \centering $h$ &  \makecell{L\\(Mpc/h)} & \centering N & \makecell{$m_{p}$\\$M_{\odot}/h$} & \makecell{$l_{soft}$\\(Kpc/h)} \\ 
\hline 
SCDM & 1.0 & \centering 0.5 & \centering 239.5 & \centering $256^{3}$ & 2.27$\times 10^{11}$ & \makecell{\centering 36} \\ 
\hline 
Illustris & 0.24    &\centering 0.704& \centering 75 & \centering $1820^3$  & 5.28$\times 10^{6}$   &  \makecell{\centering 1.4} \\

\hline 
\end{tabular}
\label{tab:1}
\end{table}

\section{Cosmic energy evolution}
\label{sec:2-2}
In this section, we analyze the temporal evolution of the global energy in a self-gravitating, collisionless dark-matter flow (SG-CFD). Although the global energy, obtained by averaging across all scales, does not resolve inter-scale energy transfer itself, its systematic decline over time provides the driving force necessary to initiate and sustain the cascade on a global scale. The corresponding scale-to-scale transfer of mass and energy is presented in Section \ref{sec:3-2} and is fundamentally distinct from the Press–Schechter formulations.

\begin{figure}
\includegraphics*[width=\columnwidth]{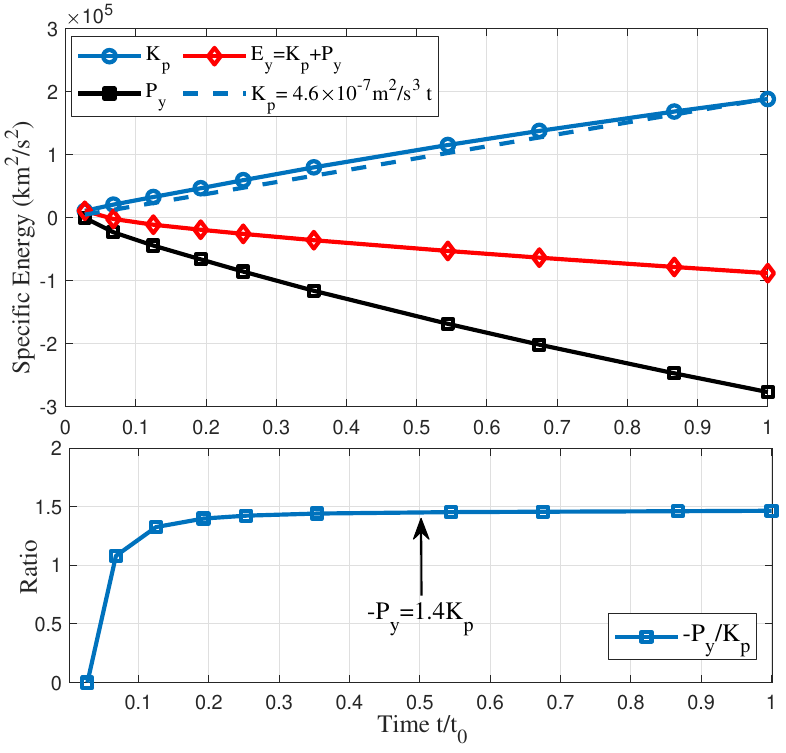}
\caption{The variation of specific kinetic energy $K_p$ (energy per unit mass), potential energy $P_y$, and total energy $E_y=K_p+P_y$ (unit: $km^2/s^2$) with time $t$ (normalized by $t_0$, the present cosmic time) from Virgo SCDM simulation. Solution in Eq. \eqref{eq:4-1-s} is also presented for comparison with $\varepsilon_u=-4.6\times 10^{-7}m^2/s^3$. Simulation confirms a linear increase of $K_p=-\varepsilon_u t$ with time and negative potential energy $P_y=-1.4K_p$. The total energy $E_y=-0.4K_p$ also decreases with time. The figure demonstrates that the total energy of the N-body system in an expanding background decreases with time as if the total energy is "dissipated" at a rate of $0.4\varepsilon_u$. This continuous decrease in total energy leads to and maintains the energy cascade in collisionless dark matter. An "effective" potential exponent $n_e=2K_p/P_y$=-10/7 can be obtained.} 
\label{fig:S1-3-2}
\end{figure}

The equations of motion for \textit{N} collisionless particles in comoving coordinates $\boldsymbol{\mathrm{x}}$ and physical time \textit{t} read \citep{Peebles:1980-The-Large-Scale-Structure-of-t}:
\begin{equation} 
\label{eq:2-2_1} 
\frac{d^{2} \boldsymbol{\mathrm{x}}_{i} }{dt^{2} } +2H\frac{d\boldsymbol{\mathrm{x}}_{i} }{dt} =-\frac{Gm_{p} }{a^{3} } \sum _{j\ne i}^{N}\frac{\boldsymbol{\mathrm{x}}_{i} -\boldsymbol{\mathrm{x}}_{j} }{\left|\boldsymbol{\mathrm{x}}_{i} -\boldsymbol{\mathrm{x}}_{j} \right|^{3} },        
\end{equation} 
where \textit{N} particles have equal mass $m_{p}$. The Hubble parameter $H\left(t\right)={\dot{a}/a}$. Here, $H$ has a "damping" effect, which leads to a decrease in total energy of the N-body system (Eq. \eqref{eq:4} and Fig. \ref{fig:S1-3-2}). In the matter-dominant era, the Hubble parameter satisfies $Ht=2/3$. 

Next, we will derive the energy evolution based on the equations of motion (Eq. \eqref{eq:2-2_1}). We first introduce a transformed time variable \textit{s} as ${ds/dt} =a^{p} $, where \textit{p} is an arbitrary exponent. In terms of the new time variable $s$, the original Eq. \eqref{eq:2-2_1} can be transformed to 
\begin{equation} 
\label{eq:2-2_2} 
\begin{split}
&\frac{d^{2} \boldsymbol{\mathrm{x}}_{i} }{ds^{2} } +\frac{d\boldsymbol{\mathrm{x}}_{i} }{ds} \left(p+2\right)a^{-p} H \equiv  a^{-(3+2p)}\frac{\boldsymbol{\mathrm{F}}_{i}}{m_p},\\
&\frac{\boldsymbol{\mathrm{F}}_{i}}{m_p} = -G m_p \sum _{j\ne i}^{N}\frac{\boldsymbol{\mathrm{x}}_{i} -\boldsymbol{\mathrm{x}}_{j} }{\left|\boldsymbol{\mathrm{x}}_{i} -\boldsymbol{\mathrm{x}}_{j} \right|^{3}} = -\frac{\partial P_s}{\partial \boldsymbol{\mathrm{x}}_{i}}, 
\end{split}
\end{equation} 
where $\boldsymbol{\mathrm{F}}_{i} $ is the resultant force on particle \textit{i} of all other particles, while $P_s$ is the total specific potential energy in the comoving coordinates. Equation \eqref{eq:2-2_2} reduces to the original Eq. \eqref{eq:2-2_1} when $p=0$. With $p=-2$, the first-order derivative vanishes in Eq. \eqref{eq:2-2_2} and \textit{s} is the time variable for integration in \textit{N-body} simulations. By setting $p=-1$, \textit{s} is the conformal time. Setting $p=-{3/2}$ along with $H_0^2=H^2a^3$ for the matter-dominant era, the equation of motion becomes 
\begin{equation} 
\label{eq:2-2_3} 
\frac{d^{2} \boldsymbol{\mathrm{x}}_{i} }{ds^{2} } +\frac{1}{2} H_{0} \frac{d\boldsymbol{\mathrm{x}}_{i} }{ds} =\frac{\boldsymbol{\mathrm{F}}_{i} }{m_{p}}.      
\end{equation} 
In this transformed equation for the matter era, the scale factor \textit{a} is not explicitly involved. The time-dependent Hubble parameter $H$ is replaced by the Hubble constant $H_0$. This transformation offers significant convenience for the evolution of cosmic energy \citep{Xu:2022-The-evolution-of-energy--momen}. 

We first identify the transformation between the velocity $\boldsymbol{\mathrm{v}}_{i}$ in the time variable $s$ and the peculiar velocity $\boldsymbol{\mathrm{u}}_{i}$,
\begin{equation} 
\label{eq:2-2_4} 
\begin{split}
&\boldsymbol{\mathrm{v}}_{i} =\frac{d\boldsymbol{\mathrm{x}}_{i} }{ds} =a^{-p} \frac{d\boldsymbol{\mathrm{x}}_{i} }{dt}=a^{-p-1}\boldsymbol{\mathrm{u}}_{i}, \quad \boldsymbol{\mathrm{u}}_{i} =a\frac{d\boldsymbol{\mathrm{x}}_{i} }{dt},  \\
&K_s = K_p a^{-2p-2}, \quad P_s = aP_y,
\end{split}
\end{equation} 
where the kinetic energy $K_s$ and the potential $P_s$ in the transformed equation can now be related to the peculiar kinetic energy $K_p$ and the potential $P_y$ in the physical coordinates.

The energy evolution of the N-body system can be obtained by multiplying ${\boldsymbol{\mathrm{v}}_{i} = d\boldsymbol{\mathrm{x}}_{i} /ds}$ on both sides of Eq. \eqref{eq:2-2_2} and adding the equation of motion for all particles together \citep{Xu:2022-The-evolution-of-energy--momen}. An exact and simple equation (in time variable $s$) for the specific kinetic energy $K_s$ and the potential energy $P_s$ can be obtained as
\begin{equation} 
\label{eq:2-2_5} 
\frac{d K_{s} }{d s}+2HK_s(p+2)a^{-p}+a^{-(3+2p)}\frac{dP_s}{ds}=0.          
\end{equation} 
Setting $p=0$ and using the relations in Eq. \eqref{eq:2-2_4}, the exact cosmic energy equation for the energy evolution of the N-body system reads as
\begin{equation} 
\label{eq:4} 
\frac{\partial E_{y} }{\partial t} +H\left(2K_{p} +P_{y} \right)=0,         
\end{equation} 
which describes the energy evolution in an expanding background. Here $K_{p}$ is the peculiar kinetic energy, $P_{y}$ is the potential energy in physical coordinates, and $E_{y}=K_{p}+P_{y} $ is the total specific energy. This is also known as the Layzer-Irvine equation in the literature \citep{Irvine:1961-Local-Irregularities-in-a-Univ,Layzer:1963-A-Preface-to-Cosmogony--I--The}. 

In the matter-dominant era with $Ht$=2/3, the cosmic energy evolution (Eq. \eqref{eq:4}) admits a linear solution
\begin{equation} 
\label{eq:4-1-s} 
\begin{split}
&K_p=-\varepsilon_u t,\quad P_y = \frac{7}{5} \varepsilon_u t \quad\textrm{and}\quad E_y=\frac{2}{5}\varepsilon_ut,
\end{split}
\end{equation} 
where $\varepsilon_u<0$ is an important physical constant (unit: m$^2$/s$^3$), the rate of the energy cascade. It represents the rate of time variation of cosmic energy. The solution in Eq. \eqref{eq:4-1-s} can be directly validated by N-body simulations. Figure \ref{fig:S1-3-2} presents the evolution of the energy from the Virgo simulation (SCDM). The (specific) kinetic energy $K_p$ and potential energy $P_y$ were calculated as the mean energy of all dark matter particles in the N-body system. Figure \ref{fig:S1-3-2} confirms the solution in Eq. \eqref{eq:4-1-s}, i.e., a linear increase of $K_p$ with time and a negative potential energy $P_y=-1.4K_p$. Two key messages are: \\

\noindent i) According to the virial theorem, in virial equilibrium, $2K_p=n_eP_y$, where $n_e$ is an exponent of the interaction potential $r^{n_e}$. For standard gravity, the exponent $n_e=-1$. Since gravity is the only interaction in the N-body system, we would expect $n_e=-1$ for an N-body system. However, the virial equilibrium can never be reached by the non-equilibrium N-body system (e.g., SG-CFD of dark matter flow). Instead of the virial equilibrium, a statistically steady state is established to continuously release system energy and maximize system entropy, which is manifested by the mass and energy cascade. In that particular state, the solution of Eq. \eqref{eq:4-1-s} is valid such that an "effective" exponent $n_e=-10/7$ is obtained, which deviates from -1; \\

\noindent ii) The total energy $E_y=-0.4K_p$ decreases with time. In analogy to hydrodynamic turbulence, the total energy of the N-body system in an expanding background decreases with time, as if "dissipated" due to the expanding background, even though there is no viscous force in collisionless dark matter. Here, we admit that cosmic energy decreases with time and estimate the rate of energy "dissipation":
\begin{equation} 
\label{eq:5} 
\begin{split}
\varepsilon_{u} =-\frac{K_{p}}{t} =-\frac{3}{2} \frac{u^{2} }{t} =-\frac{3}{2} \frac{u_{0}^{2} }{t_{0} }\approx -4.6\times 10^{-7} \frac{m^{2} }{s^{3}}, 
\end{split}
\end{equation} 
where $u_{0} \equiv u\left(t=t_{0} \right) \approx 350$km/s is the one-dimensional velocity dispersion of all dark matter particles, and $t_{0}\approx$13.7 billion years is the physical time at the present epoch or the age of the universe. Different simulations may have slightly different values of $u_0$ due to different cosmological parameters. However, the rate of energy "dissipation" $\varepsilon_u \sim -10^{-7}m^2/s^3$ should be a good estimate.

Hereafter, the quoted word "dissipation" stands for the decrease in total energy caused by the expanding background, in contrast to the actual dissipation caused by the viscous force in hydrodynamic turbulence. In the statistically steady state, the energy "dissipated" on small scales should balance the energy cascaded from large scales. Therefore, Eq. \eqref{eq:4-1-s} also suggests an inverse cascade of kinetic energy from small to large scales at a rate of $\varepsilon_u$ and a direct cascade of potential energy at a rate of $-1.4\varepsilon_u$. The total energy is directly cascaded to small scales at a rate of $-0.4\varepsilon_u$ to provide the energy that is "dissipated" on small scales. In Sections \ref{sec:3} and \ref{sec:3-3}, we illustrate how the mass and energy cascades are initiated by the global-scale energy "dissipation" and proceed in halo mass space via halo merging and in individual haloes via the random walk of particles, and their impacts on structure formation and evolution.

\section{More details on the mass cascade}
\label{sec:A1}
This section presents some additional details on the inverse mass cascade in halo mass space in the matter-dominant era. 
\subsection{Mass redistribution among halo groups}
\label{sec:A1.1}
To study the mass transfer between different mass scales, we divide the entire system into two subsystems: 1) the out-of-halo subsystem with a total mass of $M_{o} $ includes all masses that do not belong to any haloes; 2) the halo subsystem with a total mass of $M_{h} $ includes all masses contained in all haloes. The inverse mass cascade involves continuous mass and energy exchange between two subsystems. 

\begin{figure}
\includegraphics*[width=\columnwidth]{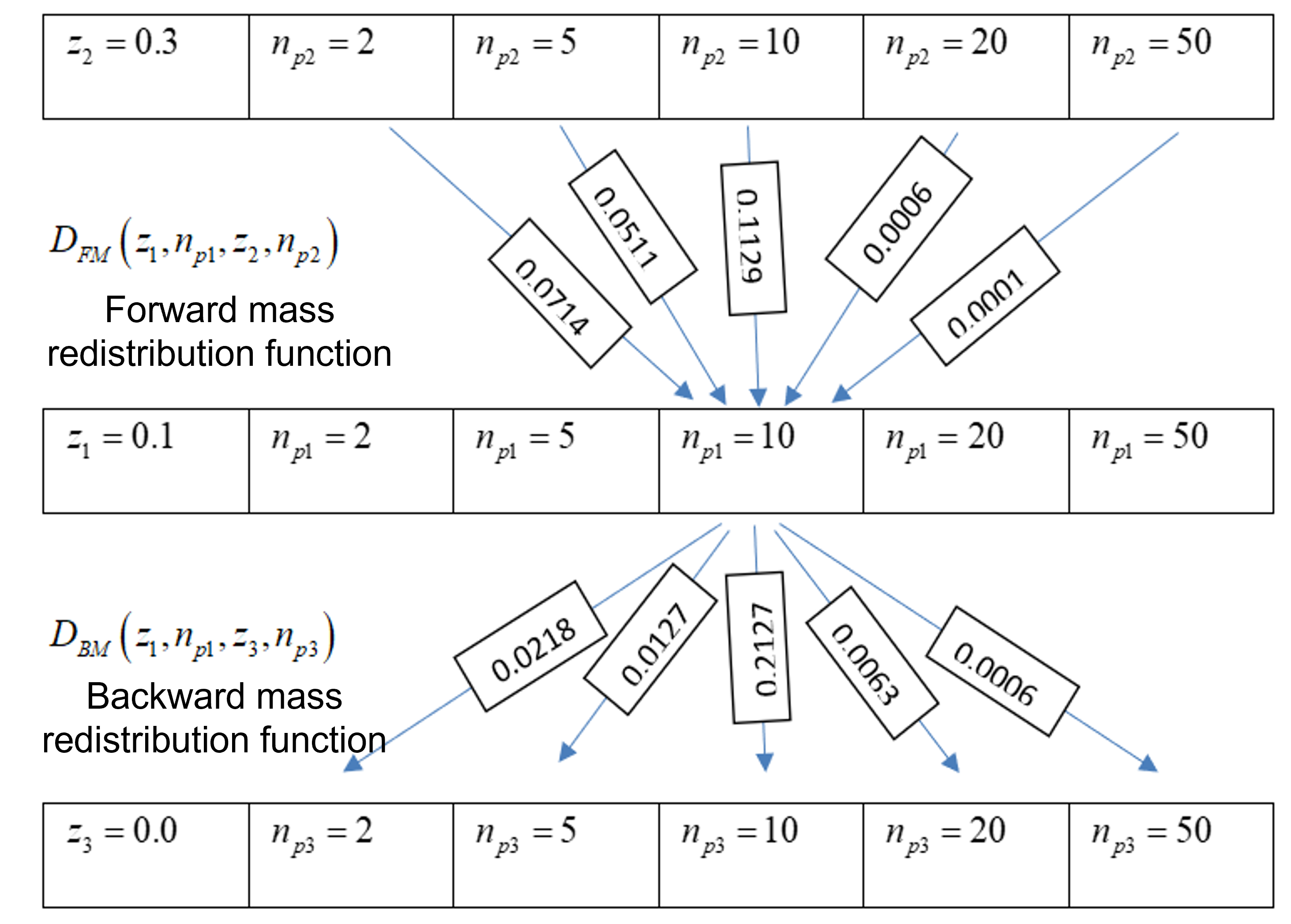}
\caption{Example of a forward redistribution function $D_{FM}$ from $z_2=0.3$ to $z_1=0.1$ and a backward redistribution function $D_{BM}$  from $z_1=0.1$ to $z_3=0$ for halo group of size $n_{p1}=10$. The figure shows the fraction of mass for a halo group of size $n_{p1}=10$ at $z_1=0.1$ inherited from halo groups of different sizes at an earlier redshift $z_2=0.3$ (forward mass redistribution). Similarly, backward redistribution gives the fraction of mass passed to halo groups of different sizes at a later redshift $z_3=0$. Halo inherits and passes most of its mass from and to haloes of similar size (locality in mass scale).}
\label{fig:A1-1}
\end{figure}

All haloes in the halo subsystem can be grouped into haloes with the same mass $m_{h} $ or particle number $n_{p}$. We focus on mass transfer (cascade) between halo groups of different sizes. Similarly to the concept of halo merger trees, the starting point is to define functions that describe the mass redistribution among halo groups at different redshifts $z$. The forward mass redistribution function $D_{FM} \left(z_{1} ,n_{p1} ,z_{2} ,n_{p2} \right)$ describes the mass fraction of a halo group of size $n_{p1} $ at the redshift $z_{1} $ that is inherited from the halo group of size $n_{p2} $ at an earlier redshift $z_{2} $. Similarly, the backward mass redistribution function $D_{BM} \left(z_{1},n_{p1},z_{3},n_{p3} \right)$ can be defined as the mass fraction of a halo group of size $n_{p1} $ at redshift $z_{1} $ that will be passed to the halo group of size $n_{p3} $ at a later redshift $z_{3} $. These functions precisely quantify the progenitor and inheritor of the total mass of a given halo group. Figure \ref{fig:A1-1} provides an example for two functions. For given $z_{1} $, $n_{p1} $, and $z_{2} $ or $z_{3} $, the normalization condition requires,
\begin{equation}
\begin{split}
&\sum _{n_{p2} }^{}D_{FM} \left(z_{1} ,n_{p1} ,z_{2} ,n_{p2} \right) =1\\
&\textrm{and}\\
&\sum _{n_{p3} }^{}D_{BM} \left(z_{1} ,n_{p1} ,z_{3} ,n_{p3} \right) =1.
\end{split}
\label{ZEqnNum418270}
\end{equation}

\begin{figure}
\includegraphics*[width=\columnwidth]{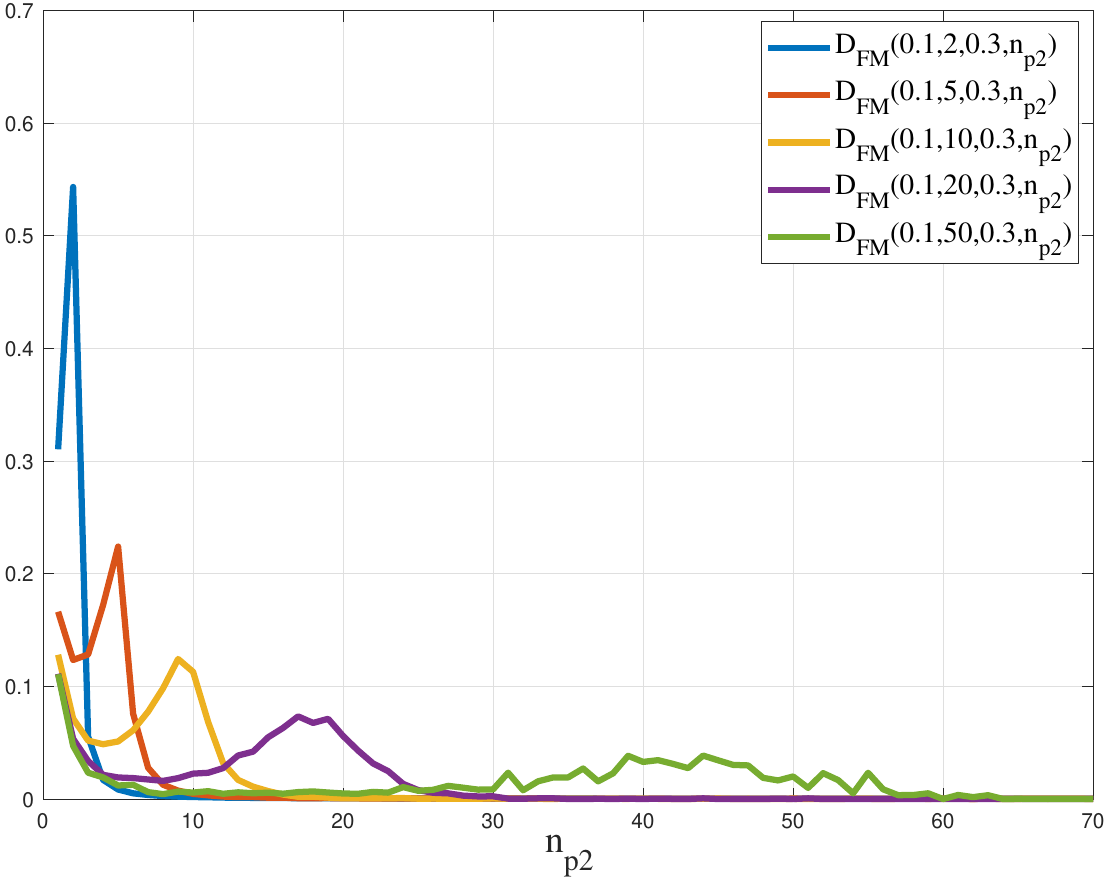}
\caption{Forward mass redistribution function $D_{FM} \left(z_{1},n_{p1},z_{2},n_{p2} \right)$ from $z_{2} =0.3$ to $z_{1} =0.1$ for five different halo groups of sizes $n_{p1} $=2, 5, 10, 20, 50, respectively. The figure shows the mass fraction of a halo group of size $n_{p1} $ at $z_{1} =0.1$ inherited from halo groups of various sizes at an earlier redshift $z_{2} =0.3$. The interaction among groups of haloes is shown to be local in the mass space, i.e., the halo group of size $n_{p1} $ inherits its mass mostly from the interaction between halo groups of similar (neighboring) sizes to $n_{p1} $ and single mergers. Note that there are two peaks at around $n_{p2} \approx n_{p1} $ (halo groups of similar size) and $n_{p2} \approx 1$ (single merger). Halo groups of larger size inherit their mass from a wider distribution in size $n_{p2}$.}
\label{fig:1}
\end{figure}

\begin{figure}
\includegraphics*[width=\columnwidth]{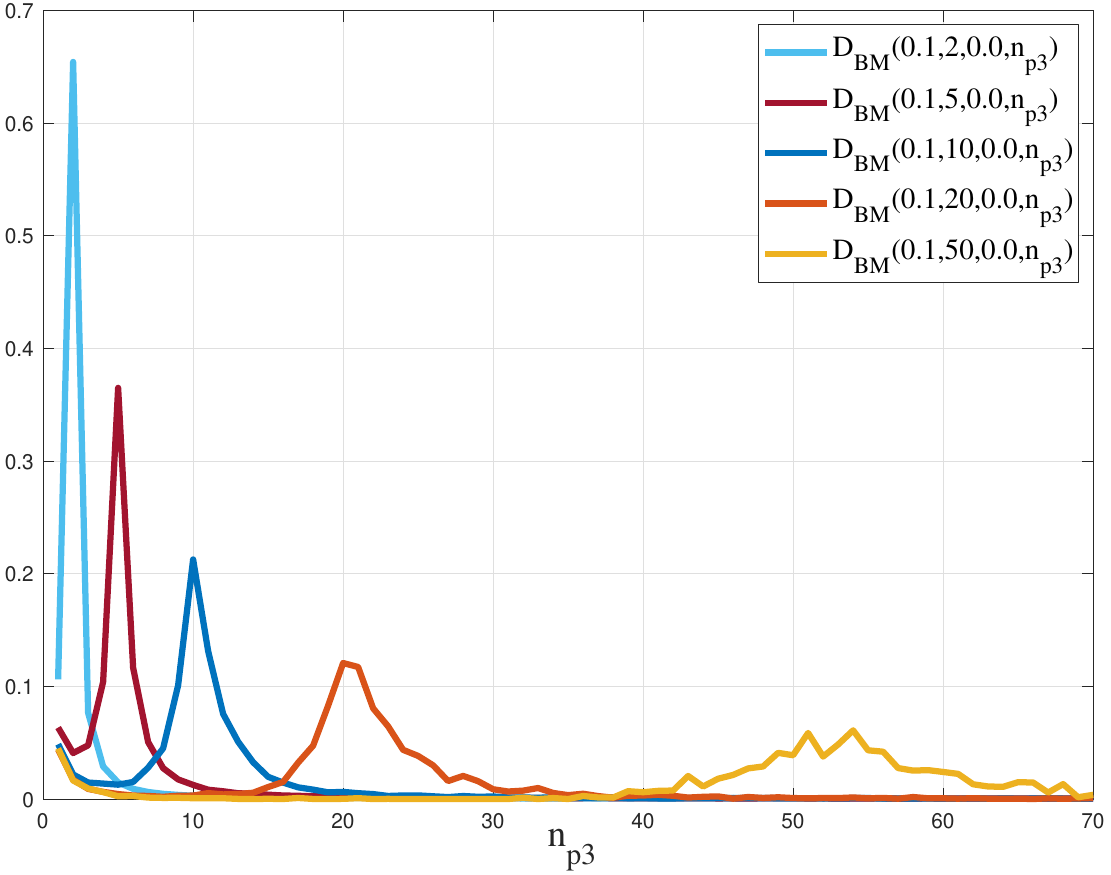}
\caption{Backward mass redistribution function $D_{BM} \left(z_{1},n_{p1},z_{3},n_{p3} \right)$ from $z_{1} =0.1$ to $z_{3} =0.0$ for five different halo groups of sizes $n_{p1} $=2, 5, 10, 20, 50, respectively. The figure shows the mass fraction of a halo group of size $n_{p1} $ at $z_{1} =0.1$ passed to halo groups of various sizes at later redshift $z_{3} =0.0$. Again, the interaction among groups of haloes is shown to be local in the mass space, i.e., the halo group of size $n_{p1} $ passes most of its mass via merging/breaking to halo groups of similar (neighboring) size to $n_{p1} $ and single mergers. Hence, there are two peaks at around $n_{p3} \approx n_{p1} $ and $n_{p3} \approx 1$. Halo groups of larger size pass their mass to a wider distribution of halo group sizes at a later redshift. In comparison, halo groups of smaller size pass their mass to a relatively narrower distribution in size $n_{p3}$.}
\label{fig:2}
\end{figure}

Figures. \ref{fig:1} and \ref{fig:2} plot the forward and backward redistribution functions as a function of the halo group size $n_{p2} $ or $n_{p3} $ for five different group sizes $n_{p1} $=2, 5, 10, 20, 50, respectively. Halo groups of five different sizes $n_{p1} $ inherit and pass their mass to a distribution of halo group sizes. The interaction among halo groups is shown to be local on the mass scale. This demonstrates that smooth and minor merging is dominant over the major merging between two haloes of comparable mass. The halo group of size $n_{p1} $ inherits or passes most of its mass via merging/breaking between haloes of similar (neighboring) size to $n_{p1}$ and single mergers. Therefore, there are two peaks for forward/backward mass redistribution functions at around $n_{pi} \approx n_{p1} $ (halo groups of similar size) and $n_{pi} \approx 1$ (single mergers), where \textit{i}=2 or 3 for forward or backward functions, respectively. 

Groups of large haloes inherit/pass their mass from/to a wider distribution of halo sizes. In comparison, groups of small haloes inherit/pass their mass from/to a relatively narrower distribution of halo sizes. Both mass redistribution functions are not symmetric about the halo size $n_{p1} $, with more mass inherited from halo groups below the size $n_{p1} $ and more mass passed to halo groups above the size $n_{p1}$, i.e., a net transfer of mass to large scales. The sharp peaks for halo groups of smaller size and the widespread distribution for halo groups of larger size in Figs. \ref{fig:1} and \ref{fig:2} indicate that small haloes have relatively longer lifespans and can exist for a longer time such that most small haloes will remain in the same group at a later redshift. Large haloes tend to have a relatively shorter lifespan. Halo lifespan will be further discussed in Section \ref{sec:3-4}.

To determine the direction of the mass cascade, we introduce a net mass redistribution function $D_{NM} $ as the difference between the backward and forward mass redistribution functions, 
\begin{equation} 
\label{ZEqnNum701103} 
\begin{split}
D_{NM}&\left(z_{1} ,n_{p1} ,z_{3} ,n_{p2} ,z_{2} \right)=\\
&D_{BM} \left(z_{1} ,n_{p1} ,z_{3} ,n_{p2} \right)-D_{FM} \left(z_{1} ,n_{p1} ,z_{2} ,n_{p2} \right). 
\end{split}
\end{equation} 
The net mass redistribution function $D_{NM} $ measures the net effect of the halo group size $n_{p1} $ at redshift $z_{1} $ on the mass cascade of halo group size $n_{p2} $ from redshift $z_{2} $ to $z_{3} $, with $D_{NM} <0$ indicating that the halo group of size $n_{p1} $ inherits more mass from the halo group size $n_{p2} $ than the mass it passes to the halo group of the same size $n_{p2} $. Obviously, from Eq. \eqref{ZEqnNum418270},
\begin{equation} 
\label{eq:3} 
\sum _{n_{p2} }^{}D_{NM} \left(z_{1} ,n_{p1} ,z_{2} ,n_{p2} \right) =0.         
\end{equation} 

\begin{figure}
\includegraphics*[width=\columnwidth]{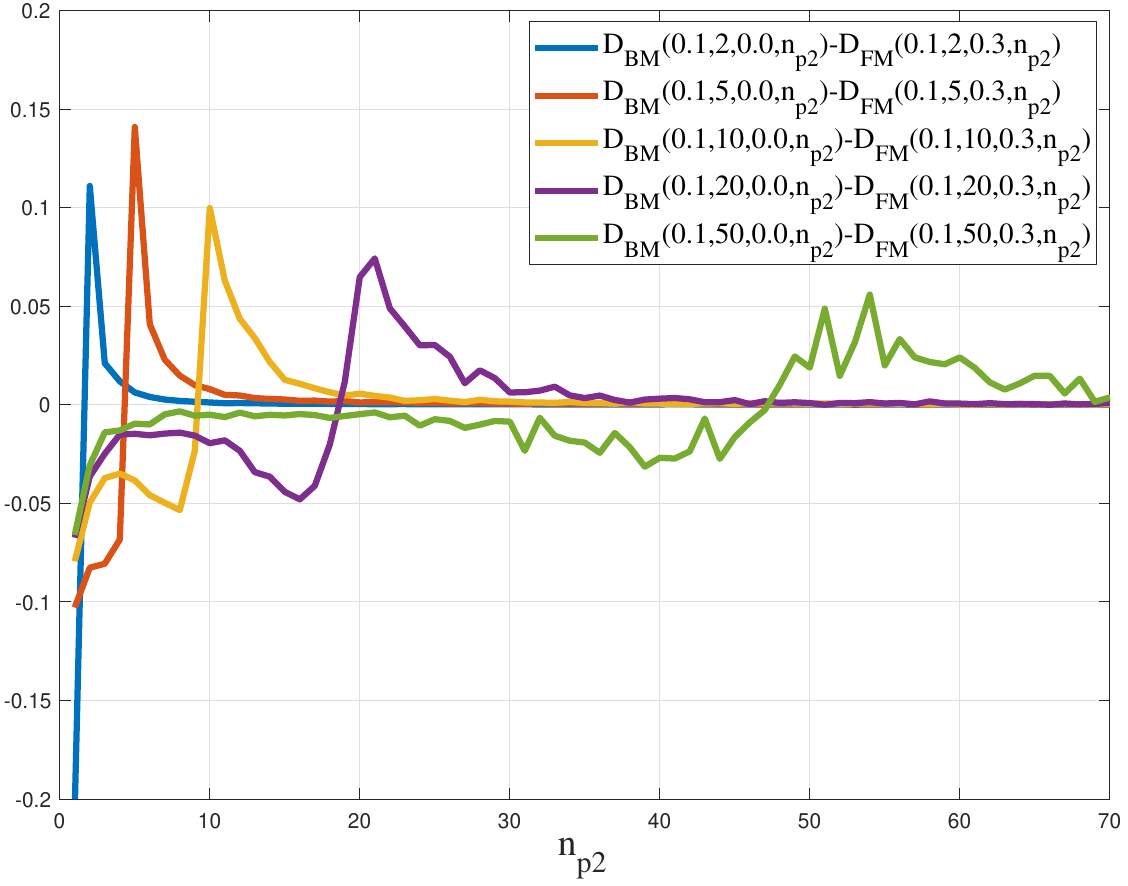}
\caption{The net mass redistribution function $D_{NM} $ as a function of halo group size $n_{p2} $ for five different halo group sizes $n_{p1} $, with $D_{NM} <0$ indicating that the halo group of size $n_{p1} $ inherits more mass from the halo group of size $n_{p2} $ than the mass it passes to the group of the same size $n_{p2} $. The net effect of mass redistribution is that haloes are transferring mass from smaller scales to larger scales, i.e., an inverse mass cascade in mass space, in contrast ot a direct mass cascade.}
\label{fig:3}
\end{figure}

Figure \ref{fig:3} plots the net mass redistribution function $D_{NM} $ for five halo group sizes $n_{p1} $, with $D_{NM} <0$ for halo groups $n_{p2} $ smaller than size $n_{p1} $ and $D_{NM} >0$ for halo groups $n_{p2} $ larger than size $n_{p1} $. The net effect is that haloes are transferring mass from small mass scales to large mass scales, i.e., an inverse mass cascade in the halo mass space. In contrast, the direct mass cascade refers to the transferring of mass from large to small mass scales. In short, three distinct features of inverse mass cascade can be identified:

\begin{enumerate}
\item \noindent Locality: the transferring of mass is local in mass space. Haloes inherit/pass their mass mostly from/to haloes of the same or similar size. The interaction among haloes is shown to be local on a mass scale. For any finite time interval $\Delta t$, the interaction (merging/breaking) between haloes can involve multiple haloes of different sizes. However, for an infinitesimal time interval $\Delta t\to 0$, the interaction is most likely between two haloes of very different sizes (haloes of a similar size and a single merger). 
\newline
\item \noindent Asymmetry: mass transfer across halo groups is a two-way process with mass cascading both upward and downward in the halo mass space. However, the net mass transfer is upward, i.e., the structure formation proceeds in a ``bottom-up'' fashion. The mass redistribution functions of a given halo size $n_{p1} $ are asymmetric about $n_{p1} $, with more mass inherited from halo groups smaller than $n_{p1} $ (via halo merging) and less mass inherited from halo groups larger than $n_{p1} $ (via halo breaking-up). Larger-size haloes accrete their mass from a relatively wider range of haloes, while smaller-size haloes accrete their mass from a narrower size range of haloes.
\newline
\item \noindent Inverse: mass cascade through halo groups of different sizes is two-way and asymmetric. The net effect is that haloes are transferring mass from smaller to larger mass scales, with halo merging being dominant over the halo breaking up, i.e., an inverse mass cascade.
\end{enumerate}

\subsection{Time scales in mass cascade}
\label{sec:A1.2}
In this section, we try to develop a few elementary ideas about the time and mass scales for inverse mass cascades. We observe that there exists a broad spectrum of halo sizes. This can be a direct result of maximizing system entropy \citep{Xu:2023-Maximum-entropy-distributions-of-dark-matter}. The smallest halo is often created by gravitational collapse at the smallest scale that merges with other haloes and passes their mass onto larger haloes. The larger haloes are themselves transitory and pass their mass to even larger haloes, and so on. At every instant \textit{t}, there is a continuous cascade of mass from the smallest to the largest mass scales that we assumed to be a characteristic mass $m_{h}^{*} $ 

Let the time scale $\tau _{h} \left(m_{h},a\right)$ be the average waiting time of a single merging event with a single merger for a halo group of mass $m_{h} <m_{h}^{*} $ at scale factor \textit{a}. The rate at which mass is passed up from this group is $\varepsilon _{m} \sim -{m_{h} /\tau _{h} } $ (negative sign for inverse mass cascade). When the system is in a statistically steady state, this rate of mass transfer must match exactly the rate of mass injecting into the halo sub-system at the smallest scales $m_{h} \to 0$ and the rate of mass dissipation at the largest mass scale $m_{h}^{*} $. If this is not the case, there would be a net accumulation of mass at some intermediate scale below $m_{h}^{*} $. We exclude this possibility because we want the statistics of halo groups to be self-similar and scale-free for halo groups of mass less than $m_{h}^{*} $ once a statistically steady state is established. This means a mass propagation range with halo mass $m_{h} <m_{h}^{*}$, 
\begin{equation} 
\label{ZEqnNum911884} 
-\varepsilon _{m} \sim {m_{h} /\tau _{h} } ={m_{h}^{*} /\tau _{h}^{*} } ,         
\end{equation} 
where the mass flux $\varepsilon _{m} $ is independent of the halo mass $m_{h}$. 

With the system in the statistically steady state, halo groups with mass below $m_{h}^{*} $ ($m_{h} <m_{h}^{*} $) propagate the mass to larger scales without any net accumulation of mass in that group. The total mass in the group $m_{g} =m_{h} n_{h} $ should be time-invariant, where $n_{h} $ is the number of haloes in that group that should also be time-invariant. Mass cascade in this range does not contribute to growing the halo group mass $m_{g} $.  
The average waiting time $\bar{\tau }_{h} $ (halo lifespan) for a given halo in the group can be calculated,
\begin{equation} 
\label{ZEqnNum342616} 
\begin{split}
\bar{\tau }_{h} \left(m_{h} ,a\right)&=\sum _{k=1}^{\infty }\frac{k\tau _{h} }{n_{h} } \left(\frac{n_{h} -1}{n_{h} } \right)^{k-1} \\
&= \frac{\tau _{h} }{n_{h} } +\frac{n_{h} -1}{n_{h} } \frac{\left(2\tau _{h} \right)}{n_{h} } +...=n_{h} \tau _{h},
\end{split}
\end{equation} 
where \textit{k} is the number of time intervals $\tau _{h}$ for that halo to merge with a single merger. All haloes in the same group are assumed to merge with a single merger with the same probability during the time interval of $\tau _{h} $. A second time scale $\tau _{g} \left(m_{h} ,a\right)$ reads
\begin{equation} 
\label{ZEqnNum458510} 
\tau _{g} \left(m_{h} ,a\right)=\bar{\tau }_{h} \left(m_{h} ,a\right)=n_{h} \tau _{h}=-{m_{g} /\varepsilon _{m} } ,         
\end{equation} 
which is the average waiting time (lifespan) for a given halo in a halo group of mass scale $m_{h} $, or equivalently, the time required to cascade the entire mass $m_{g} $ of that halo group. The time scale $\tau _{g} $ should decrease with $m_{h}$ due to faster mass accretion of larger haloes. 

 Let $M_{h} \left(a\right)$ be the total mass in the halo sub-system at physical time \textit{t }or scale factor \textit{a.} The third time scale $\tau _{M} \left(a\right)$ is introduced as the time required to cascade the entire mass in the halo sub-system,
\begin{equation} 
\label{ZEqnNum782676} 
\tau _{M} \left(a\right)=-{M_{h} \left(a\right)/\varepsilon _{m} } \left(a\right)\sim t,         
\end{equation} 
which is expected to be on the order of the current cosmic time \textit{t}. 

We are now ready to determine the characteristic mass scale $m_{h}^{*} $ for the mass cascade. Let $\tau _{g} \left(m_{h},a\right)$ be the average waiting time for a halo of mass $m_{h} $ to merge with a single merger of mass $m_{p} $ at physical time \textit{t}. The fourth time scale $\tau _{f} \left(m_{h} ,a\right)$ that we will introduce is  
\begin{equation} 
\label{ZEqnNum400186} 
\tau _{f} \left(m_{h} ,a\right)=\tau _{g} \left(m_{h} ,a\right)n_{p} ={\tau _{g} \left(m_{h} ,a\right)m_{h} /m_{p} } ,      
\end{equation} 
where $n_{p} $ is the number of particles in that given halo. The time scale $\tau _{f} $ approximately represents the time the entire halo of mass $m_{h}$ was formed via a sequence of merging events ($n_{p} $ times) with single mergers of mass $m_{p} $. Let's assume a typical halo of mass $m_{h}^{L} \left(t\right)$ that is constantly growing with the waiting time exactly to be $\tau _{g} $ for every single merging event during its entire mass accretion history. The actual waiting time of haloes can be random and either less or greater than $\tau _{g} $. The mass accretion of that typical halo should read
\begin{equation} 
\label{ZEqnNum943264} 
\frac{dm_{h}^{L} }{dt} =\frac{m_{p} }{\tau _{g}^{L} } =\frac{n_{p}^{L} m_{p} }{n_{p}^{L} \tau _{g}^{L} } =\frac{m_{h}^{L} }{n_{p}^{L} \tau _{g}^{L} } ,          
\end{equation} 
where $\tau _{g}^{L} \left(a\right)=\tau _{g} \left(m_{h}^{L} ,a\right)$. We further have (from Eq. \eqref{ZEqnNum943264}),
\begin{equation}
\label{ZEqnNum714805} 
\frac{d\ln m_{h}^{L} }{d\ln t} =\frac{t}{n_{p}^{L} \tau _{g}^{L} } =\frac{t}{\tau _{f} \left(m_{h}^{L} ,a\right)} ,         
\end{equation} 
where we should expect $\tau _{f} \left(m_{h}^{L} ,a\right)\sim t$ if haloes of mass $m_{h}^{L} \left(t\right)\approx m_h^*$ is the largest halo at present epoch. Here $n_{p}^{L} ={m_{h}^{L} /m_{p} } $ is the number of particles in that typical halo. It turns out that this is the case (Eq. \eqref{ZEqnNum877578}). We expect that large haloes require more time to form, and the time scale $\tau _{f} $ increases with the halo size $m_{h} $. Obviously, the four time scales we introduced satisfy the inequality
\begin{equation} 
\label{eq:11} 
\tau _{M} \left(a\right)\ge \tau _{f} \left(m_{h} ,a\right)\ge \tau _{g} \left(m_{h} ,a\right)\ge \tau _{h} \left(m_{h} ,a\right).       
\end{equation} 

 For small haloes with mass $m_{h} <m_{h}^{*} $, we expect the time scale $\tau _{f} \left(m_{h} ,a\right)\ll t$ to have sufficient time to form these haloes. For haloes with mass $m_{h} >m_{h}^{*} $, the time required to form that halo $\tau _{f} \left(m_{h},a\right)\gg t$ and these haloes are very rare to find at the current time \textit{t}. The time required to form haloes of a characteristic mass $m_{h}^{*} $ should be exactly on the order of the current physical time \textit{t}, i.e.$\tau _{f} \left(m_{h}^{*},a\right)\sim t\sim \tau _{M} \left(a\right)$, from which we can derive,
\begin{equation} 
\label{eq:12} 
-\frac{m_{h}^{*} }{\varepsilon _{m} } n_{h}^{*} n_{p}^{*} \sim -\frac{M_{h} \left(a\right)}{\varepsilon _{m} } \sim \frac{1}{H},
\end{equation} 
and
\begin{equation}
m_{h}^{*} \sim \frac{M_{h} \left(a\right)}{n_{h}^{*} n_{p}^{*} } \sim -\frac{\varepsilon _{m} \left(a\right)}{Hn_{h}^{*} n_{p}^{*} } \quad \textrm{or} \quad \frac{M_{h} \left(a\right)}{m_{g}^{*} } \sim \frac{m_{h}^{*} }{m_{p} } =n_{p}^{*}.    
\label{ZEqnNum811636}
\end{equation}

\noindent Here $n_{p}^{*} $, $n_{h}^{*} $, and $m_{g}^{*} $ are the number of particles in haloes of the characteristic mass $m_{h}^{*} $, number of haloes in halo group of mass $m_{h}^{*} $, and the total mass of that group, respectively. There is not enough time to form haloes larger than the characteristic mass $m_{h}^{*} $. This does not exclude the existence of these large haloes because of the random nature of waiting time. In \textit{N}-body simulations, the total number of particles in the system scales as $N\sim {M_{h} /m_{p} } \sim n_{h}^{*} n_{p}^{*2} $ from Eq. \eqref{ZEqnNum811636}. A dimensionless number $z_{h} $ can be defined for each halo group to reflect the competition between the local rate of mass transfer (${1/\tau _{f} } $) and the Hubble constant $H$,
\begin{equation}
z_{h} =\frac{M_{h} \left(a\right)}{m_{g} n_{p} } \sim -\frac{\varepsilon _{m} \left(a\right)}{m_{g} n_{p} H} \sim \frac{t}{\tau _{f} } \quad \textrm{and} \quad z_{h}^{*} =\frac{M_{h} \left(a\right)}{m_{g}^{*} n_{p}^{*} },    
\label{ZEqnNum595654}
\end{equation}

\noindent where $z_{h} $ decreases with halo size and $z_{h}^{*} $ for haloes with characteristic mass $m_{h}^{*} $ should be on the order of one. The exact value of $z_{h}^{*} $ can be determined with Eq. \eqref{ZEqnNum956499} ($z_{h}^{*} \approx {1/\beta _{0} } $).

In short, two distinct ranges can be identified for inverse mass cascade from time/mass scales: 1) mass propagation range with $m_{h} <m_{h}^{*}$, where the system is in a statistically steady state with a scale-independent mass flux $\varepsilon _{m} $ and a time-invariant group mass $m_{g} =m_{h} n_{h}$ (Fig. \ref{fig:S4}); 2) mass deposition range with $m_{h} >m_{h}^{*} $, where mass cascaded from small scales is actively consumed to grow haloes. The mass of the halo group $m_{g} $ increases with time in this range.

\subsection{Mass flux and mass transfer functions}
\label{sec:A1.3}
To quantify the mass cascade across halo groups, we introduce the real-space mass flux function that quantifies the net transfer of mass from all haloes smaller than the size $m_{h} $ to all haloes greater than $m_{h} $. The mass flux function $\Pi _{m} \left(m_{h} ,a\right)$  can be defined as
\begin{equation} 
\label{ZEqnNum400994} 
\Pi _{m} \left(m_{h} ,a\right)=-\frac{\partial }{\partial t} \left[M_{h} \left(a\right)\int _{m_{h} }^{\infty }f_{M} \left(m,m_{h}^{*} \right) dm\right].      
\end{equation} 
Here $M_{h} \left(a\right)$ is the total mass in the halo sub-system that increases with scale factor \textit{a}. The halo mass function $f_{M} \left(m_{h} ,a\right)$ is the probability distribution of the total mass $M_{h} \left(a\right)$ with respect to the halo mass $m_{h} $. 

Since the halo mass $m_{h} $ and the scale factor \textit{a} are the only two independent variables, the mass function can be written as a function of $m_{h} $ and \textit{a}, i.e., $f_{M} \left(m_{h},a\right)=f_{M} \left(m_{h},m_{h}^{*} \left(a\right)\right)$. The characteristic mass scale $m_{h}^{*} \left(a\right)$ varies with the scale factor \textit{a} only, a monotonically increasing function reflecting the fact that larger haloes emerge at a later time. The mass flux function $\Pi _{m} $ across the halo groups should be independent of halo size $m_{h} $ for halo groups smaller than $m_{h}^{*} \left(a\right)$, where the mass flux function reduces to
\begin{equation}
\varepsilon _{m} \left(a\right)=\Pi _{m} \left(m_{h} ,a\right) \quad \textrm{for} \quad m_{h} \ll m_{h}^{*}.      
\label{eq:16}
\end{equation}

\begin{figure}
\includegraphics*[width=\columnwidth]{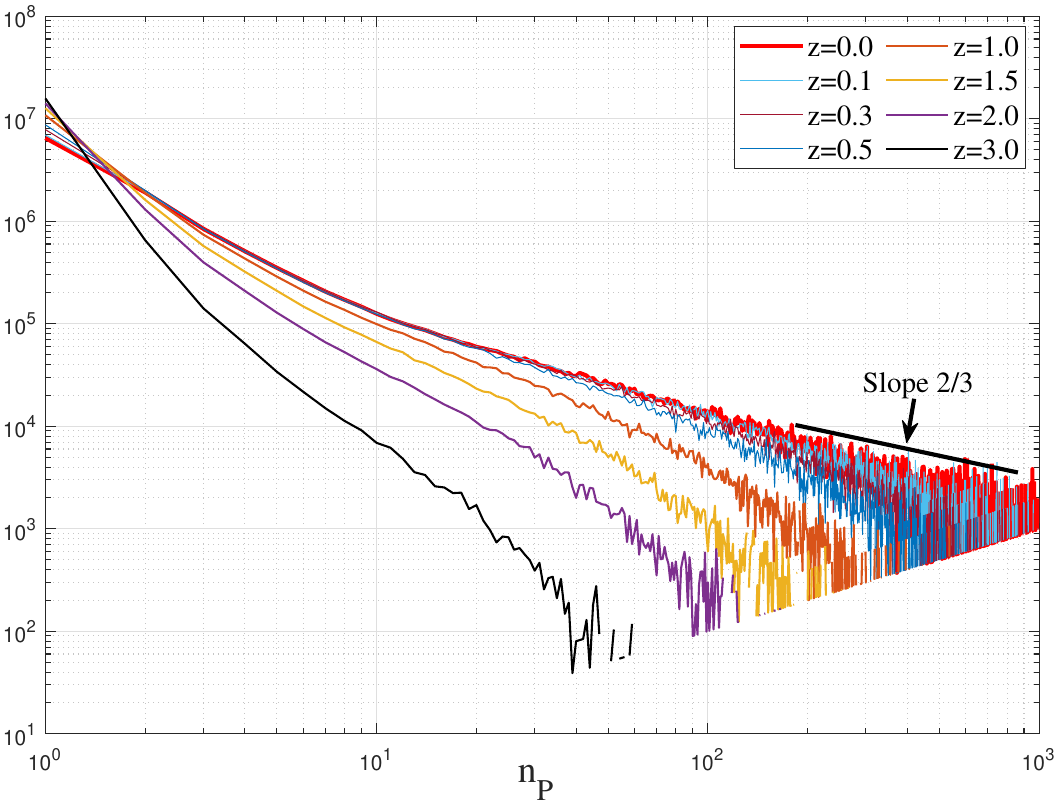}
\caption{The variation of halo group mass $m_g$ (normalized by particle mass $m_p$) with group size $n_p$ at different redshift $z$. At a statistically steady state, $m_g$ does not vary with time for halo groups with mass below $m_h*$, where mass propagation is dominant. Data for $m_g$ is from halo-based statistical dataset for dark matter flow \citep{Xu:2022-Dark_matter-flow-dataset-part1} and is also used to compute rate of mass cascade $\Pi_m$ in Fig. \ref{fig:3-3}.}
\label{fig:S4}
\end{figure}

The constant mass flux (or the mass dissipation rate $\varepsilon _{m} $ that is independent of mass scale $m_{h} $) cascades mass from the smallest mass scale to the characteristic scale ($0\ll m_{h} <m_{h}^{*} $) in the mass propagation range. The total mass of the halo group $m_{h} $ is
\begin{equation}
\label{ZEqnNum444911} 
m_{g} \left(m_{h} ,a\right)=M_{h} \left(a\right)f_{M} \left(m_{h} ,m_{h}^{*} \right)m_{p} .        
\end{equation} 
A direct result of the scale-independent mass flux is that the group mass $m_{g} \left(m_{h},a\right)$ of a halo group of size $m_{h} $ reaches a steady state (Not varying with time, see Eq. \eqref{ZEqnNum690015}). The total mass injected at the smallest mass scale (mass continuously injected from the out-of-halo sub-system into the halo sub-system) passes through the propagation range. It is consumed to grow the mass of halo groups above the characteristic mass $m_{h}^{*}$ (Fig. \ref{fig:S4}). 

The real-space mass transfer function can be defined as the derivative of the mass flux function with respect to the halo mass; 
\begin{equation} 
\label{ZEqnNum980409} 
\begin{split}
T_{m} \left(m_{h} ,a\right)&=\frac{\partial \Pi _{m} \left(m_{h} ,a\right)}{\partial m_{h} }\\
&=\frac{\partial \left[M_{h} \left(a\right)f_{M} \left(m_{h} ,m_{h}^{*} \right)\right]}{\partial t} =\frac{\partial m_{g} \left(m_{h} ,a\right)}{m_{p} \partial t},
\end{split}
\end{equation} 
which quantifies the rate of change in the mass of the group $m_{g} \left(m_{h},a\right)$. For the mass propagation range, 
\begin{equation}
T_{m} \left(m_{h} ,a\right)=0 \quad \textrm{and} \quad\frac{\partial m_{g} \left(m_{h} ,a\right)}{\partial t} =0 \quad \textrm{for} \quad m_{h} <m_{h}^{*}.     
\label{ZEqnNum690015}
\end{equation}

\noindent The mass transfer function $T_{m} \left(m_{h},a\right)$ describes the removal of mass on a small scale and the deposition of mass on a large scale ($T_{m} \left(m_{h},a\right)>0$ for $m_{h} >m_{h}^{*} $). 

Since the mass dissipation rate $\varepsilon _{m} \left(a\right)$ is independent of halo size for $m_{h} <m_{h}^{*} $, we may compute the mass flux function at the smallest scales using Eq. \eqref{ZEqnNum400994} with $m_{h} =0$, 

\begin{equation}
\varepsilon _{m} \left(a\right)=\Pi _{m} \left(m_{h} =0,a\right)=-\frac{\partial M_{h} \left(a\right)}{\partial t} \quad \textrm{for} \quad m_{h} <m_{h}^{*}.   
\label{ZEqnNum512327}
\end{equation}

\noindent Let the time scale $\tau _{h} \left(m_{h},a\right)$ be the average time for a single merging event in a halo group of mass $m_{h} $, or equivalently an event frequency $f_{h} \left(m_{h},a\right)$. The rate of mass transfer from the scale below $m_{h} $ to the scale above $m_{h} $ is 
\begin{equation}
\varepsilon _{m} \left(a\right)=-\alpha _{0} m_{h} f_{h} \left(m_{h} ,a\right) \quad \textrm{for} \quad m_{h} \ll m_{h}^{*},     
\label{ZEqnNum379703}
\end{equation}

\noindent where $\alpha _{0} $ is a numerical factor on the order of unity. The frequency of the event $f_{h} \left(m_{h},a\right)$ should be proportional to the number of haloes in the group (term 1 in Eq. \eqref{ZEqnNum133775}) and the surface area of the haloes (term 2). Because the halo interactions in mass space are local, we can assume that the mass cascade involves merging events between a halo of similar size and a single merger (Figs. \ref{fig:1} and \ref{fig:2}). The halo group with more haloes (term 1 in Eq. \eqref{ZEqnNum133775}) and haloes with a larger surface area (proportional to $m_{h}^{\lambda } $, i.e., term 2 in Eq. \eqref{ZEqnNum133775}) have a greater probability of merging with a single merger, 
\begin{equation} 
\label{ZEqnNum133775} 
f_{h} \left(m_{h} ,a\right)=f_{0} \left(a\right)\underbrace{M_{h} \left(a\right)f_{M} \left(m_{h} ,m_{h}^{*} \right)\frac{m_{p} }{m_{h} } }_{1}\underbrace{\left(\frac{m_{h} }{m_{p} } \right)^{\lambda } }_{2},      
\end{equation} 
where $f_{0} \left(a\right)$ is a fundamental frequency for the merging between two single mergers at a given redshift \textit{z} or scale factor \textit{a} and may be used to determine the mass of dark matter particles $m_p$ (Eq. \eqref{ZEqnNum476562}). The characteristic time of a single merging event can be written as $\tau _{h} \left(m_{h},a\right)={1/\left(\alpha _{0} f_{h} \right)} $.

Without loss of generality, the exponent $\lambda $ is a halo geometry parameter that represents the effect of the halo surface area on the merging frequency $f_{h} \left(m_{h},a\right)$. For two haloes of very different sizes (merging between a large halo and a single merger), it is estimated that $\lambda ={2/3} $ with $m_{h} \propto r_{h}^{3} \propto A_{h}^{{3/2} } $, where $A_{h} $ is the surface area of the halo. For small haloes, merging is more likely between two haloes of comparable sizes, where $\lambda $ can deviate from 2/3 and approach 1. 

Substitution of Eq. \eqref{ZEqnNum133775} for the event frequency into Eq. \eqref{ZEqnNum379703} leads to the mass flux,
\begin{equation}
\varepsilon _{m} \left(a\right)=-\alpha _{0} f_{0} \left(a\right)M_{h} \left(a\right)f_{M} \left(m_{h} ,m_{h}^{*} \left(a\right)\right)m_{p} \left(\frac{m_{h} }{m_{p} } \right)^{\lambda }. 
\label{ZEqnNum590150}
\end{equation}

\noindent For self-similar gravitational clustering in the mass propagation range ($m_{h} <m_{h}^{*} $), the halo mass $m_{h} $ and the characteristic mass scale $m_{h}^{*} $ are the only two controlling variables. We can simply express the mass function as $f_{M} \left(m_{h},m_{h}^{*} \right)\sim \left(m_{h} \right)^{x} \left(m_{h}^{*} \right)^{-x-1} $. Now using dimensional analysis, the only possible form of the mass function $f_{M} $ that satisfies Eq. \eqref{ZEqnNum590150} is ($f_{M} $ should have a unit of 1/kg in SI units and $\varepsilon _{m} $ is a function of \textit{a} alone and is independent of $m_{h} $),
\begin{equation}
f_{M} \left(m_{h} ,m_{h}^{*} \right)=\beta _{0} m_{h}^{-\lambda } \left(m_{h}^{*} \right)^{\lambda -1} \quad \textrm{for} \quad m_{h} <m_{h}^{*},    
\label{ZEqnNum972525}
\end{equation}
\noindent where $\beta _{0} \sim O\left(1\right)$ is a numerical constant. The mass flux and event frequency in the mass propagation range can be expressed as (after substituting Eq. \eqref{ZEqnNum972525} into \eqref{ZEqnNum590150}),
\begin{equation}
\varepsilon _{m} \left(a\right)=-\alpha _{0} \beta _{0} \lambda _{0} Nm_{p} f_{0} \left(a\right) \quad \textrm{for} \quad m_{h} <m_{h}^{*}, 
\label{ZEqnNum696670}
\end{equation}
and
\begin{equation}
f_{h} \left(m_{h} ,a\right)=\beta _{0} \lambda _{0} f_{0} \left(a\right){Nm_{p} /m_{h} } \quad \textrm{for} \quad m_{h} <m_{h}^{*}, 
\label{eq:4-4-9}
\end{equation}

\noindent where a dimensionless constant $\lambda _{0} $ is defined as
\begin{equation} 
\label{ZEqnNum480255} 
\lambda _{0} =\frac{M_{h} \left(a\right)}{Nm_{p}^{} } \left(\frac{m_{h}^{*} }{m_{p}^{} } \right)^{\lambda -1} ,          
\end{equation} 
which is a constant invariant in time and dependent only on the mass resolution $m_{p}^{} $. For $m_{p} =2.27\times 10^{11} {M_{\odot } /h} $ from Table \ref{tab:1} and $m_{h}^{*} \left(z=0\right)\approx 2\times 10^{13} {M_{\odot } /h} $, $\lambda _{0} $ is on the order of 0.13. 

The halo group mass in the propagation range (Eq. \eqref{ZEqnNum972525}) reads
\begin{equation}
m_{g} \left(m_{h} ,a\right)\equiv m_{g} \left(m_{h} \right)=\lambda _{0} \beta _{0} Nm_{p} \left({m_{p} /m_{h} } \right)^{\lambda }.  
\label{ZEqnNum744409}
\end{equation}

\noindent Equivalently, we have
\begin{equation}
m_{g} \left(m_{p} \right)=\lambda _{0} \beta _{0} Nm_{p} \textrm{,} \quad m_{g} \left(m_{h} \right)=m_{g} \left(m_{p} \right)\left({m_{p} /m_{h} } \right)^{\lambda }    \label{ZEqnNum488540}
\end{equation}

\noindent  for the group mass of single mergers with $m_{h} =m_{p} $. The group mass $m_{g} $ is proportional to $m_{h}^{-\lambda } $ (Fig. \ref{fig:S4}). The relation between the rate of change of mass $M_{h} $ and $m_{h}^{*} $ can be found from Eq. \eqref{ZEqnNum480255},  
\begin{equation} 
\label{eq:30} 
\frac{\partial \ln M_{h} }{\partial \ln a} =\left(1-\lambda \right)\frac{\partial \ln m_{h}^{*} }{\partial \ln a} .         
\end{equation} 
With Eqs. \eqref{ZEqnNum512327} and \eqref{ZEqnNum480255}, we may derive the mass dissipation rate as a function of $m_{h}^{*} $,
\begin{equation} 
\label{ZEqnNum794927} 
\begin{split}
\varepsilon _{m} \left(a\right)&=-\left(1-\lambda \right)\frac{\partial \ln m_{h}^{*} }{\partial \ln a} H\left(a\right)M_{h} \left(a\right)\\
&=-\lambda _{0} \left(1-\lambda \right)\frac{\partial \ln m_{h}^{*} }{\partial \ln a} H\left(a\right)Nm_{p} \left(\frac{m_{h}^{*} }{m_{p} } \right)^{1-\lambda }.
\end{split}
\end{equation} 
With Eqs. \eqref{ZEqnNum696670} and \eqref{ZEqnNum794927}, we find the fundamental frequency $f_{0} \left(a\right)$ as a function of $m_{h}^{*} $,
\begin{equation} 
\label{ZEqnNum476562} 
f_{0} \left(a\right)=\frac{\left(1-\lambda \right)}{\alpha _{0} \beta _{0} } \frac{\partial \ln m_{h}^{*} }{\partial \ln a} H\left(a\right)\left(\frac{m_{h}^{*} }{m_{p} } \right)^{1-\lambda },  
\end{equation} 
that is also related to the Hubble constant $H\left(a\right)$ and the mass resolution $m_{p}^{} $. The characteristic time scale $\tau _{h}^{*} =\tau _{h}^{} \left(m_{h}^{*} ,a\right)$ associated with the characteristic mass $m_{h}^{*} $ is   
\begin{equation} 
\label{ZEqnNum378741} 
\tau _{h}^{*} \left(a\right)=-\frac{m_{h}^{*} }{\varepsilon _{m} } =\frac{1/(\lambda _{0} \left(1-\lambda \right))}{N\frac{\partial \ln m_{h}^{*} }{\partial \ln a} H} \left(\frac{m_{h}^{*} }{m_{p} } \right)^{\lambda } .    
\end{equation} 

The fundamental frequency $f_{0} \left(a\right)$ is the frequency for the elementary merging of two single mergers and is expected to decrease with time. Without loss of generality, let us assume a power-law of $f_{0} \left(a\right)\propto a^{-\tau _{0} } $ that leads to $\varepsilon _{m} \left(a\right)\propto a^{-\tau _{0} } $ (Eq. \eqref{ZEqnNum696670}), $\left(m_{h}^{*} \right)^{1-\lambda } \propto a^{{3/2} -\tau _{0} } $ (Eq. \eqref{ZEqnNum794927}), and $M_{h} \left(a\right)\propto a^{{3/2} -\tau _{0} } $ (Eq. \eqref{ZEqnNum480255}). Once the statistically steady state is established for inverse mass cascade, the total mass of the halo sub-system increases as $M_{h} \left(a\right)\propto a^{{3/2} -\tau _{0} } $ regardless of the value of $\lambda $. Obviously,  
\begin{equation} 
\label{ZEqnNum200370} 
\frac{\partial \ln M_{h} }{\partial \ln a} =\left(1-\lambda \right)\frac{\partial \ln m_{h}^{*} }{\partial \ln a} =\frac{3}{2} -\tau _{0} >0.       
\end{equation} 
With the total mass in the halo sub-system $M_{h} \left(a\right)$ increasing with the scale factor \textit{a}, it requires $0<\tau _{0} <{3/2} $. With $\tau _{0} >0$ and $a\to \infty $, the mass flux $\varepsilon _{m} \left(a\right)$ approaches zero when the whole system approaches the limiting thermodynamic equilibrium but cannot reach. In this regard, the inverse mass cascade with a constant rate is a key feature of the statistically (intermediate) steady state as the system evolves toward the limiting equilibrium.

The number of haloes $n_{h} \left(m_{h} \right)$ in the halo group with a given mass $m_{h} $ is
\begin{equation}
n_{h} \left(m_{h} \right)\equiv {m_{g} \left(m_{h} \right)/m_{h} } =\lambda _{0} \beta _{0} Nm_{h}^{-\lambda -1} m_{p} {}^{\lambda +1}, 
\label{ZEqnNum955316}
\end{equation}

\noindent, which does not vary with time once the statistically steady state is established. Substituting Eq. \eqref{ZEqnNum200370} into Eq. \eqref{ZEqnNum476562}, we can express the fundamental frequency $f_{0} \left(a\right)$ as  
\begin{equation} 
\label{ZEqnNum846416} 
f_{0} \left(a\right)=\frac{1}{\alpha _{0} \beta _{0} } \left(\frac{3}{2} -\tau _{0} \right)H\left(a\right)\left(\frac{m_{h}^{*} }{m_{p} } \right)^{1-\lambda } =\frac{b_{0} }{\alpha _{0} \beta _{0} } H_{0} a^{-\tau _{0} } ,     
\end{equation} 
where the mass resolution parameter $b_{0} $ can be related to a fixed characteristic mass $m_{h}^{*} \left(a\right)$ at \textit{a}=1,
\begin{equation} 
\label{eq:37} 
b_{0} =\left(\frac{3}{2} -\tau _{0} \right)\left(\frac{m_{h}^{*} \left(a=1\right)}{m_{p} } \right)^{1-\lambda } ,        
\end{equation} 
which is dependent on the mass resolution $m_{p} $ only (the mass of dark matter particle), with a smaller particle mass $m_{p} $ giving rise to a higher fundamental frequency $f_{0} \left(a\right)$ in \textit{the} N-body simulation. For the mass resolution of $m_{p} =2.27\times 10^{11} {M_{\odot } /h} $ from Table \ref{tab:1} and $m_{h}^{*} \left(a=1\right)=2\times 10^{13} {M_{\odot } /h} $, $b_{0} \approx 2.2$ with $\tau _{0} =1$. In other words, the mass of dark matter particles can be determined if the fundamental frequency $f_{0} \left(a\right)$ can be precisely measured.

Now, we can introduce a numerical constant, 
\begin{equation} 
\label{eq:38} 
c_{0} =b_{0} \lambda _{0} =\left(\frac{3}{2} -\tau _{0} \right)\frac{M_{h} \left(a\right)}{Nm_{p} } a^{\tau _{0} -{3/2} } ,
\end{equation} 
which is the mass fraction of halo sub-system $M_{h} \left(a\right)$ and is not dependent on the mass resolution $m_{p} $ and the scale factor \textit{a}. We estimate $c_{0} \approx 0.29$ with $\tau _{0} =1$, i.e. $M_{h} \approx 0.58Nm_{p} $ when \textit{a}=1. This information is used to study the density distributions for particles in haloes and out-of-haloes, respectively \citep{Xu:2022-Two-thirds-law-for-pairwise-ve}. 

We present simplified expressions for time and mass scales and mass flux function that can be fully described as functions of halo mass $m_{h} $, scale factor \textit{a}, and mass resolution $m_{p}$ with four numerical constants $\beta _{0} $ (pre-factor for the halo mass function $f_{M} $), $c_{0} $ (the mass fraction of the halo sub-system), $\lambda $ (the halo geometry parameter) and $\tau _{0} $ (the exponent of the fundamental frequency $f_{0} $), and the characteristic mass $m_{h}^{*} \left(a=1\right)$ and $H_{0} $, with four
\begin{equation} 
\label{ZEqnNum233035} 
\varepsilon _{m} \left(a\right)=-\left(\frac{3}{2} -\tau _{0} \right)H\left(a\right)M_{h} \left(a\right)=-c_{0} H_{0} Nm_{p} a^{-\tau _{0} } ,      
\end{equation} 
\begin{equation} 
\label{eq:40} 
\tau _{h}^{*} \left(a\right)=\frac{1/(N\lambda _{0} H\left(a\right))}{\left({3}/{2} -\tau _{0} \right)} \left(\frac{m_{h}^{*} }{m_{p} } \right)^{\lambda } =\frac{m_{h}^{*} \left(a=1\right)}{c_{0} H_{0} Nm_{p} } a^{\frac{\left({3/2} -\lambda \tau _{0} \right)}{\left(1-\lambda \right)} } ,     
\end{equation} 
\begin{equation} 
\label{ZEqnNum763048} 
m_{h}^{*} \left(a\right)=\left(\frac{b_{0} }{{3/2} -\tau _{0} } \right)^{\frac{1}{1-\lambda } } a^{\frac{\left({3/2} -\tau _{0} \right)}{\left(1-\lambda \right)} } m_{p} =m_{h}^{*} \left(a=1\right)a^{\frac{\left({3/2} -\tau _{0} \right)}{\left(1-\lambda \right)} } ,     
\end{equation} 
\begin{equation} 
\label{ZEqnNum815746} 
M_{h} \left(a\right)=\frac{c_{0} }{{3/2} -\tau _{0} } Nm_{p} a^{\left({3/2} -\tau _{0} \right)} .        
\end{equation} 

 Recall the time scales we introduced in Section \ref{sec:A1.2}. The time scale $\tau _{M} \left(a\right)$ is 
\begin{equation} 
\label{ZEqnNum797746} 
\tau _{M} \left(a\right)=-\frac{M_{h} \left(a\right)}{\varepsilon _{m} \left(a\right)} =\frac{{3/2} }{{3/2} -\tau _{0} } a^{{3/2} } t_{0} =\frac{{3/2} }{{3/2} -\tau _{0} } t,      
\end{equation} 
which is the time it takes to cascade all mass in the halo sub-system at a given scale factor \textit{a} and $\tau _{M} \left(a\right)$ is on the order of $t$. This relation might be used to determine the value of $\tau _{0} $ from \textit{N}-body simulations. 

The time scale $\tau _{h} $ for a single merging in a halo group of mass $m_{h}$ is,
\begin{equation} 
\label{eq:44} 
\tau _{h} \left(m_{h} ,a\right)=\frac{m_{h} }{\varepsilon _{m} } =\frac{m_{h}/(\alpha _{0} \beta _{0} \lambda _{0} ) }{f_{0} \left(a\right)Nm_{p} } =\frac{m_{h} }{c_{0} H_{0} Nm_{p} } a^{\tau _{0} } .    
\end{equation} 

The time scale $\tau _{g} $ that takes to cascade the group mass $m_{g} $ for the halo group of mass $m_{h} $ is
\begin{equation} 
\label{ZEqnNum673645}
\begin{split}
\tau _{g} \left(m_{h} ,a\right)&=\frac{m_{h} n_{h} }{\varepsilon _{m} } =\frac{1}{\alpha _{0} f_{0} \left(a\right)} \left(\frac{m_{h} }{m_{p} } \right)^{-\lambda }\\
&=\frac{\beta _{0} }{\left({3/2} -\tau _{0} \right)H_{0} } \frac{m_{h}^{-\lambda } m_{p} }{\left[m_{h}^{*} \left(a=1\right)\right]^{1-\lambda } } a^{\tau _{0} }. 
\end{split}
\end{equation} 
As expected, the mean waiting time (lifespan) $\tau _{g} $ of a given halo decreases with halo mass as $\tau _{g} \propto m_{h}^{-\lambda } $ and increases with \textit{a}. Larger haloes have a shorter life. Extremely large haloes have very fast mass accretion and an infinitesimal lifespan. 

The time scale $\tau _{f} $ is introduced as the average time it takes to form the halo of mass $m_{h} $ that reads 
\begin{equation} 
\label{ZEqnNum630905} 
\begin{split}
\tau _{f} \left(m_{h} ,a\right)&=\frac{m_{h} n_{h} n_{p} }{\varepsilon _{m} }=\frac{1}{\alpha _{0} f_{0} \left(a\right)} \left(\frac{m_{h} }{m_{p} } \right)^{1-\lambda } \\
&=\frac{\beta _{0} }{\left({3/2} -\tau _{0} \right)H_{0} } \left(\frac{m_{h} }{m_{h}^{*} \left(a=1\right)} \right)^{1-\lambda } a^{\tau _{0} },
\end{split}
\end{equation} 
which increases with halo mass as $\tau_f\propto m_{h}^{1-\lambda }a^{\tau_0}$ such that Smaller haloes were formed earlier with smaller $\tau_f$. The relation between time scales $\tau _{f} \left(m_{h}^{*} ,a\right)=\beta _{0} \tau _{M} \left(a\right)$ can be easily obtained from Eqs. \eqref{ZEqnNum797746} and \eqref{ZEqnNum630905} that is consistent with our analysis in Section \ref{sec:A1.2} (Eq. \eqref{ZEqnNum811636}), i.e. $\beta _{0} M_{h} \left(a\right)=m_{g}^{*} n_{p}^{*} =n_{h}^{*} n_{p}^{*2} m_{p} $ and $z_{h}^{*} \beta _{0} =1$ (from Eq. \eqref{ZEqnNum595654}). Different time scales can be related to the mass flux as
\begin{equation} 
\label{ZEqnNum161813} 
\varepsilon _{m} =\frac{-M_{h} \left(a\right)}{\tau _{M} \left(a\right)} =\frac{-m_{g} \left(m_{h} \right)}{\tau _{g} \left(m_{h} ,a\right)} =\frac{-m_{g} n_{p} }{\tau _{f} \left(m_{h} ,a\right)} =\frac{-m_{h} }{\tau _{h} \left(m_{h} ,a\right)} .     
\end{equation} 

The corresponding time scales at the characteristic mass $m_{h}^{*} $ are
\begin{equation} 
\label{eq:48} 
\tau _{g}^{*} \left(a\right)=\frac{\beta _{0} m_{p} }{\left({3/2-\tau _{0} } \right)H_{0} m_{h}^{*} \left(a=1\right)} a^{\frac{\tau _{0} -{3\lambda /2} }{1-\lambda } } ,       
\end{equation} 
\begin{equation} 
\label{ZEqnNum929904} 
\tau _{f}^{*} \left(a\right)=\frac{\beta _{0} }{\left({3/2-\tau _{0} } \right)H} .          
\end{equation} 
The two time scales $\tau _{g} \propto a^{\tau _{0} } m_{h}^{-\lambda } $ (Eq. \eqref{ZEqnNum673645}) and $\tau _{f} \propto a^{\tau _{0} } m_{h}^{1-\lambda } $ (Eq. \eqref{ZEqnNum630905}), where larger haloes have a shorter lifespan but take more time to form. For $\lambda ={2/3} $ and $\tau _{0} =1$, the lifespan $\tau _{g}^{*} $ of characteristic haloes is independent of time which is consistent with Eq. \eqref{eq:3-9-2}.

\begin{figure}
\includegraphics*[width=\columnwidth]{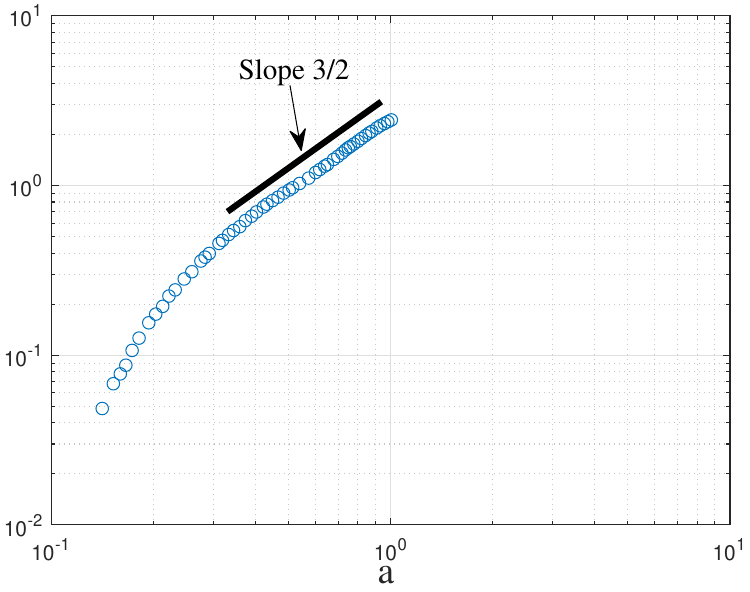}
\caption{The halo mass (normalized by $10^{12}M_{\odot }$) accretion history for type II haloes, i.e. the dominant type for large haloes \citep[see][Fig. 2]{McBride:2009-Mass-accretion-rates-and-histo}, exhibits a power law scaling $\propto a^{{3/2} }$.}
\label{fig:4}
\end{figure}

Now we can track the growth of typical haloes by integrating Eq. \eqref{ZEqnNum943264} with respect to the scale factor \textit{a} and using expression of time scale $\tau _{g} $ (Eq. \eqref{ZEqnNum673645}) with initial condition $m_{h}^{L} \left(a=0\right)=0$ to obtain
\begin{equation}
\begin{split}
&\frac{m_{h}^{L} \left(a\right)}{m_{h}^{*} \left(a=1\right)} =\left(\frac{1-\lambda }{\beta _{0} } \right)^{{1/\left(1-\lambda \right)} } a^{\frac{{3/2} -\tau _{0} }{1-\lambda } }\\ &\textrm{and}\\
&\frac{m_{h}^{L} \left(a\right)}{m_{h}^{*} \left(a\right)} =\left(\frac{1-\lambda }{\beta _{0} } \right)^{{1/\left(1-\lambda \right)} },
\end{split}
\label{ZEqnNum184238}
\end{equation}

\noindent which follows the same scaling as characteristic mass scale $m_{h}^{*} \left(a\right)$. For large haloes with $\tau _{0} =1$ and $\lambda ={2/3} $, $m_{h}^{L} \left(a\right)\sim m_{h}^{*} \left(a\right)\sim a^{{3/2} } $. This scaling matches the mass growth of type II haloes, i.e., the dominant type for large haloes \citep[see][Fig. 2]{McBride:2009-Mass-accretion-rates-and-histo}, as shown in Fig. \ref{fig:4}. 

With the help of Eq. \eqref{ZEqnNum714805}, we confirm that the time scale $\tau _{f} $ (Eq. \eqref{ZEqnNum630905}) to form the typical halo is on the same order of \textit{t} (Eq. \eqref{ZEqnNum714805}),
\begin{equation} 
\label{ZEqnNum877578} 
\tau _{f} \left(m_{h}^{L} ,a\right)=\frac{1-\lambda }{1-{2\tau _{0} /3} } t\sim t.         
\end{equation} 
Finally, it can be easily confirmed that (from Eq. \eqref{ZEqnNum794927}),
\begin{equation} 
\label{ZEqnNum907189} 
\varepsilon _{m} \left(a\right)=-\frac{dM_{h} \left(a\right)}{dt} =-\left(\frac{3}{2} -\tau _{0} \right)HM_{h} \left(a\right). \end{equation} 
The relations between total mass of all haloes $M_{h} $, typical halo mass $m_{h}^{L} $, and mass scale $m_{h}^{*} $ are (from Eqs. \eqref{ZEqnNum480255}, \eqref{ZEqnNum955316} and \eqref{ZEqnNum184238}),
\begin{equation} 
\label{ZEqnNum956499} 
M_{h} \left(a\right)=\frac{1}{1-\lambda } m_{h}^{L} n_{h}^{L} n_{p}^{L} =\frac{1}{\beta _{0} } m_{h}^{*} n_{h}^{*} n_{p}^{*}. 
\end{equation} 

The mass flux function $\varepsilon _{m} \left(a\right)$ can be interpreted as the rate of mass change of a typical halo $m_{h}^{L} $ or $m_{h}^{*} $ multiplied by the equivalent number of that halo in the system, 
\begin{equation} 
\label{ZEqnNum781767} 
\begin{split}
\varepsilon _{m} \left(a\right)=-\frac{dm_{h}^{L} }{dt} n_{h}^{L} n_{p}^{L} &=-\frac{1}{\left(1-\lambda \right)}\frac{d\left(m_{h}^{L} n_{h}^{L} n_{p}^{L} \right)}{dt} \\
&=-\lambda _{0} \beta _{0} N\frac{dm_{h}^{L} }{dt} \left(\frac{m_{h}^{L} }{m_{p} } \right)^{-\lambda }
\end{split}
\end{equation} 
\noindent{or}
\begin{equation} 
\label{eq:55} 
\begin{split}
\varepsilon _{m} \left(a\right)&=-\frac{1-\lambda }{\beta _{0} } \frac{dm_{h}^{*} }{dt} n_{h}^{*} n_{p}^{*}=-\frac{1}{\beta_0}\frac{d\left(m_{h}^{*} n_{h}^{*} n_{p}^{*} \right)}{dt}\\
&=-\lambda _{0} \left(1-\lambda \right)N\frac{dm_{h}^{*} }{dt} \left(\frac{m_{h}^{*} }{m_{p} } \right)^{-\lambda }.
\end{split}
\end{equation} 

In summary, the mathematical model for inverse mass cascade provides the complete dependence of time/mass scales and mass flux/transfer functions on the scale factor \textit{a}, halo mass $m_{h} $, and mass resolution $m_{p} $. An interesting case is that $\tau _{0} =1$, where the fundamental frequency $f_{0} \left(a\right)\propto a^{-1} $. Note that this is the same scaling as the photon frequency decaying due to the cosmological redshift. Table \ref{tab:2} lists the scaling exponents with respect to \textit{a} for different values of $\tau _{0} $ and $\lambda $. The scaling of $\varepsilon _{m} \left(a\right)$, $f_{0} \left(a\right)$ and $M_{h} \left(a\right)$ are only dependent on $\tau _{0} $, while $m_{h}^{*} \left(a\right)$ and $\tau _{h}^{*} \left(a\right)$ depend on both $\tau _{0} $ and $\lambda $. Table \ref{tab:3} summarizes the dependence on the halo size $m_{h} $, where $\tau _{g} \sim m_{h}^{-\lambda } $ and $\tau _{f} \sim m_{h}^{1-\lambda } $. Table \ref{tab:4} presents the dependence of relevant parameters on the mass resolution $m_{p} $. 

\begin{table*}
\centering
\begin{tabular}{p{0.3in}p{0.32in}p{0.3in}p{0.3in}p{0.4in}p{0.4in}p{0.6in}p{0.6in}p{0.6in}p{0.6in}p{0.6in}} 
\hline 
$\lambda $ & $\tau _{0} $ & $f_{0} $ & $\varepsilon _{m} $  & $M_{h} $ & $f_{M} $ & $m_{h}^{*} $ & $\tau _{h}^{*} $  & $n_{h}^{*} $ & $m_{g}^{*} $ & $\tau _{g}^{*} $ \\ 
\hline 
$\lambda $ & $\tau _{0} $ & $a^{-\tau _{0} } $ & $a^{-\tau _{0} } $ & $a^{{3}/{2}-\tau _{0}}$  & $a^{\tau _{0}-{3}/{2}} $ & $a^{\frac{\left({3/2} -\tau _{0} \right)}{\left(1-\lambda \right)} } $ & $a^{\frac{\left({3/2} -\lambda \tau _{0} \right)}{\left(1-\lambda \right)} } $ & $a^{-\left(\frac{3}{2} -\tau _{0} \right)\frac{\left(1+\lambda \right)}{\left(1-\lambda \right)} } $ & $a^{-\left(\frac{3}{2} -\tau _{0} \right)\frac{\lambda }{\left(1-\lambda \right)} } $ & $a^{\frac{\left(\tau _{0} -{3\lambda /2} \right)}{\left(1-\lambda \right)} } $ \\ 
\hline 
${2}/{3} $\textbf{} & $1$\textbf{} & $a^{-1} $\textbf{ } & $a^{-1} $\textbf{} & $a^{{1/2} } $\textbf{} & $a^{{-1/2} } $\textbf{} & $a^{{3/2} } $\textbf{} & $a^{{5/2} } $\textbf{} & $a^{{-5/2} } $\textbf{} & $a^{-1} $\textbf{} & $a^{0} $\textbf{} \\ 
\hline 
${2}/{3} $\textbf{} & ${1}/{2} $\textbf{} & $a^{-{1}/{2} } $\textbf{} & $a^{-{1/2} } $\textbf{} & $a^{1} $\textbf{} & $a^{-1} $\textbf{} & $a^{3} $\textbf{} & $a^{{7/2} } $\textbf{} & $a^{-5} $\textbf{} & $a^{-2} $\textbf{} & $a^{{-3/2} } $\textbf{} \\ 
\hline 
${3}/{4} $ & $1$ & $a^{-1} $ & $a^{-1} $ & $a^{{1/2} } $ & $a^{{-1/2} } $ & $a^{2} $ & $a^{3} $ & $a^{-{7/2} } $ & $a^{{-3/2} } $ & $a^{{-1/2} } $ \\ \hline 
\end{tabular}
\caption{Dependence on the scale factor \textit{a} for different values of $\tau _{0} $ and $\lambda $}
\label{tab:2}
\end{table*}

\begin{table}
\centering
\begin{tabular}{p{0.25in}p{0.25in}p{0.25in}p{0.25in}p{0.25in}p{0.25in}p{0.2in}p{0.3in}} 
\hline 
$f_{M} $ & $\varepsilon _{m} $ & $f_{h} $ & $\tau _{h} $ & $\tau _{g} $\textbf{} & $\tau _{f} $ & $m_{g} $  & $n_{h} $ \\ 
\hline 
$m_{h}^{-\lambda } $ & $m_{h}^{0} $ & $m_{h}^{-1} $ & $m_{h}^{1} $ & $m_{h}^{-\lambda } $ & $m_{h}^{1-\lambda } $ & $m_{h}^{-\lambda } $ & $m_{h}^{-1-\lambda } $  \\ 
\hline 
\end{tabular}
\caption{Dependence on the halo mass $m_{h}$}
\label{tab:3}
\end{table}

\begin{table}
\centering
\begin{tabular}{p{0.17in}p{0.17in}p{0.17in}p{0.17in}p{0.17in}p{0.17in}p{0.17in}p{0.17in}p{0.17in}p{0.17in}} 
\hline 
$f_{M} $ & $f_{h} $ & $\tau _{f} $ & $\varepsilon _{m} $\textbf{} & $\tau _{h}^{*} $ & $\tau _{h} $ & $m_{h}^{*} $ & $M_{h} $ & $c_{0} $ & $\beta _{0} $ \\ \hline 
$m_{p}^{0} $ & $m_{p}^{0} $ & $m_{p}^{0} $ & $m_{p}^{0} $ & $m_{p}^{0} $ & $m_{p}^{0} $ & $m_{p}^{0} $ & $m_{p}^{0} $ & $m_{p}^{0} $ & $m_{p}^{0} $ \\ \hline 
 &  &  &  &  &  &  &  &  &  \\ \hline 
$n_{h} $ & $\tau _{g} $ & $m_{g} $  & $n_{p} $ & $f_{0} $ & $b_{0} $ & $\lambda _{0} $ & $n_{h}^{*} $ & $m_{g}^{*} $ & $n_{p}^{*} $ \\ \hline 
$m_{p}^{} $  & $m_{p}^{} $ & $m_{p}^{} $ & $m_{p}^{-1} $  & $m_{p}^{\lambda -1} $ & $m_{p}^{\lambda -1} $ & $m_{p}^{1-\lambda } $ & $m_{p}^{} $ & $m_{p}^{} $ & $m_{p}^{-1} $ \\ \hline 
\end{tabular}
\caption{Dependence on the mass resolution $m_{p}$}
\label{tab:4}
\end{table}

\subsection{The probability distribution of waiting time}
\label{sec:4.2}

This section discusses the probability distribution of the waiting time for a given halo to merge with a single merger. The mean waiting time for a merging event in a halo group of mass $m_{h} $ is $\tau_{h}=\tau_g/n_{h} $ from Eq. \eqref{ZEqnNum458510}, where $\tau_g$ is the mean waiting time for the merging of a given halo. Let the actual time interval of a single merging for a given halo be a random variable $\tau _{g}^r $ with its mean given by $\tau _{g} =\left\langle \tau _{gr} \right\rangle =n_{h} \tau _{h} \gg \tau _{h} $ (from Eq. \eqref{ZEqnNum342616}). Typical haloes accreting mass at a fixed waiting time have a direct delta distribution with a deterministic $\tau _{g}^r \equiv \tau_{g}^L $ (Eq. \eqref{ZEqnNum943264}). The probability distribution of time $\tau _{g}^r =k\tau _{h} $ can be described by a discrete distribution $P\left(k,n_{h} \right)$, where \textit{k} is the number of time interval $\tau _{h} $ for a given halo to wait till the first merging with a single merger (See Eq. \eqref{ZEqnNum342616}),  
\begin{equation}
\begin{split}
&P\left(k,n_{h} \right)=P_{r} \left(X=k\right)=\frac{1}{n_{h} } \left(1-\frac{1}{n_{h} } \right)^{k-1}\\
&\textrm{with} \quad \sum _{k=1}^{\infty }P\left(k,n_{h} \right) =1.   
\label{ZEqnNum697223}
\end{split}
\end{equation}

\noindent The cumulative function and moments of the probability mass function $P\left(k,n_{h} \right)$ are given by, 
\begin{equation} 
\label{eq:58} 
Q\left(k,n_{h} \right)=\sum _{m=1}^{k}P\left(m,n_{h} \right) =1-\left(1-\frac{1}{n_{h} } \right)^{k}  
\end{equation} 
\begin{equation} 
\label{eq:59} 
\left\langle k\right\rangle =\sum _{k=1}^{\infty }\left[P\left(k,n_{h} \right)k\right] =n_{h} ,         
\end{equation} 
\begin{equation} 
\label{eq:60} 
\left\langle k^{2} \right\rangle =\sum _{k=1}^{\infty }\left[P\left(k,n_{h} \right)k^{2} \right] =n_{h} \left(2n_{h} -1\right),       
\end{equation} 
\begin{equation} 
\label{eq:61} 
\left\langle k^{m} \right\rangle =\sum _{k=1}^{\infty }\left[P\left(k,n_{h} \right)k^{m} \right] =\frac{PolyLog(-m,1-{1/n_{h} } )}{n_{h} -1} .      
\end{equation} 
The probability distribution of time interval $\tau _{g}^r $ of a given halo finally reads (from Eq. \eqref{ZEqnNum697223}), 
\begin{equation} 
\label{ZEqnNum438177} 
P\left(\tau _{gr} ,n_{h} \right)=\frac{1}{n_{h} } \left(1-\frac{1}{n_{h} } \right)^{\frac{n_{h} \tau _{g}^r }{\tau _{g} } -1} \approx \frac{1}{n_{h} } \exp \left(-\frac{\tau _{g}^r }{\tau _{g} } \right),       
\end{equation} 
with the mean $\tau _{g} \sim m_{h}^{-\lambda } $. The exponential distribution of waiting time $\tau _{g}^r $ is obtained by taking the limit $n_{h} \to \infty $. The probability density function of the continuous random waiting time $\tau _{g}^r $ reads   
\begin{equation}
P\left(\tau _{g}^r \right)=\frac{1}{\tau _{g} } \exp \left(-\frac{\tau _{g}^r }{\tau _{g} } \right).       
\label{ZEqnNum310627}
\end{equation}
\noindent Clearly, the waiting time $\tau _{g}^r $ for a given halo follows an exponential distribution that is dependent on its mean value $\tau _{g} $ (Eq. \eqref{ZEqnNum673645}). Therefore, the distribution of $\tau _{g}^r $ is position-dependent (i.e., dependent on the halo mass $m_{h}$) and scale factor \textit{a}.

\subsection{Heterogeneous diffusion model and halo mass functions}
\label{sec:4.3}
The heterogeneous diffusion with spatially dependent diffusivity plays an important role in many physical problems. Examples are mass transport in porous, inhomogeneous media, and plasmas. The transport in these examples involves the waiting time that is explicitly dependent on the position. Particularly, the power-law dependence of the diffusivity is natural for many systems exhibiting self-similarity, for example, the disordered materials, the diffusion on fractals, and the mass cascade of self-gravitating collisionless dark matter flow (SG-CFD) in this work. Here, a heterogeneous diffusion model can be established for inverse mass cascade with waiting time explicitly dependent on the halo mass ($\tau _{g} \sim m_{h}^{-\lambda } $). First, the group mass $m_{g} $ for the halo groups reads
\begin{equation}
\label{ZEqnNum539079} 
m_{g} \left(m_{h} ,a\right)=M_{h} \left(a\right)f_{M} \left(m_{h} ,a\right)m_{p} ,        
\end{equation} 
where $M_{h} $ is the total mass in the halo sub-system, $f_{M} \left(m_{h} ,a\right)$ is the probability distribution with respect to the mass of the halo $m_{h} $ (mass function) and $m_{p} $ is the mass resolution (particle mass). The time variation of $m_{g} $ has two contributions from Eq. \eqref{ZEqnNum539079},
\begin{equation} 
\label{eq:73} 
\frac{\partial m_{g} }{\partial a} =\underbrace{M_{h} \left(a\right)m_{p} \frac{\partial f_{M} }{\partial a} }_{1}+\underbrace{f_{M} \left(m_{h} ,a\right)m_{p} \frac{\partial M_{h} }{\partial a} }_{2},      
\end{equation} 
where term 1 is due to the time variation of $f_{M} \left(m_{h} ,a\right)$ and term 2 is due to the variation of total halo mass $M_{h} $. For position-dependent power-law diffusivity$D_{md} =D_{m0} m_{h}^{2\lambda } $, the dynamics of $m_{g} $ can be described by the heterogeneous diffusion model and transforming the derivative from \textit{t} to \textit{a}),
\begin{equation} 
\label{ZEqnNum612269} 
\frac{\partial m_{g} }{\partial a} =\underbrace{\frac{\partial }{\partial m_{h} } \left[\sqrt{D_{m0} } m_{h}^{\lambda } \frac{\partial }{\partial m_{h} } \left(\sqrt{D_{m0} } m_{h}^{\lambda } m_{g} \right)\right]}_{1}+\underbrace{\frac{\partial \ln M_{h} }{\partial \ln a} \frac{m_{g} }{a} }_{2},     
\end{equation} 
where term 1 describes the heterogeneous diffusion of $m_{g} $ in mass space and term 2 describes the source term due to the increasing total mass $M_{h} $ in all haloes. The boundary conditions are:
\begin{equation} 
\label{eq:75} 
\left. \frac{\partial m_{g} }{\partial a} \right|_{m_{h} =0} =\frac{m_{p} }{Ha} \left. T_{m} \right|_{m_{h} =0} =0,         
\end{equation} 
\begin{equation} 
\label{ZEqnNum405128} 
-\frac{1}{m_{p} } \frac{\partial }{\partial a} \int _{0}^{\infty }m_{g} \left(m_{h} ,a\right)dm_{h} \equiv \frac{\varepsilon _{m} }{Ha} =-\frac{\partial M_{h} }{\partial a} .     
\end{equation} 
The governing equation for the mass function $f_{M} \left(m_{h},a\right)$ can eventually be found with the substitution of Eq. \eqref{ZEqnNum539079} into Eq. \eqref{ZEqnNum612269},
\begin{equation} 
\label{ZEqnNum619234} 
\frac{\partial f_{M} \left(m_{h} ,a\right)}{\partial a} =D_{m0} \frac{\partial }{\partial m_{h} } \left[m_{h}^{\lambda } \frac{\partial }{\partial m_{h} } \left(m_{h}^{\lambda } f_{M} \right)\right].       
\end{equation} 
The solution to Eq. \eqref{ZEqnNum619234} is a stretched Gaussian function that has an exponential cut-off for a large halo mass $m_{h} $ and a power-law behavior for small $m_{h} $,
\begin{equation}
\label{ZEqnNum162907} 
f_{M} \left(m_{h} ,a\right)=\frac{m_{h}^{-\lambda } }{\sqrt{\pi D_{m0} a} } \exp \left[-\frac{m_{h}^{2-2\lambda } }{\left(2-2\lambda \right)^{2} D_{m0} a} \right].      
\end{equation} 
The mean square displacement of $m_{h} $ in the mass space can be defined 
\begin{equation} 
\label{ZEqnNum433658} 
\begin{split}
\left\langle m_{h}^{2} \right\rangle &=\int _{0}^{\infty }f_{M} \left(m_{h} ,a\right) m_{h}^{2} dm_{h}\\
&=\frac{1}{\sqrt{\pi } } \Gamma \left(\frac{3-\lambda }{2-2\lambda } \right)\left(2-2\lambda \right)^{\frac{2}{1-\lambda } } \left(D_{m0} a\right)^{\frac{1}{1-\lambda } } \equiv \gamma _{0} m_{h}^{*2}.
\end{split}
\end{equation} 
where $m_{h}^{*} $ is the characteristic mass scale and $\gamma _{0} $ is a proportional constant. With the exponent of ${1/\left(1-\lambda \right)} \ge 1$ in Eq. \eqref{ZEqnNum433658}, diffusion in mass space is of a super-diffusion nature. The solution of $f_{M} \left(m_{h} ,a\right)$ (Eq. \eqref{ZEqnNum162907}) can be expressed in terms of $m_{h}^{*} $,
\begin{equation}
\label{ZEqnNum663085} 
f_{M} \left(m_{h} ,a\right)=\frac{\left(1-\lambda \right)}{\sqrt{\pi \eta _{0} } } \left(\frac{m_{h}^{*} }{m_{h} } \right)^{\lambda } \frac{1}{m_{h}^{*} } \exp \left[-\frac{1}{4\eta _{0} } \left(\frac{m_{h} }{m_{h}^{*} } \right)^{2-2\lambda } \right],     
\end{equation} 
where the dimensionless constant $\eta _{0} $ 
\begin{equation} 
\label{eq:81} 
\eta _{0} =\frac{1}{4} \left[\frac{\gamma _{0} \sqrt{\pi } }{\Gamma \left({\left(3-\lambda \right)/\left(2-2\lambda \right)} \right)} \right]^{1-\lambda } .        
\end{equation} 
For $\lambda ={2/3} $ and $\gamma _{0} =15$, we have $\eta _{0} =0.5$. We also found the constant $\beta _{0} $ in the mass propagation range (in Eq. \eqref{ZEqnNum972525}) as,
\begin{equation}
\label{ZEqnNum991280} 
\beta _{0} ={\left(1-\lambda \right)/\sqrt{\pi \eta _{0} } } \approx 0.266.          
\end{equation} 
Since $m_{h}^{*} $ is related to the mean square displacement of the halo mass, the diffusivity in the halo mass space can be expressed in terms of $m_{h}^{*} $,
\begin{equation} 
\label{eq:83} 
D_{m0} =\frac{\eta _{0} m_{h}^{*\left(2-2\lambda \right)} }{\left(1-\lambda \right)^{2} a} \sim a^{2-2\tau _{0} } .         
\end{equation} 
For a constant diffusivity $D_{m0} $ with respect to \textit{a}, we would also expect $\tau _{0} =1$, i.e. the fundamental frequency $f_{0} 
\left(a\right)\sim a^{-1} $. 

The \textit{k}th moments of the mass function $f_{M} $ can be obtained as, 
\begin{equation} 
\label{eq:85} 
\begin{split}
\int _{0}^{\infty }f_{M} \left(m_{h} ,a\right)&\left(m_{h} \right)^{k} dm_{h}\\
&=\frac{1}{\sqrt{\pi } } \left(4\eta _{0} \right)^{\frac{k}{\left(2-2\lambda \right)} } \Gamma \left(\frac{k-\lambda +1}{2-2\lambda } \right)\left(m_{h}^{*} \right)^{k},
\end{split}
\end{equation} 
where, in particular, the mean halo mass is proportional to the characteristic mass scale $m_{h}^{*} $,
\begin{equation} 
\label{ZEqnNum604996} 
\left\langle m_{h} \right\rangle =\frac{1}{\sqrt{\pi } } \left(4\eta _{0} \right)^{\frac{1}{\left(2-2\lambda \right)} } \Gamma \left(\frac{2-\lambda }{2-2\lambda } \right)m_{h}^{*} .        
\end{equation} 
Finally, solution of the group mass $m_{g} \left(m_{h} ,a\right)$ is (from Eq. \eqref{ZEqnNum539079}),
\begin{equation} 
\label{eq:87} 
m_{g}=M_{h} \left(a\right)\frac{\left(1-\lambda \right)}{\sqrt{\pi \eta _{0} } } \left(\frac{m_{h}^{*} }{m_{h} } \right)^{\lambda } \frac{m_{p} }{m_{h}^{*} } \exp \left[-\frac{1}{4\eta _{0} } \left(\frac{m_{h} }{m_{h}^{*} } \right)^{2-2\lambda } \right].    
\end{equation} 
The mass flux and transfer functions can be obtained from the definitions (Eqs. \eqref{ZEqnNum400994}, \eqref{ZEqnNum980409}) in Section \ref{sec:A1.3}. The mass transfer function reads 
\begin{equation} 
\label{ZEqnNum622939} 
T_{m}=\frac{1-\lambda }{2\eta _{0} \sqrt{\pi \eta _{0} } } \frac{1}{\tau _{h}^{*} \left(a\right)} \left(\frac{m_{h} }{m_{h}^{*} } \right)^{2-3\lambda } \exp \left[-\frac{1}{4\eta _{0} } \left(\frac{m_{h} }{m_{h}^{*} } \right)^{2-2\lambda } \right].    
\end{equation} 
Here $T_{m} \left(m_{h} ,a\right)\propto m_{g} \left({m_{h} /m_{p} } \right)^{2-2\lambda } $ is a typical feature of heterogeneous diffusion. The mass flux function is finally given by
\begin{equation} 
\label{ZEqnNum632692} 
\begin{split}
\Pi _{m} \left(m_{h} ,a\right)&=-\frac{m_{h}^{*} \left(a\right)}{\tau _{h}^{*} \left(a\right)} \left\{\underbrace{\textrm{erfc}\left[\frac{1}{2\sqrt{\eta _{0} } } \left(\frac{m_{h} }{m_{h}^{*} } \right)^{1-\lambda } \right]}_{I}\right.\\
&\left.+\underbrace{\frac{1}{\sqrt{\eta _{0} \pi } } \left(\frac{m_{h} }{m_{h}^{*} } \right)^{1-\lambda } \exp \left[-\frac{1}{4\eta _{0} } \left(\frac{m_{h} }{m_{h}^{*} } \right)^{2-2\lambda } \right]}_{II}\right\},
\end{split}
\end{equation} 
With term I (complementary error function) dominating for small $m_{h} $ and term II (exponential function) dominating for large $m_{h} $. Additionally, $\Pi _{m} \left(m_{h} ,a\right)=\varepsilon _{m} $ with $m_{h} \to 0$ satisfies the boundary conditions (Eq. \eqref{ZEqnNum405128}). 

\begin{figure}
\includegraphics*[width=\columnwidth]{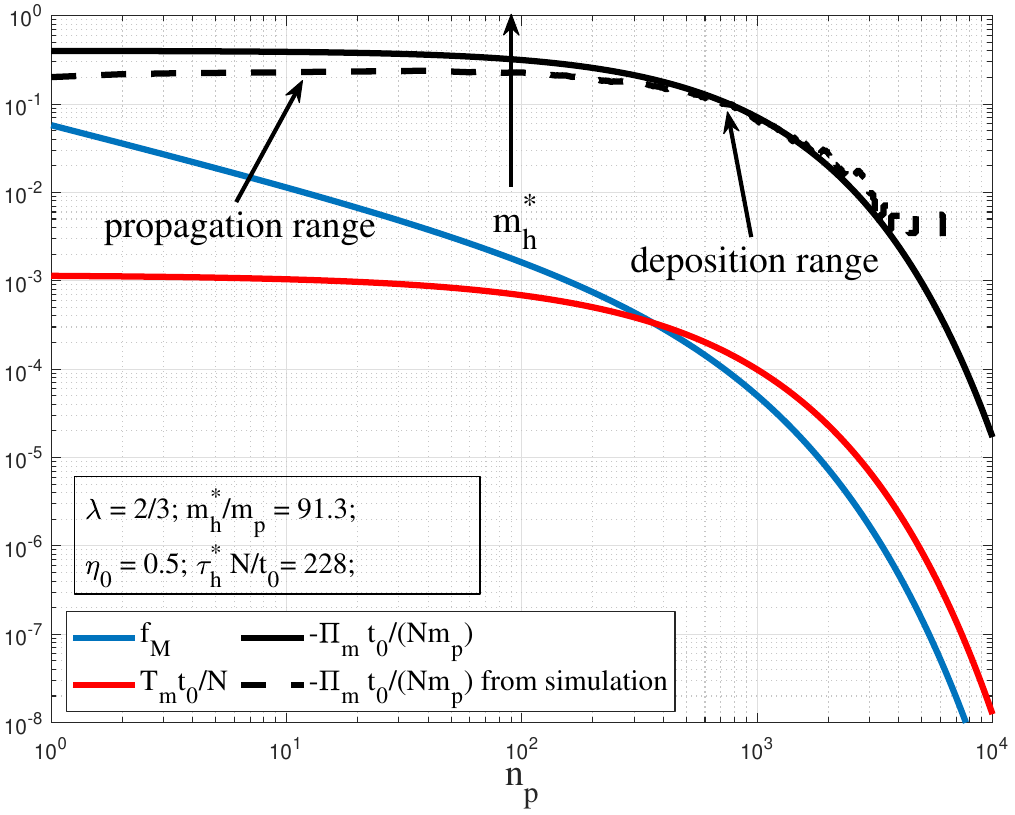}
\caption{The variation of halo mass function $f_{M} $ (Eq. \eqref{ZEqnNum663085}), mass transfer function $\Pi _{m} $ (Eq. \eqref{ZEqnNum632692}) and mass flux function $T_{m} $ (Eq. \eqref{ZEqnNum622939}) with halo size $n_{p} ={m_{h} /m_{p} } $ for a given set of parameters at the current physical time $t_{0} $. Two distinct ranges can be clearly identified: a mass propagation range with a constant mass flux $\varepsilon _{m} =\Pi _{m} $ for $m_{h} <m_{h}^{*} $ and a mass deposition range for $m_{h} >m_{h}^{*} $. For comparison, the mass transfer function $\Pi _{m} $ obtained using N-body simulation data at two different redshifts, $z=0$ and $z=0.1$, is also presented in the same plot as the dashed line.}
\label{fig:6}
\end{figure}

Figure \ref{fig:6} plots the variation of the halo mass function $f_{M} $ (Eq. \eqref{ZEqnNum663085}), the mass transfer function $-\Pi _{m} $ (Eq. \eqref{ZEqnNum632692}) and the mass flux function $T_{m} $ (Eq. \eqref{ZEqnNum622939}) with the halo size $n_{p} $ for a given set of parameters at present $t_{0} $. Two distinct ranges can be clearly identified from Fig. \ref{fig:6}: the mass propagation range with a constant mass flux $\varepsilon _{m} =\Pi _{m} $ for $m_{h} <m_{h}^{*} $ and the mass deposition range for $m_{h} >m_{h}^{*} $.

\subsection{Existing halo mass functions}
\label{sec:A3}
The abundance of haloes, i.e., halo mass function $f_{M} $, is one of the most fundamental quantities for analytically or semi-analytically modeling of structure formation and evolution. The Press-Schechter (PS) formalism is one of the first landmarks on the halo mass function \citep{Press:1974-Formation-of-Galaxies-and-Clus,Bower:1991-The-Evolution-of-Groups-of-Gal} that can be used to predict the shape and evolution of the halo mass distribution, 
\begin{equation} 
\label{ZEqnNum297590} 
\begin{split}
f_{PS} \left(m_{h} \right)&=\frac{1}{\sqrt{2\pi } } \left(1+\frac{n_{ps} }{3} \right)\frac{1}{m_{h} }\\
&\cdot \left(m_{h}/m_{h}^{*} \right)^{\frac{\left(3+n_{ps} \right)}{6} }\exp \left[-\frac{1}{2} \left(m_{h}/m_{h}^{*}\right)^{\frac{\left(3+n_{ps} \right)}{3}} \right], 
\end{split}
\end{equation} 
where $n_{ps} $ is the effective index of the power spectrum of density fluctuation. A normalized dimensionless variable $v$ can be introduced to simplify the halo mass function, 
\begin{equation} 
\label{ZEqnNum549327-2} 
v=\frac{\delta _{c}^{2} \left(a\right)}{\sigma _{\delta }^{2} \left(m_{h} \right)} =\left(\frac{m_{h} }{m_{h}^{*} \left(a\right)} \right)^{1+\frac{n_{ps} }{3} } =\left[\frac{\sigma _{v}^{2} \left(m_{h} ,a\right)}{\sigma _{v}^{2} \left(m_{h}^{*} \right)} \right]^{\frac{3+n_{ps} }{3+n} } ,      
\end{equation} 
where $\delta _{c}^{} \left(a\right)\sim a^{-1} $ is the critical density that has to be determined from a spherical collapse model or a two-body collapse model \citep{Xu:2021-A-non-radial-two-body-collapse}. Here $\sigma _{\delta }^{2} \left(m_{h} \right)$ is the variance of the initial density fluctuation when smoothed with a tophat filter of size $R=\left({3m_{h} /4\pi \bar{\rho }_{0} } \right)^{{1/3} } $. Here $\bar{\rho }_{0} $ is the physical background density at the current epoch $a=1$. The term $\sigma _{v}^{2} $ is the halo virial velocity dispersion. The second equality in Eq. \eqref{ZEqnNum549327-2} comes from the linear theory prediction of $\sigma _{\delta }^{2} \left(m_{h} \right)\sim m_{h}^{-\left(1+{n_{ps} /3} \right)} $ for a power-law power spectrum with an effective index of $n_{ps} $. The third equality in Eq. \eqref{ZEqnNum549327-2} comes from the virial theorem for haloes of mass $m_{h} $. Here $\sigma _{v}^{2} \left(m_{h} \right)\propto {Gm_{h} /r_{h}^{{}^{-n} } } \propto m_{h}^{{1+n/3} }$. The parameter $n$ is the exponent of gravitational potential $V_{p} \left(r\right)\sim r^{n} $. Since $\delta _{c}^{} \left(a\right)\sim a^{-1} $ from the spherical collapse model, linear theory predicts that $\sigma _{v}^{2} \left(m_{h} ,a\right)\sim a^{-1} m_{h}^{1+{n_{ps} /3} } $, $\sigma _{v}^{2} \left(m_{h}^{*} \right)\sim a$ , $m_{h}^{*} \sim a^{{6/\left(3+n_{ps} \right)} } $ and $v\sim a^{-2} m_{h}^{1+{n_{ps} /3} } $.

With the dimensionless variable $v$ introduced in Eq. \eqref{ZEqnNum549327-2}, the equivalent dimensionless PS mass function in Eq. \eqref{ZEqnNum297590} is 
\begin{equation} 
\label{ZEqnNum197880-2} 
f_{PS} \left(\nu \right)=\frac{1}{\sqrt{2\pi } } \nu ^{-{1/2} } \exp \left(-\frac{\nu }{2} \right).         
\end{equation} 

Further improvement was achieved by extending the PS formalism with the elliptical collapse model \citep{Sheth:2001-Ellipsoidal-collapse-and-an-im,Sheth:1999-Large-scale-bias-and-the-peak-}. The modified PS model (ST model, hereafter ST) is:
\begin{equation} 
\label{eq:93-2} 
f_{ST} \left(\nu \right)=\frac{\left(1+{1}/{\left(q\nu \right)^{p}} \right)\sqrt{2q} }{\Gamma \left({1/2} \right)+2^{-p} \Gamma \left({1/2} -p\right)}\frac{1}{2\sqrt{\nu } } e^{-{q\nu/2} } .     
\end{equation} 
The best-fitted parameters from large-scale \textit{N}-body simulations are $q=0.75$ and $p=0.3$ \citep{Sheth:2002-An-excursion-set-model-of-hier}. Many other forms of empirical mass functions were also proposed by fitting to the high-resolution simulation data. For example, a universal JK mass function was proposed to cover a wide range of simulation data with different cosmologies and redshifts \citep{Jenkins:2001-The-mass-function-of-dark-matt},
\begin{equation} 
\label{eq:94-2} 
f_{JK} \left(\nu \right)=\frac{0.315}{2\nu } \exp [-\left|\ln \left({\sqrt{v} /\delta _{c} } \right)+0.61\right|^{3.8} ] .       
\end{equation} 
With $\delta _{c} =1.6865$ at $z=0$. It should be noted that the empirical mass function does not satisfy the normalization constraint (Integral of mass function in mass space should give unity) and cannot extrapolate beyond the range of fitting data. 

Now, let us return to our halo mass function from the inverse mass cascade (Eq. \eqref{ZEqnNum663085}), which does not rely on any particular collapse model (spherical or elliptical). If we introduce $v=\left({m_{h} /m_{h}^{*} } \right)^{2-2\lambda } $, the halo mass function of Eq. \eqref{ZEqnNum663085} can be simplified to 
\begin{equation} 
\label{ZEqnNum170065-2} 
f_{\nu } \left(\nu \right)=\frac{1}{2\sqrt{\pi \eta _{0} } } \nu ^{-{1/2} } \exp \left[-\frac{\nu }{4\eta _{0} } \right].   
\end{equation} 
Clearly, Eq. \eqref{ZEqnNum663085} reduces to the Press-Schechter (PS) mass function if $\lambda ={\left(3-n_{ps} \right)/6} $ and $\eta _{0} ={1/2} $ (See Eq. \eqref{ZEqnNum549327-2}). However, it should be noted that Eq. \eqref{ZEqnNum663085} is more general and the parameter $\eta _{0} $ is related to the parameter $\beta _{0} $ (Eq. \eqref{ZEqnNum991280}), where $\beta _{0} $ is the prefactor for power-law scaling in Eq. \eqref{ZEqnNum972525}. The halo geometry exponent $\lambda $ has a fundamental connection to the effective index of the power spectrum $n_{ps} $. In principle, the halo geometry exponent $\lambda $ should be smaller than one, such that $n_{ps} >-3$. 

A universal halo mass function like Eq. \eqref{ZEqnNum170065-2} is clearly a manifestation of a statistically steady state involving a mass and energy cascade with scale-independent rates. All these results provide insights into a fundamental question: how does the non-equilibrium system maximize its entropy and approach the limiting equilibrium via a cascade process? Two typical examples are, of course, the hydrodynamic turbulence and SG-CFD (self-gravitating collisionless dark matter flow). Both examples exhibit an intermediate and statistically steady state, where a direct energy (or inverse mass) cascade process is well established.

\label{lastpage}
\end{document}